\documentclass[
    aps,pre,
    twocolumn,
    reprint,
    superscriptaddress,
    nofootinbib,
    floatfix,
    amssymb,
    long bibliography
]{revtex4-1}

\usepackage[utf8]{inputenc}
\usepackage{amsmath}
\usepackage{amsfonts}
\usepackage{array}
\usepackage{graphicx}
\usepackage{bm}
\usepackage{dcolumn}
\usepackage{overpic}
\usepackage{mathtools}
\usepackage{dsfont}
\usepackage{nicefrac}
\usepackage[normalem]{ulem}
\usepackage{xcolor}
\definecolor{linkColor}{rgb}{0,0.3,0.7}
\usepackage{hyperref}
\hypersetup{colorlinks=true,
            allcolors=linkColor,
            pdfborder={0 0 0},
            pdfencoding = auto
            }

\definecolor{myGreen}{RGB}{19,132,23}

\begin{document}
\title{Non-equilibrium phase coexistence in conserved chemically active mixtures}

\author{Florian Raßhofer}
\affiliation{Arnold Sommerfeld Center for Theoretical Physics and Center for NanoScience, Department of Physics, Ludwig-Maximilians-Universit\"at M\"unchen, Theresienstra\ss e 37, D-80333 Munich, Germany}

\author{Erwin Frey}
\email[Corresponding author: ]{frey@lmu.de}
\affiliation{Arnold Sommerfeld Center for Theoretical Physics and Center for NanoScience, Department of Physics, Ludwig-Maximilians-Universit\"at M\"unchen, Theresienstra\ss e 37, D-80333 Munich, Germany}
\affiliation{Max Planck School Matter to Life, Hofgartenstraße 8, D-80539 Munich, Germany}

\date{\today}


\begin{abstract}
Chemical activity is known to affect phase coexistence and coarsening in liquid mixtures, most commonly through reaction-induced changes of intermolecular interactions.
Here, we analyze a scenario in which chemical reactions regulate particle transport while leaving thermodynamic interactions unchanged.
We study an incompressible mixture of thermodynamically identical solutes with unequal diffusivities that interconvert through driven chemical reactions.
Using linear stability analysis and finite-element simulations, we show that the system can phase-separate into solute-rich and solute-poor domains via two qualitatively different pathways.
When interactions are too weak to induce phase separation, patterns arise through a generalized mass-redistribution instability and coarsen uninterruptedly.
When interactions favor phase separation, coarsening can be arrested if chemical activity locally enriches faster-diffusing solutes within dense domains.
In the limit of fast chemical turnover, the system always coarsens, and phase coexistence is governed by an effective free energy that explicitly depends on kinetic parameters.
Beyond this limit, we develop a sharp-interface theory that predicts the onset of arrested coarsening, stationary droplet sizes, and nucleation conditions under chemical driving.
Taken together, our results establish kinetic regulation as a minimal and robust mechanism to control phase coexistence and coarsening in chemically active mixtures.
\end{abstract}

\maketitle
Phase separation is a ubiquitous mechanism of spatial self-organization across physical\@~\cite{Chen.2002}, biological\@~\cite{Hyman.2014,Banani.2017}, and even ecological systems\@~\cite{Siteur.2023,Koppel.2008}.
Whenever local interactions or transport processes favor aggregation over mixing, extended systems can spontaneously segregate into coexisting dense and dilute regions.
Such pattern formation appears in a wide variety of contexts, ranging from intracellular liquid-like compartments\@~\cite{Brangwynne.2009,Feric.2016}, to aggregating bacteria\@~\cite{Ben-Jacob.2000} and synthetic particle assemblies\@~\cite{Liebchen.2018,Palacci.2013}.

In thermodynamic equilibrium systems, phase separation of multi-component mixtures is a well-understood process.
It arises when the energetic gain associated with demixing into coexisting phases of different composition outweighs the corresponding loss of mixing entropy\@~\cite{Doi.2013}.
In the absence of long-range interactions\@~\cite{Liu.1989,Muratov.2002,Kumar.2023,Winter.2025} or an externally imposed length scale, e.g., due to confinement in an elastic medium\@~\cite{Style.2018,Zhang.2021,Vidal-Henriquez.2021}, phase-separated domains do not exhibit a characteristic size.
Instead, they undergo continual \emph{coarsening}, whereby larger domains grow at the expense of smaller ones through diffusive mass transport.
This process, known as \textit{Ostwald ripening}\@~\cite{Ostwald.1897,Wagner.1961,Lifshitz.1961}, is driven by interfacial tension, which favors the reduction of total interfacial area and ultimately leads to complete macroscopic phase separation.

Beyond thermodynamic equilibrium, phase separation also arises in a wide range of driven and active systems, where it is sustained or modified by continuous energy consumption.
Prominent examples include biomolecular condensates in living cells\@~\cite{Weber.2019,Zwicker.2025}, motility-induced phase separation (MIPS)\@~\cite{Schnitzer.1993,Tailleur.2008,Cates.2014}, and chemotactic bacterial populations\@~\cite{Murray.2003,Weyer.2025}.
Closely related phenomena are even found in mass-conserving reaction-diffusion systems (mcRD)\@~\cite{Turing.1952,Halatek.2018b,Frey.2022}, which, despite being dilute, exhibit a deep formal analogy to classical phase separation dynamics\@~\cite{Brauns.2020,Brauns.2021}.

In contrast to equilibrium phase separation, such non-equilibrium systems exhibit a far richer phenomenology, including stationary patterns\@~\cite{Glotzer.1994,Carati.1997,Cates.2010,Zwicker.2015,Wurtz.2018,Donau.2023,Tjhung.2018}, morphological instabilities\@~\cite{Zwicker.2017,Rasshofer.2025}, traveling structures\@~\cite{Haefner.2024,Demarchi.2023,Jambon-Puillet.2024,Agudo-Canalejo.2019,Saha.2020}, and non-equilibrium steady states that cannot be described by a free-energy minimization principle\@~\cite{Bauermann.2023}.

Understanding how activity modifies, arrests, or replaces classical coarsening dynamics has, therefore, become a central challenge in non-equilibrium pattern formation.
In this context, it is useful to distinguish systems in which phase separation originates from thermodynamic interactions and is subsequently modified by active processes, from systems in which phase separation itself is an intrinsically active phenomenon driven by non-equilibrium transport phenomena.

For example, biomolecular condensates are commonly thought to form through equilibrium liquid–liquid phase separation\@~\cite{Brangwynne.2009,Boeynaems.2018,Fritsch.2021,Alberti.2019l6i}.
However, in living cells they persist under conditions far from thermodynamic equilibrium, owing to the presence of continuous energy-consuming processes\@~\cite{Hondele.2020}.
In this context, considerable attention has been given to enzymatic post-translational modifications of condensate constituents\@~\cite{Rai.2018,Hofweber.2019,Owen.2019,Snead.2019}.
Such reactions can alter intermolecular interactions, thereby actively tuning the propensity of molecules to phase separate.
This mechanism is thought to provide cells with dynamic control over condensate size, composition, and lifetime\@~\cite{Soeding.2020,Wurtz.2018l4k}.

By contrast, in intrinsically active systems phase separation is driven by non-equilibrium processes themselves.
For example, motility-induced phase separation arises due to a feedback between local density and motility suppression\@~\cite{Cates.2014,Schnitzer.1993}.
Similar density–transport feedback arises in quorum-sensing bacterial populations\@~\cite{Fu.2012}, chemotactic systems\@~\cite{Keller.1970}, and mass-conserving reaction–diffusion models\@~\cite{Frey.2022,Brauns.2020}, where chemical signaling or reaction kinetics regulate effective mass transport.
In these systems, phase separation does not originate from short-range interactions alone, but from activity-controlled coupling between density and transport.

In Ref.\@~\cite{prl}, we introduced a system that interpolates between these two limiting classes of non-equilibrium phase separation. 
We proposed a minimal model in which chemical activity alters only the kinetic transport properties of an interacting mixture, while leaving the thermodynamic driving forces for phase separation unchanged. 
The model describes an incompressible ternary mixture with two thermodynamically identical solute species that differ only in their diffusivity and are coupled by driven chemical interconversion. 
Within this framework, we identified two qualitatively distinct regimes of pattern formation. 
When interactions are too weak to induce phase separation, patterns arise via a reaction driven mass-redistribution instability~\cite{Brauns.2020,Frey.2022} and coarsen into a single phase-separated domain.
By contrast, when interactions favor phase separation, chemical activity can arrest coarsening by locally enriching the faster-diffusing species within dense domains.

In the present work, we develop a comprehensive theoretical framework that underlies and extends the results of Ref.\@~\cite{prl}. 
We generalize the theory to mixtures with an arbitrary number of components and formulate a systematic description of the resulting dynamics. 
We analyze the onset of pattern formation through a linear stability analysis of the homogeneous steady state, providing the full derivation of the instability criteria previously summarized in Ref.\@~\cite{prl}. 
Beyond the linear regime, we study phase coexistence in the limit of macroscopically large domains and show that, within the coarsening regime, the coexisting phases are governed by an effective free-energy functional that depends explicitly on the kinetic transport parameters. 
Finally, to address the arrested regime, we perform sharp-interface calculations that predict the onset of arrested coarsening, determine the stationary droplet size, and describe droplet nucleation under chemical driving.

This work is organized as follows.
In Sec.\@~\ref{sec:Model}, we introduce the model and derive the governing field equations.
In Sec.\@~\ref{sec:LSA}, we perform a linear stability analysis of the homogeneous steady state to determine the conditions under which patterns form.
Building on this, in Sec.\@~\ref{sec:Coexistence}, we investigate phase coexistence in the limit of macroscopically large domains. 
Using adiabatic elimination of the composition fields, we show that the dynamics can be mapped onto a purely relaxational form governed by an effective free energy that explicitly depends on the kinetic parameters of the system.
In Sec.\@~\ref{sec:singleDroplet}, we develop a sharp-interface theory to identify the conditions under which coarsening is arrested and to characterize the resulting steady states.
Finally, in Sec.\@~\ref{sec:Summary}, we summarize our findings and discuss them in the context of the existing literature.

\section{Model}
\label{sec:Model}
\begin{figure}
    \centering
    \includegraphics[]{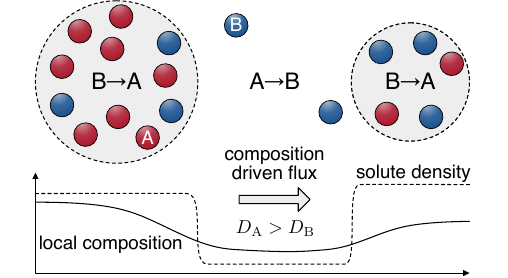}
    \caption{\textbf{Chemically active mixture}. A mixture consisting of two solutes,  A (red) and B (blue), and an inert solvent (not shown) phase separates into solute-rich (gray) and solute-poor domains (white). The solutes are thermodynamically identical but differ in their diffusivities (${D_\mathrm{A}>D_\mathrm{B}}$).
    Density-dependent chemical reactions convert A and B (small arrows), leading to preferential enrichment of the faster-diffusing species A within solute-rich domains.
    Bottom panel:
    Since chemical conversion competes with diffusive transport, the relative composition of A and B inside a droplet depends on its size, with smaller droplets remaining closer to the background composition.
    The resulting composition gradients generate non-equilibrium mass fluxes that redistribute solute from regions of high to low effective mobility, i.e., from larger to smaller droplets.
}
    \label{fig:model}
\end{figure}
In this section, we derive the dimensionless dynamical equations used throughout this work.
Starting from the Flory–Huggins free energy of a multicomponent mixture, we use linear non-equilibrium thermodynamics to obtain the mass-conserving dynamics, which we extend to include slow chemical conversion between solute species.
Specializing to a ternary mixture, which serves as a minimal model, we then reformulate the dynamics in terms of the total solute volume fraction and local composition fields.
\subsection{Flory-Huggins free energy functional}
\label{sec:freeEnergy}
We consider an isothermal system of volume $V$ which contains an incompressible fluid mixture consisting of ${N-1}$ solutes, and a solvent S. Each component has a fixed particle number~$N_i$.
For simplicity, we assume that each component ${i\in \{1,\dots,N\}}$ occupies the same molecular volume $\nu$.
For a well-mixed system, the Flory-Huggins free energy density is then given by\@~\cite{Berry.2018,Zwicker.2025,Mao.2018}
\begin{align}
    f(\{\phi_i\}) = 
    \frac{k_\text{B} T}{\nu} \bigg(
    &\sum_{i=1}^N \phi_i\log(\phi_i)
    + \sum_{i=1}^N w_i \phi_i \notag \\
    &+\frac{1}{2}\sum_{i,j=1}^N \varepsilon_{ij}\phi_i\phi_j
    \bigg) \, ,
    \label{app:FreeEnergyDensity}
\end{align}
where ${\phi_i=\nu N_i/V}$ denotes the volume fraction of component $i$.
These volume fractions satisfy the incompressibility constraint 
\begin{align}
    \sum_{i=1}^N \phi_i = 1 \, .
    \label{eq:Incompressibility}
\end{align}
The first term in Eq.\@~\eqref{app:FreeEnergyDensity} accounts for the mixing entropy, and the second and third terms represent enthalpic contributions from single-particle offsets $w_i$ and pairwise interactions $\varepsilon_{ij}$, both in units of the thermal energy scale $k_\text{B} T$ and the molecular volume\@~$\nu$.

To describe spatially inhomogeneous states, we consider local volume fractions\@~$\phi_i(\boldsymbol{r},t)$ rather than global averages. 
Following the approach of Cahn and Hilliard\@~\cite{Cahn.1958}, we treat $f(\{\phi_i\})$ as a local free energy density and incorporate short-range interactions via a gradient expansion, yielding the free energy functional
\begin{align}
    \mathcal{F}\left[\{\phi_i\}\right] =
    \int_V\mathrm{d} \boldsymbol{r} \left(
    \frac{k_\text{B} T}{\nu}\sum_{i,j} \frac{\tilde \kappa_{ij}}{2}\boldsymbol{\nabla}\phi_i\boldsymbol{\nabla}\phi_j
    + f(\{\phi_i\}) \right) \, ,
    \label{app:FreeEnergy}
\end{align}
where the phenomenological parameters $\tilde \kappa_{ij}$ quantify the energetic cost of spatial gradients.

Our goal is to analyze a situation in which chemical reactions selectively modify transport properties without affecting local interactions.
To isolate this kinetic effect, we assume that all solutes are thermodynamically identical, implying
\begin{align}
    w_i = w_j\,, \qquad \forall \,i,j< N \, .
\end{align}
Since only differences in internal energy are physically meaningful, we set ${w_i = 0}$ without loss of generality. 
Moreover, we only consider three types of interactions: solute-solute\@~(PP), solute-solvent\@~(SP), and solvent-solvent\@~(SS).
Accordingly, we define the interaction and stiffness parameters as follows: 
\begin{align}
    \varepsilon_{ij}, \tilde \kappa_{ij} &\;=\; 
    \begin{cases}
        \varepsilon_\mathrm{PP}, \, \tilde \kappa_\mathrm{PP}\,, \  &\text{if} \quad i,j < N \,,\\
        \varepsilon_\mathrm{SS}, \, \tilde \kappa_\mathrm{SS}\, , \  &\text{if} \quad i=j=N \,,\\
        \varepsilon_\mathrm{SP}, \, \tilde \kappa_\mathrm{SP}\,, \ &\text{else.} 
    \end{cases} 
\end{align}

We proceed by introducing the \emph{total solute volume fraction}
\begin{align}
    \rho(\boldsymbol{r},t) = \sum_{i=1}^{N-1}\phi_i \, , \qquad
\end{align}
and the local composition variables ${s_i(\boldsymbol{r},t)=\phi_i/\rho}$.
Note that although $\rho$ formally defines a volume fraction, we will often refer to it as a density, since the two differ only by a constant factor $1/\nu$.
Up to irrelevant linear terms, which do not enter the dynamics, the free energy of mixing is then given by
\begin{align}
    \mathcal{F}\left[\{\phi_i\}\right] = \frac{k_\text{B}T}{\nu}
    \int_V &\mathrm{d}\boldsymbol{r} \,\bigg(
    \frac{\tilde\kappa}{2}\left(\boldsymbol{\nabla}\rho\right)^2 
    + f(\rho) \notag \\
    &+ \rho \sum_{i=1}^{N-1}s_i\log(s_i) \bigg)\,,
    \label{app:FreeEnergy2}
\end{align}
where
\begin{align}
    f(\rho) &\equiv \rho\log(\rho)+(1-\rho)\log(1-\rho)+\chi\,\rho  (1-\rho) \label{app:FHdensity}\, ,
\end{align}
and we introduced the effective parameters 
\begin{subequations}
\begin{align}
    \chi &= (2\varepsilon_\mathrm{SP}-\varepsilon_\mathrm{PP}-\varepsilon_\mathrm{SS})/2 \, , \label{app:FloryHugginsParameter} \\[3mm]
    \tilde \kappa &= (\tilde\kappa_\mathrm{PP}+\tilde\kappa_\mathrm{SS}-2\tilde\kappa_\mathrm{SP}) \,.
\end{align}
\end{subequations}

The first two terms in Eq.\@~\eqref{app:FreeEnergy2} correspond to the free energy of a \emph{symmetric binary mixture} where we do not distinguish between the different solutes:
Density gradients are penalized by the stiffness coefficient\@~$\tilde\kappa$, and the local free energy density ${f(\rho)}$ includes entropic and energetic contributions, with $\chi$ denoting the Flory–Huggins parameter\@~\cite{Rubinstein.2003}.
Throughout this manuscript, we relate the stiffness and interaction parameter via ${\tilde \kappa = \text{max}(2\chi \nu^{2/d}/z,0)}$, which is obtained from coarse-graining a lattice model of equally sized particles with lattice spacing $\nu^{1/d}$ and coordination number~$z$ (App.\@~\ref{sec:coarseGraining}).
For ${\chi>2}$, the function $f(\rho)$ loses convexity and develops two local minima $\rho_\pm$, indicating that the system's free energy can be minimized by phase separating into a solute-rich phase (${\rho\approx \rho_+}$) and a solute-poor phase\@~(${\rho\approx \rho_-}$).
The last term in Eq.\@~\eqref{app:FreeEnergy2} accounts for the entropy of mixing between the different solute species, weighted by the local solute volume fraction $\rho$.
It is convex for all parameter choices, indicating thermodynamic stability against compositional fluctuations.
\subsection{Mass-conserving relaxational dynamics}
We describe phase separation using the mass-conserving gradient-flow dynamics for an incompressible multicomponent mixture~\cite{Groot.1984,Onuki.2002}:
\begin{align}
    \partial_t\phi_i(\boldsymbol{r},t) =
    \boldsymbol{\nabla} \cdot\sum_{j =1}^N M_{ij} \boldsymbol{\nabla}\mu_j
    \, , \quad
    \mu_j = \nu \, \frac{\delta \mathcal{F}}{\delta\phi_j} \, ,
    \label{app:DynamicsStart}
\end{align}
where $\mu_i$ denotes the chemical potential of species $i$, with ${i \in \{1,\dots,N\}}$, and $M_{ij}(\{\phi_i\})$ is a mobility matrix that generally depends on the local volume fractions.
To ensure thermodynamic consistency, the mobility matrix must satisfy the Onsager reciprocity relation ${M_{ij} = M_{ji}}$ and be positive semi-definite to ensure a positive dissipation rate ${\sum_{i,j} M_{ij} \,(\boldsymbol\nabla \mu_i) \cdot  (\boldsymbol\nabla \mu_j)}$\@~\cite{Onsager.1930,Onsager.1931}. 
The incompressibility constraint, Eq.\@~\eqref{eq:Incompressibility}, further implies that the mobility matrix must satisfy the null-space condition 
\begin{align}
    \sum_{i = 1}^N M_{ij} = 0 \, ,
    \label{eq:NullSpace}
\end{align}
which fixes the diagonal elements to ${M_{jj} = -\sum_{i \ne j} M_{ij}}$.

To describe the independent dynamics of the local solute volume fractions, it is convenient to introduce the \emph{exchange chemical potentials} relative to the solvent:
\begin{align}
    \bar{\mu}_i(\boldsymbol{r},t) \equiv \mu_i - \mu_N, \qquad i \in \{1,\dots,N-1\} \, .
\end{align}
Using this definition, the chemical potential gradients can be decomposed as ${\boldsymbol{\nabla} \mu_j = \boldsymbol{\nabla} \bar{\mu}_j + \boldsymbol{\nabla} \mu_\text{S} }$, so that the volume fraction flux of species $i$ [Eq.\@~\eqref{app:DynamicsStart}] becomes
\begin{align}
    -\sum_{j = 1}^N M_{ij} \boldsymbol{\nabla} \mu_j
    &= 
    -\sum_{j = 1}^{N-1} M_{ij} \boldsymbol{\nabla} \bar{\mu}_j 
    -
    \sum_{j = 1}^N M_{ij}\boldsymbol{\nabla} \mu_\text{S} \notag \\
    &= -\sum_{j = 1}^{N-1} M_{ij} \boldsymbol{\nabla} \bar{\mu}_j \, ,
\end{align}
where the second term vanishes due to the null space condition, Eq.\@~\eqref{eq:NullSpace}.
As a result, the dynamics for the ${N-1}$ solute components read
\begin{align}
    \partial_t \phi_i(\boldsymbol{r},t) = \boldsymbol{\nabla} \cdot  \sum_{j=1}^{N-1}  M_{ij} \, \boldsymbol{\nabla} \bar{\mu}_j 
    \, .
    \label{eq:ReducedDynamics}
\end{align}
This reduced formulation governs the solute dynamics in an incompressible mixture and implies that only gradients of exchange chemical potentials contribute to diffusive mass redistribution. 
The solvent acts as a passive background, and its chemical potential appears only as a reference.

For the mobility matrix, we adopt the common form ${M_{i \ne j} = -D_{ij} \phi_i \phi_j / (k_\text{B} T)}$\@~\cite{Bo.2021,Cotton.2022}, where ${D_{ij} = D_{ji}}$ due to Onsager reciprocity. 
This choice ensures that, in the dilute and non-interacting limit (${\phi_{i\neq j} \ll 1}$, ${\phi_j \approx 1}$), the dynamics reduce to an ordinary diffusion equation: ${\partial_t \phi_\text{i} = D_{ij} \boldsymbol{\nabla}^2 \phi_i}$, where $D_{ij}$ denotes the diffusivity of species $i$ in a solution that consists predominantly of species $j$.

From the free energy functional\@~[Eqs.\@~\eqref{app:FreeEnergy}, \eqref{app:FreeEnergy2}], the exchange chemical potentials $\bar\mu_i$ take the form
\begin{align}
    \frac{\bar \mu_i (\rho,s) }{k_\text{B} T}
    = \mu(\rho) + \log\left(s_i\right) \, ,
    \label{app:chemicalExchangePotentials}
\end{align}
where we introduce the non-dimensionalized chemical potential of an ordinary binary mixture
\begin{align}
    \mu(\boldsymbol{r},t) = - \tilde \kappa \,  \boldsymbol{\nabla}^2\rho + f^\prime(\rho) \,.
    \label{eq:binaryPotential}
\end{align} 

Substituting the specific form of the mobility matrix, we find that the solute volume fractions obey dynamical equations of the form 
\begin{align}
    \partial_t \phi_i(\boldsymbol{r},t) =
    \boldsymbol{\nabla}\cdot\bigg[&
    D_i\,\rho (1-\rho) \left(s_i\boldsymbol{\nabla}\mu  + \boldsymbol{\nabla} s_i\right) \notag\\ 
    &+ \sum_{j\neq i}^{N-1} D_{ij} \, \phi_i \phi_j \boldsymbol{\nabla}\log\left(\frac{s_i}{s_j}\right) 
    \bigg] \, ,
    \label{eq:PhiDynamics}
\end{align}
where we abbreviate ${D_i = D_{iN}}$ as the diffusion constant of species $i$ with respect to the solvent.
Notably, the cross-diffusive fluxes encoded in the second term only depend on local composition ratios ${s_i/s_j}$ but not on $\mu$.
They arise purely from entropic mixing and are unaffected by short-range interactions.
As a consequence, the time evolution of the total solute volume fraction $\rho$ is given by
\begin{align}
    \partial_t \rho(\boldsymbol{r},t) =
    \boldsymbol{\nabla}\cdot\left[
    \left( \sum_{i=1}^{N-1} D_i s_i\right) \rho (1-\rho) \,  \boldsymbol{\nabla} \eta\right] \, ,
    \label{eq:SoluteDynamics}
\end{align}
where we introduced the \emph{mass-redistribution} potential
\begin{align}
    \eta(\boldsymbol{r},t) = \mu + \log \left( \sum_{i=1}^{N-1} D_i s_i \right) \, .
    \label{eq:MassRedistribution}
\end{align}
Gradients of $\eta$ act as the generalized thermodynamic force that drives redistribution of total solute mass.
Besides the chemical potential $\mu$ of a binary mixture, which encodes interaction-driven demixing, $\eta$ contains an additional contribution proportional to the logarithm of the effective local diffusivity 
\begin{align}
    D(\{s_i\}) = \sum_{i=1}^{N-1}D_i s_i \, .
    \label{eq:mobility}
\end{align}

As we will show in the following, driven chemical reactions can sustain spatial variations in the local composition $\{s_i\}$.
By locally changing the mass redistribution potential $\eta$, these variations generate genuinely non-equilibrium mass fluxes that modify phase-coexistence and coarsening dynamics.

\subsection{Chemical conversion}
In addition to diffusive mass redistribution, we allow for chemical reactions that interconvert the ${N-1}$ solute species while conserving the total solute density,
\begin{align}
    i \rightleftarrows j,
    \qquad i,\,j \in \{1,\dots,N-1\}\, .
\end{align}
Accordingly, we augment the mass-conserving dynamics in Eq.\@~\eqref{eq:PhiDynamics} by local reaction-rate densities $\tilde{\mathcal R}_i(\boldsymbol r,t)$, which quantify the net change of the local volume fraction of species $i$ per unit time:
\begin{align}
    \tilde{\mathcal R}_i
    =
    \sum_{j\neq i}^{N-1}
    \left(
        \tilde{\mathcal R}_{j\to i}
        -
        \tilde{\mathcal R}_{i\to j}
    \right),
    \label{eq:Reactions}
\end{align}
with $\tilde{\mathcal R}_{j\to i}$ denoting the local reactive flux converting species~$j$ into~$i$.
The evolution equations for the solute volume fractions then take the form
\begin{align}
    \partial_t \phi_i(\boldsymbol r,t)
    =
    -\boldsymbol{\nabla}\cdot\boldsymbol J_i
    + \tilde{\mathcal R}_i \, ,
    \label{eq:PhiDynamicsWithReactions}
\end{align}
with $\boldsymbol J_i$ the conserved flux given in Eq.~\eqref{eq:PhiDynamics}.
Conservation of the total solute mass requires that the reactions satisfy
\begin{align}
    \sum_{i=1}^{N-1} \tilde{\mathcal R}_i = 0 \, .
    \label{eq:Conservation}
\end{align}

In the absence of external chemical driving, thermodynamic consistency further implies that the reaction rates obey local \emph{detailed balance of the rates}\@~\cite{Groot.1984,Kondepudi.2014}.
Thus, the ratio of forward and backward reaction rates is fixed by the difference in chemical potentials,
\begin{align}
    \frac{\tilde{\mathcal R}_{j\to i}}{\tilde{\mathcal R}_{i\to j}}
    =
    \exp\!\left(
        \frac{\mu_j - \mu_i}{k_\text{B}T}
    \right)
    =
    \frac{s_j}{s_i},
    \label{eq:detailedBalance}
\end{align}
where the second equality follows from the explicit form of the exchange chemical potentials [Eq.\@~\eqref{app:chemicalExchangePotentials}].
This shows that close to thermodynamic equilibrium, the ratio of forward and backward fluxes depends solely on the local composition variables $s_i$.

As a result, chemical equilibrium (${\tilde{\mathcal R}_{j\to i}=\tilde{\mathcal R}_{i\to j}}$) enforces equal local compositions ${s_i=s_j}$ for all pairs of interconverting species, independent of the local solute density~$\rho$.
For a well-connected reaction network, this implies the stronger condition
\begin{align}
    s_i=s_j=\frac{1}{N-1} \,, \quad \forall i,\, j \in \{1,\dots,N-1\} \, .
\end{align}

In this limit, the mass-redistribution potential~$\eta$ [Eq.\@~\eqref{eq:MassRedistribution}] effectively reduces to the chemical potential~$\mu$ of a binary mixture [Eq.\@~\eqref{eq:binaryPotential}].
Consequently, once chemical equilibrium is established, the system behaves identically to an equilibrium binary mixture.
As a result, interesting behavior is only expected when considering driven reactions that are not constrained by Eq.\@~\eqref{eq:detailedBalance}.

\subsection{Ternary mixtures}
To allow for analytical progress, throughout most of this manuscript and in Ref.\@~\cite{prl} we focus on a minimal setting: 
An incompressible ternary mixture consisting of two solute species, A and B, and a solvent S.
This choice is special in two respects.
First, the solute composition can be characterized by a single scalar field,
\begin{align}
    s(\boldsymbol r,t) = \frac{\phi_{\mathrm A}-\phi_{\mathrm B}}{\phi_{\mathrm A}+\phi_{\mathrm B}} \, ,
\end{align}
which measures the local imbalance between the two solutes.
Moreover, conservation of the total solute volume fraction $\rho$ implies [Eq.\@~\eqref{eq:Conservation}]
\begin{align}
    \tilde{\mathcal R}_{\mathrm A}
    =
    -\tilde{\mathcal R}_{\mathrm B}
    \, .
\end{align}
That is, the chemical dynamics are described by a single reactive flux ${\tilde{\mathcal{R}}\equiv\tilde{\mathcal R}_{\mathrm A}}$, which quantifies the net local conversion between species\@~A and\@~B.
Without loss of generality, we write
\begin{align}
    \tilde{\mathcal R} = k_{\rightarrow} \,  \phi_\mathrm{B} - k_{\leftarrow} \, \phi_\mathrm{A}\, ,
\end{align}
where $k_\to$ and $k_\leftarrow$ denote the effective forward and backward conversion rates, which are generally not constrained by local detailed balance.
Reformulating this expression in terms of the solute density $\rho$ and local composition $s$ gives
\begin{align}
    \tilde{\mathcal R} 
    =
    - \frac {\rho}{2} \, \big( k_\to+k_\leftarrow \big ) \, 
    \left(s-\frac{k_\to-k_\leftarrow}{k_\to+k_\leftarrow}\right) \, .
\end{align}

We find that for constant rates $k_{\leftrightarrows}$, the steady state composition is homogeneous throughout the system, ${s=(k_\to-k_\leftarrow)/(k_\to+k_\leftarrow)}$, independent of the overall solute distribution $\rho(\boldsymbol{r})$.
As a result, the solute dynamics [Eq.\@~\eqref{eq:SoluteDynamics}], again, reduce to that of an ordinary binary mixture.
This is distinct from other examples of chemically active mixtures where already linear conversion allows for interesting non-equilibrium phenomena\@~\cite{Glotzer.1994,Wurtz.2018,Zwicker.2015}.

In the following, most of our analysis applies to a general reactive flux $\tilde{\mathcal R}(\rho,s)$.
However, whenever we present explicit analytical results or numerical simulations, we focus on a specific reaction scheme that violates local detailed balance [Eq.\@~\eqref{eq:detailedBalance}] as it features a \emph{density-dependent steady state}.
Specifically, we consider a set of reactions where species B is converted to A via a bimolecular reaction at rate ${k_\to = \rho/\tau}$, and A is converted to B via a unimolecular reaction at rate ${k_\leftarrow = \rho_0/\tau}$.
Here \@~${\tau}$ denotes the characteristic reaction timescale and, if not stated otherwise, ${\rho_0=0.5}$ throughout the manuscript.
That is, conversion from B to A is catalyzed by the presence of other solutes, whereas conversion from A to B occurs spontaneously. 
This leads to an enrichment of species\@~A in regions of high solute density.
The corresponding reactive flux in terms of ${\rho}$ and ${s}$ reads
\begin{align}
    \tilde{\mathcal R} (\rho,s) = -\frac{\rho  (\rho+\rho_0)}{2\tau}\left(s-\frac{\rho-\rho_0}{\rho+\rho_0}\right) \,,
    \label{app:Reaction1}
\end{align}
which has a unique and stable fixed point 
\begin{align}
    \bar s (\rho) = \frac{\rho-\rho_0}{\rho+\rho_0} \, .
\end{align}

\subsection{Non-Dimensionalization}
\label{sec:NonDimensionalization}
We non-dimensionalize the dynamical equations [Eq.\@~\eqref{eq:PhiDynamicsWithReactions}]  for a $d$-dimensional system by rescaling space and time as
\begin{align}
    \boldsymbol{r} \rightarrow a \, \boldsymbol{r}^\prime \, , \qquad
    t \rightarrow \frac{ a^2}{\bar D}\, t^\prime \, \, ,
    \label{app:Rescalings}
\end{align}
where ${a = \nu^{1/d}}$ is the typical particle diameter, and 
\begin{align}
    \bar D = \frac{1}{N-1} \sum_{i=1}^{N-1} D_i
\end{align}
is the mean solute diffusivity. 
The energy scale is fixed by the choice ${k_\mathrm{B}T=1}$.
In the following, we drop the primes and understand $\boldsymbol r$ and\@~$t$ as dimensionless variables.

For a ternary mixture, we can rewrite the dynamics in terms of the conserved solute density ${\rho=\phi_\mathrm{A}+\phi_\mathrm{B}}$ and ${\rho s=\phi_\mathrm{A}-\phi_\mathrm{B}}$.
The resulting dimensionless equations take the form
\begin{subequations}
\label{eq:RescaledDynamics}
\begin{align}
    \partial_t \rho (\boldsymbol{r},t) =& \, 
    \boldsymbol{\nabla} 
    \cdot \big[ 
     \rho  (1-\rho) (1+D \, s)
     \boldsymbol{\nabla}\eta
    \big]
    \, ,  \label{eq:RescaledRhoDynamics}\\
      \partial_t (\rho \, s)(\boldsymbol{r},t) =& \, 
    \boldsymbol{\nabla} \cdot \bigg\{ \rho  (1-\rho) \Big[
    \Big(1 + \bar D_\mathrm{AB} \frac{\rho}{1-\rho}
    \Big)\boldsymbol{\nabla}s  \notag \\
    &\phantom{\boldsymbol{\nabla} \cdot \bigg\{ }+(D +s) \boldsymbol{\nabla}\mu\Big]\bigg\} 
    + \frac{\mathcal{R}(\rho,s)}{\ell^2} \, ,
    \label{eq:RescaledSDynamics}
\end{align}
\end{subequations}
with ${\mathcal{R}= \tau \mathcal{\tilde R}}$ denoting the non-dimensionalized reactive flux, and ${\bar D_\mathrm{AB}=2 D_\mathrm{AB}/(D_\mathrm{A}+D_\mathrm{B})}$.
The rescaled mass-redistribution potential [Eq.\@~\eqref{eq:MassRedistribution}] is given by
\begin{align}
    \eta(\boldsymbol{r},t) = \mu + \log(1+D\, s) \, ,
    \label{eq:TernaryMassRedistribution}
\end{align}
where we retain the symbol ${\mu(\rho) = - \kappa \,\boldsymbol{\nabla}^2\rho + f^\prime(\rho) }$ for the dimensionless chemical potential of a binary mixture, with ${\kappa = \tilde \kappa /a^2}$.

While the chemical potential\@~$\mu$ depends only on the conserved density $\rho$ and is therefore unaffected by chemical activity, the local composition $s$ is directly controlled by the form and strength of the reactive flux\@~$\mathcal{R}$.
Consequently, the mass-redistribution potential $\eta$ [Eq.\@~\eqref{eq:TernaryMassRedistribution}] combines equilibrium chemical-potential gradients with genuinely non-equilibrium, composition-induced contributions.
The deviation from equilibrium binary-mixture dynamics is governed by two dimensionless control parameters,
\begin{align}
    \ell^2 
    = \frac{\bar D\tau}{2a^2} 
    \, , \quad
    D   
    = \frac{D_\text{A} - D_\text{B}}{D_\text{A} + D_\text{B}} \, .
    \label{app:EffectiveParameters}
\end{align}   

The parameter ${\ell^2=l_\tau^2/(2a^2)}$ quantifies the ratio of the characteristic length scale ${l_{\tau}=\sqrt{\bar D \tau}}$, over which particles redistribute during the reaction timescale\@~${\tau}$, to the microscopic length scale ${a}$. 
It, thus, captures the relative rates of local solute interconversion and spatial redistribution by diffusion. 
In the limit ${\ell^2 \ll 1}$, reactions are fast compared to diffusion, and the system rapidly approaches the local reactive steady state. 
Conversely, for ${\ell^2 \gg 1}$, reactions are slow and composition changes are dominated by diffusive mass transport.

The parameter ${D}$ captures the asymmetry in diffusivities of the two solute species. 
It varies between ${-1 < D < 1}$, with positive values indicating that species A diffuses faster than B, while negative values indicate the reverse. 
For ${D=0}$ species A and B diffuse equally fast and the two solutes are indistinguishable.
In this limit, ${\eta=\mu}$, and the solute dynamics\@~[Eq.\@~\eqref{eq:RescaledRhoDynamics}] reduces exactly to the conserved dynamics of an ordinary binary mixture.

Finally, we note that solute cross-diffusion (${D_{\mathrm{AB}}\neq0}$) enters only as an additive correction to the composition-gradient term in Eq.\@~\eqref{eq:RescaledSDynamics}.
In particular, it contributes a term proportional to $\bar D_{\mathrm{AB}}\,\rho^2\,\boldsymbol{\nabla}s$ and thus only renormalizes the effective diffusivity of the composition field\@~$s$.
For this reason, and to simplify the analysis, we set $\bar D_{\mathrm{AB}}=0$ in the remainder of this manuscript.
\section{Linear stability analysis}
\label{sec:LSA}
\begin{figure*}
    \centering
    \includegraphics[]{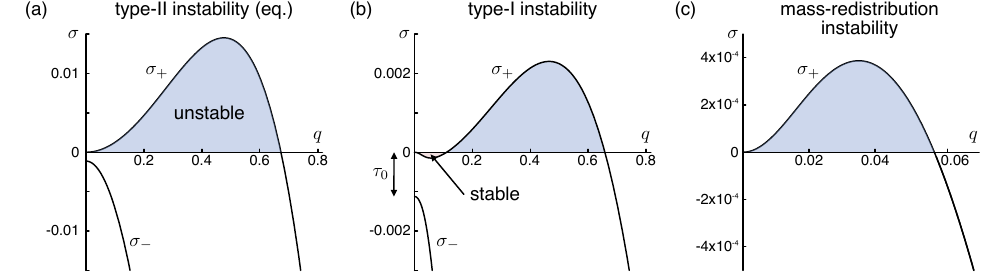}
    \caption{\textbf{Linear stability analysis}. (a--c) Linear spectrum $\sigma(q)$ of the homogeneous steady state ${(\bar \rho,\bar s)}$ for a ternary mixture in a two-dimensional system (${d=2}$), with ${\ell=20}$, ${\bar\rho=0.4}$, and ${(\chi, D)=}$ \mbox{(a) ${(2.35,0)}$}, \mbox{(b) ${(2.35,0.9)}$}, \mbox{(c) ${(1.8,-0.8)}$}. Panels\@~(a,b) display both branches $\sigma_\pm(q)$; in all shown cases, the spectrum is real-valued.}
    \label{figAPP:LSA1}
\end{figure*}
For ${\chi>2}$, the homogeneous free-energy density $f(\rho)$ [Eq.\eqref{app:FreeEnergy2}] is no longer convex, indicating a spinodal instability in the corresponding thermodynamic system. 
To assess how this instability is altered by non-equilibrium chemical reactions, we perform a linear stability analysis of the homogeneous steady state\@~${(\bar \rho,\bar s)}$. 
For concreteness, we consider a ternary mixture governed by the non-dimensionalized equations in Eq.\@~\eqref{eq:RescaledDynamics}.
The reaction term $\mathcal R$ is left unspecified, except for the assumption that it admits a unique, locally stable fixed point $\bar s(\rho)$.

We consider small perturbations of the form 
\begin{align}
    (\delta \rho, \delta s)^T 
    =
    \exp(i \boldsymbol{q} \cdot \boldsymbol{r} + \sigma t) \,  
    (\delta \rho_0, \delta s_0)^T 
    \, .
\end{align}
Inserting the ansatz ${(\rho,s)^T=(\bar \rho+\delta \rho,\bar s+\delta s)^T}$ into the dynamical equations, Eq.\@~\eqref{eq:RescaledDynamics}, and expanding to linear order in $\delta\rho_0$ and $\delta s_0$, yields the eigenvalue problem 
\begin{align}
    \sigma 
    \begin{pmatrix}
        \delta \rho_ 0 \\[2mm]
        \delta s_0 
    \end{pmatrix}
    =
    \boldsymbol{\hat J}
    \begin{pmatrix}
        \delta \rho_ 0 \\[2mm]
        \delta s_0 
    \end{pmatrix}
    \, ,
\end{align}
where the Jacobian is given by
\begin{align}
    \boldsymbol{\hat J} =
    &- M(\bar \rho) q^2 \begin{pmatrix}
        (1+D\bar s)J(q) & D \\[2mm]
         D (1-\bar s^2)J(q)/\bar\rho &  (1-D \bar s)/\bar\rho \, 
    \end{pmatrix}
    \notag \\
    &+ 
    \frac{1}{\bar \rho \ell^2}
    \begin{pmatrix}
        0 & 0 \\[2mm]
        \partial_\rho \mathcal R & \partial_s \mathcal R
    \end{pmatrix} \, .
    \label{eq:Jacobian}
\end{align}
The first term accounts for the contributions from the mass-conserving dynamics, while the second term accounts for the local chemical reactions.

Linear instability occurs when at least one eigenvalue of the Jacobian ${\hat{\boldsymbol{J}}(q)}$ has a positive real part, indicating exponential growth of small perturbations. 
In the absence of chemical reactions, the stability of the homogeneous steady state is determined by the term  
\begin{align}
    J(q)=\kappa q^2+f^{\prime \prime}(\bar \rho) \, ,
    \label{app:EquilibriumInstability}
\end{align}
with ${J(q)<0}$ marking the spinodal regime of a binary mixture\@~\cite{Bray.2002}.
For the general case, the eigenvalues of the Jacobian\@~$\boldsymbol{\hat J}$ can be written as
\begin{align}
    \sigma_\pm(q) = \dfrac{\tau(q) \pm \sqrt{\tau(q)^2-4\delta(q)}}{2} \, ,
    \label{app:2x2Eigenvalues}
\end{align}
where ${\tau(q)=\mathrm{tr}(\boldsymbol{\hat J})}$ and ${\delta(q)=\mathrm{det}(\boldsymbol{\hat J})}$ denote the Jacobian's trace and determinant, respectively.
From Eq.\@~\eqref{eq:Jacobian}, both quantities are polynomials in ${q^2}$,
\begin{subequations}
\begin{align}
    \tau(q) &= \tau_0 + \tau_2 q^2 + \tau_4 q^4 \,, \label{eq:Trace}\\
    \delta(q) &= \delta_2 q^2 + \delta_4 q^4 + \delta_6 q^6 \, .
    \label{eq:Determinant}
\end{align}
\end{subequations}
The corresponding coefficients are collected below:
\begin{subequations}
\label{eq:Coefficients}
\begin{align}
    \tau_0 =&\;  \frac{\partial_s \mathcal R(\bar\rho,\bar s)}{\ell^2\bar\rho}~\label{app:tau0} \, ,\\
    \tau_2 =&\;  - M(\bar\rho)\left[(1+D\bar s)f^{\prime \prime}(\bar \rho) + \frac{(1-D \bar s)}{\bar\rho}\right]  \, , \label{app:tau2}\\[3mm]
    \tau_4 =&\;  - M(\bar\rho)(1+D\bar s)\kappa  \, ,\\[3mm]
    \delta_2 =&\;  -\frac{M(\bar\rho)}{\ell^2\bar \rho}\big[(1+D\bar s)f^{\prime\prime}(\bar \rho)\partial_s \mathcal R(\bar\rho,\bar s) \notag\\
    &\qquad \qquad \ -D\partial_\rho \mathcal R(\bar\rho,\bar s)\big] \, ,
    \label{app:delta2}\\
    \delta_4 =&\;  M(\bar\rho)\bigg[ \frac{M(\bar\rho)}{\bar\rho}(1-D^2)f^{\prime \prime}(\bar \rho) \\ \notag
    & \qquad  \quad -(1+D\bar s)\kappa \frac{\partial_s \mathcal R(\bar\rho,\bar s)}{\ell^2\bar\rho} \bigg] \label{app:delta4} \, ,\\
    \delta_6 =&\;  \frac{M(\bar\rho)^2}{\bar\rho}(1-D^2)\kappa \, .
\end{align}
\end{subequations}

Although the Jacobian is generally non-symmetric and may admit complex eigenvalues, we initially restrict attention to the case of a purely real spectrum; a detailed analysis of oscillatory instabilities is deferred to Sec.\@~\ref{app:oscillatory}.
Under this assumption, the eigenvalues ${\sigma_\pm(q)}$ are ordered such that ${\sigma_+ > \sigma_-}$~[Fig.\@~\ref{figAPP:LSA1}].
Stability of the homogeneous steady state is therefore determined by the larger branch~$\sigma_+(q)$.

For a homogeneous perturbation (${q = 0}$), the growth rates reduce to ${\sigma_\pm(0) \in \{0, \tau(0)\}}$. 
We assume that for a well-mixed system the chemical kinetics admits a stable fixed point ${\bar s(\rho)}$, which requires ${\sigma(0)\leq0}$, or equivalently ${\partial_s\mathcal R(\bar\rho,\bar s)<0}$ [Eq.\@~\eqref{app:tau0}].
Consequently, the lower branch $\sigma_-$ is strictly negative at the origin, whereas the larger eigenvalue $\sigma_+$ is marginal at ${q=0}$~[Fig.\@~\ref{figAPP:LSA1}(a–b)].
This marginal mode reflects conservation of the solute volume fraction~${\rho}$.
Although mathematically well defined, such a perturbation is dynamically suppressed as it violates global mass conservation.

From Eq.\@~\eqref{app:2x2Eigenvalues}, we see that for finite wave-length perturbations ${\sigma_+(q) > 0}$ when either the trace is positive, ${\tau(q) > 0}$, or the determinant is negative, ${\delta(q) < 0}$. 
Thus, the homogeneous steady state is linearly unstable if either condition is met for any wavenumber\@~${q}$.

\subsection{Generalized mass-redistribution instability}
\label{app:instability_condition}
We begin by analyzing the case ${\tau(q)>0}$, which provides a sufficient condition for linear instability.
Since ${\tau_4\leq0}$ for all parameter values, a necessary condition for ${\tau(q)>0}$ is ${\tau_2>0}$.
For ${\chi<0}$, where ${\kappa=0}$\@~(Sec.\@~\ref{sec:freeEnergy}), this condition is also sufficient because ${\tau_4=0}$.
For ${\chi>0}$ and hence ${\kappa>0}$, completing the square yields
\begin{align}
    \tau(q)
    =
    \tau_4\!\left[q^2-\!\left(-\frac{\tau_2}{2\tau_4}\right)\right]^2
    + \tau_0
    - \tau_4\!\left(\frac{\tau_2}{2\tau_4}\right)^2 ,
\end{align}
which describes a downward-opening parabola in ${q^2}$.
A sufficient condition for ${\tau(q)>0}$ is therefore
\begin{align}
    \tau_0 - \tau_4\left(\frac{\tau_2}{2\tau_4}\right)^2 
    > 0 \,.
    \label{app:instabilityCon1}
\end{align}

We now turn to the determinant ${\delta(q)}$.
Even when ${\tau(q)<0}$, the leading eigenvalue ${\sigma_+(q)}$ can become positive if ${\delta(q)<0}$ [Eq.\@~\eqref{app:2x2Eigenvalues}].
For large wavenumbers, the determinant is dominated by the sixth-order term ${\delta_6 q^6}$.
When ${\kappa>0}$, one has ${\delta_6>0}$ for all parameters, implying ${\delta(q)>0}$ at large ${q}$ and hence ${\lim_{q\to\infty}\sigma_+(q)<0}$.
Equivalently, a positive stiffness ${\kappa}$ guarantees stability against short-wavelength perturbations.
If ${\kappa=0}$, short-wavelength stability instead requires ${\delta_4>0}$.

Consequently, for ${\delta(q)}$ to become negative at finite ${q}$, either ${\delta_2<0}$ or ${\delta_4<0}$ must hold.
When ${\delta_2<0}$ and ${\delta_4>0}$, ${\delta(q)}$ is negative near ${q=0}$, corresponding to a long-wavelength instability.
We therefore identify ${\delta_2<0}$ as a sufficient condition for instability of the homogeneous steady state.
Using Eq.\@~\eqref{app:delta2} together with the steady-state condition ${\mathcal R(\bar\rho,\bar s)=0}$, yields the instability criterion
\begin{align}
    \big( 
    1 + D \bar s
    \big) \,
    f^{\prime\prime}(\bar\rho) 
    + D\, \frac{\mathrm{d} \bar s}{\mathrm{d} \rho}(\bar\rho)
    < 0
     \, .
    \label{app:instabilityCon2}
\end{align}
This condition generalizes the \emph{mass-redistribution instability} known from two-component mass-conserving reaction-diffusion (McRD) systems\@~\cite{Brauns.2020,Frey.2022}.

To illustrate its physical meaning, we consider the limit in which the system remains close to local chemical equilibrium, ${s\approx\bar s(\rho)}$.
Linearizing Eq.\@~\eqref{eq:RescaledRhoDynamics} around a homogeneous background yields
\begin{align}
\label{eq_app:effective-diffusion}
    \partial_t \delta\rho  (\boldsymbol{r},t)
    = \bar\rho  (1 - \bar\rho) \left[ (1 + D \bar s)\, f^{\prime\prime}(\bar\rho) + D\, \frac{\mathrm{d} \bar s}{\mathrm{d} \rho} \right] \boldsymbol{\nabla}^2 \delta \rho \, ,
\end{align}
where higher-order spatial derivatives ${\sim\nabla^4\delta\rho}$, have been neglected.
This approximation is valid for weakly modulated perturbations.

When Eq.\@~\eqref{app:instabilityCon2} is satisfied, the effective diffusion coefficient in Eq.\@~\eqref{eq_app:effective-diffusion} becomes negative, leading to an amplification of density fluctuations and destabilization of the homogeneous state.
The slope criterion, Eq.\@~\eqref{app:instabilityCon2}, thus marks the onset of a diffusion-driven instability induced by the coupling between local composition and chemical turnover.

Although Eq.\@~\eqref{app:instabilityCon2} superficially resembles the equilibrium spinodal condition ${f^{\prime\prime}(\bar\rho)<0}$, this analogy is misleading.
In the limit of vanishing chemical activity, ${\delta_2}$ vanishes identically, and instability is instead controlled by higher-order terms in ${\delta(q)}$.
As in classical McRD systems, Eq.\@~\eqref{app:instabilityCon2} therefore represents a genuinely non-equilibrium instability criterion, reflecting the interplay between short-range interactions and the slope of the reactive nullcline ${\bar s(\rho)}$.
Depending on the sign of ${D}$ and the curvature ${f^{\prime\prime}(\bar\rho)}$, a sufficiently steep positive or negative nullcline slope is required to overcome diffusive relaxation.

Finally, we consider the complementary regime ${\delta_4<0}$, where ${\delta(q)}$ may become negative even if ${\delta_2>0}$.
Completing the square yields
\begin{align}
    \frac{\delta(q)}{q^2} 
    = \delta_6 \left[q^2-\left(-\frac{\delta_4}{2\delta_6}\right)\right]^2 + 
    \delta_2 - \delta_6 \left(\frac{\delta_4}{2\delta_6}\right)^2 \, ,
\end{align}
corresponding to an upward-opening parabola in $q^2$ with a minimum at finite $q$. 
A sufficient condition for ${\delta(q) < 0}$ is then given by
\begin{align}
    \delta_2 - \delta_6 \left(\frac{\delta_4}{2\delta_6}\right)^2 < 0 \, .
    \label{app:instabilityCon3}
\end{align}
While more involved than Eq.~\eqref{app:instabilityCon2}, this condition is automatically satisfied when ${\delta_2 < 0}$, consistent with the argument above that ${\delta_2 < 0}$ always implies instability, regardless of the sign of ${\delta_4}$.

\subsection{Short- vs.\@~long-wavelength instabilities}
\label{app:instability_type}
To classify the different instabilities, we consider the leading order expansion of the largest eigenvalue\@~$\sigma_+(q)$ around the origin (${q=0}$).
\begin{align}
    \sigma_+(q) = - \frac{\delta_2}{\left|\tau_0\right|}q^2+ \mathcal{O}(q^4) \, .
    \label{app:sigmaExpansion}
\end{align}
The sign of $\delta_2$ determines the slope of $\sigma_+$ near ${q = 0}$ and, thus, the behavior of long-wavelength modes.
The resulting instability types are summarized in Fig.\@~\ref{figAPP:LSA1}.

For ${\delta_2<0}$, the slope of $\sigma_+$ near ${q=0}$ is positive, implying instability of arbitrarily long-wavelength modes\@~[Fig.\@~\ref{figAPP:LSA1}(a,c)].
The band of unstable modes extends continuously to the origin, and fluctuations grow without an intrinsic length scale.
This defines a \emph{long-wavelength} (type-II) instability\@~\cite{Cross.1993}, typically associated with coarsening or phase-separation dynamics.
In the present ternary mixture, type-II instabilities arise generically in the limit of vanishing chemical activity [Fig.\@~\ref{figAPP:LSA1}(a)], recovering classical Cahn-Hilliard dynamics~\cite{Cahn.1958}.

For ${\delta_2>0}$, the slope of $\sigma_+$ at the origin is negative, and long-wavelength perturbations are linearly stable\@~[Fig.\@~\ref{figAPP:LSA1}(b)].
Any instability must therefore occur at a finite wavenumber ${q_c>0}$, selecting a characteristic pattern scale.
This corresponds to a \emph{short-wavelength} \mbox{(type-I)} instability\@~\cite{Cross.1993}, commonly associated with micro phase-separation or stationary pattern formation.

\subsection{Physical interpretation of the instability regimes}
\label{app:instability_discussion}
\begin{figure}
    \centering
    \includegraphics[]{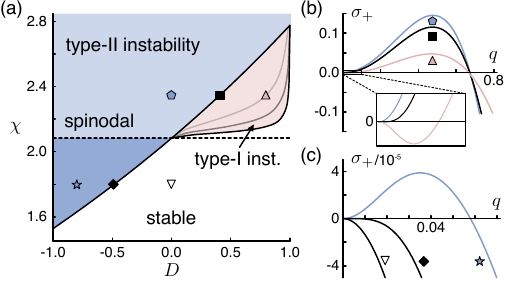}
    \caption{\textbf{Bifurcation diagram}. 
    The homogeneous steady state is stable (white) or displays a \mbox{type-I} (red/light gray) or \mbox{type-II} (blue/gray) instability.
    (a) Bifurcation diagram as a function of $D$ and $\chi$ for fixed ${\bar\rho=0.4}$, ${d=2}$, and ${\ell\in\{5,10,20\}}$ (light gray, gray, and black lines). 
    In the presence of driven chemical reactions, the boundary of the type-II regime (black line) is determined by Eq.\@~\eqref{app:instabilityCon2}. 
    In the equilibrium limit (${\ell \to \infty}$), the homogeneous state is linearly unstable above the (dashed) spinodal line ${\chi=1/[2\bar{\rho}(1 - \bar{\rho})]}$, corresponding to ${f^{\prime\prime}(\bar\rho)<0}$.
    (b,\@~c) Upper branch of the dispersion relation, $\sigma_+(q)$, for parameters corresponding to the different markers in subpanel (a) and ${\ell=20}$.
    The figure is adapted from Ref.\@~\cite{prl}.
    }
    \label{fig:bifurcationDiagram}
\end{figure}

The distinction between finite-wavelength (type-I) and long-wavelength (type-II) instabilities, together with the corresponding criteria
[Eqs.\@~\eqref{app:instabilityCon1}, \eqref{app:instabilityCon2}, and \eqref{app:instabilityCon3}],
allows us to construct a representative bifurcation diagram.
This diagram is discussed in Ref.\@~\cite{prl} and reproduced in Fig.\@~\ref{fig:bifurcationDiagram}.

In the limit of vanishing chemical activity (${\ell\to\infty}$) we recover the equilibrium spinodal criterion Eq.\@~\eqref{app:EquilibriumInstability}, corresponding to the dashed black line in Fig.\@~\ref{fig:bifurcationDiagram}(a).
This is the classical condition for phase separation in a binary mixture and is independent of both the composition $\bar s$ and the relative diffusivity $D$.
Near ${q=0}$, the growth rate obeys
\begin{align}
    \sigma_+(q)
    =
    -M(\bar\rho)(1+D\bar s) f^{\prime\prime}(\bar\rho)\, q^2
    + \mathcal{O}(q^4) ,
\end{align}
which is positive in the spinodal regime.
Equilibrium phase separation therefore proceeds via a long-wavelength type-II instability~[Fig.\@~\ref{figAPP:LSA1}(a)].
That is, from the perspective of linear stability analysis, a ternary mixture, where the species A and B only differ in their transport properties, behaves like an ordinary binary mixture.

In the presence of chemical reactions, the system exhibits a richer instability landscape that depends on whether short-range interactions alone can induce phase separation.
We distinguish three physically distinct regimes, corresponding to the different colored regions in Fig.\@~\ref{fig:bifurcationDiagram}(a):

\textit{(i) Reaction-driven instability (dark blue).}
When ${f^{\prime\prime}(\bar\rho)>0}$, the system lies outside the spinodal regime, and short-range interactions alone cannot destabilize the homogeneous state.
In this case, ${\tau_2<0}$ [Eq.\@~\eqref{app:tau2}] and ${\delta_4>0}$ [Eq.\@~\eqref{app:delta4}], so instability is possible only if ${\delta_2<0}$, corresponding to the mass-redistribution criterion in Eq.\@~\eqref{app:instabilityCon2}.
Equation~\eqref{app:sigmaExpansion} then implies that the instability is necessarily of type-II.
Thus, when interactions are insufficient to drive phase separation on their own, pattern formation can occur only via a reaction-driven mass-redistribution mechanism, with unstable modes extending to ${q=0}$\@~[Figs.\@~\ref{figAPP:LSA1}(c), \ref{fig:bifurcationDiagram}(c)].
Nevertheless, short-range interactions still affect the instability threshold through ${f^{\prime\prime}(\bar\rho)}$: 
The further the system is from the spinodal regime, the steeper the reactive nullcline must be.

\textit{(ii) Interaction-driven instability (light blue and red).}
When ${f^{\prime\prime}(\bar\rho)<0}$, short-range interactions favor phase separation.
If ${\delta_2>0}$ and either Eq.\@~\eqref{app:instabilityCon1} or Eq.\@~\eqref{app:instabilityCon3} holds, instability arises from higher-order terms in the dispersion relation [red region in Fig.\@~\ref{fig:bifurcationDiagram}(a)].
Chemical activity then suppresses long-wavelength modes, resulting in a type-I instability with a finite band of unstable modes centered at ${q_c}$\@~[Fig.\@~\ref{figAPP:LSA1}(b)].
If instead ${\delta_2<0}$, both interactions and reaction--diffusion coupling promote instability [light blue region in Fig.\@~\ref{fig:bifurcationDiagram}(a)], leading to a type-II instability\@~[Fig.\@~\ref{fig:bifurcationDiagram}(b)].

\textit{(iii) Suppression of pattern formation (white).}
Finally, sufficiently strong chemical activity can stabilize the homogeneous state even within the spinodal regime.
Avoiding a mass-redistribution instability requires ${\delta_2>0}$, while stability against higher-order modes further demands ${\tau(q)<0}$ and ${\delta(q)>0}$ for all ${q}$.
These conditions are satisfied when ${\partial_s\mathcal R(\bar\rho,\bar s)}$ is sufficiently negative, ensuring linear stability of the reactive fixed point.
In this regime, all perturbations decay and pattern formation is suppressed.

\subsection{Absence of oscillatory modes}
\label{app:oscillatory}
\begin{figure}
    \centering
    \includegraphics[]{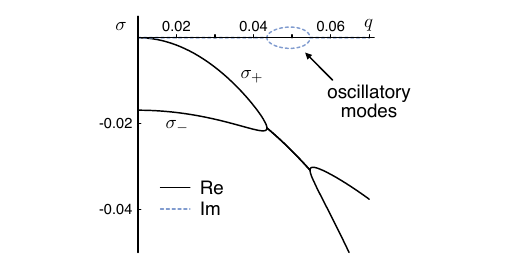}
    \caption{\textbf{Oscillatory modes.} Spectrum\@~$\sigma(q)$ of the homogeneous steady state\@~${(\bar \rho,\bar s)}$ obtained from linear stability analysis of a ternary mixture in a two-dimensional system (${d=2}$) with ${\ell=20}$, ${\bar\rho=0.85}$, ${\chi=2.35}$, and ${D=0.35}$. Solid and dashed lines indicate the real [Re$(\sigma)$] and imaginary [Im$(\sigma)$] part of $\sigma$, respectively.}
    \label{figAPP:LSA2}
\end{figure}
Throughout the preceding analysis, we have assumed that the eigenvalues $\sigma_\pm(q)$ are real, which requires further justification.
A nonvanishing imaginary part corresponds to oscillatory modes and can arise only if the discriminant $\tau(q)^2-4\delta(q)$ in Eq.\@~\eqref{app:2x2Eigenvalues} becomes negative.
While a fully general analysis would require specifying the reaction term $\mathcal R$, the situation simplifies when restricting attention to the linearly unstable regime.

If ${\tau(q)<0}$, the homogeneous steady state can become unstable only when ${\delta(q)<0}$.
In this case, the discriminant is strictly positive, and oscillatory modes are excluded.
Conversely, if ${\tau(q)>0}$, oscillatory modes may occur when ${\delta(q)>0}$.
From Eq.\@~\eqref{eq:Coefficients}, one finds that ${\tau(q)>0}$ requires ${J(q)<0}$, implying that unstable oscillatory modes can arise only within the spinodal regime.

Inspecting Eqs.\@~\eqref{eq:Determinant} and \eqref{eq:Coefficients}, we find that within the spinodal regime, the only contribution that can render ${\delta(q)}$ positive is proportional to
${D\,\partial_\rho\mathcal R(\bar\rho,\bar s)\,q^2}$, which vanishes for ${D=0}$.
Hence, for vanishing or moderate ${D}$, the unstable part of the spectrum is guaranteed to be purely real.
For large\@~$D$, oscillatory instabilities cannot be ruled out in general and would depend on the sign and magnitude of $\partial_\rho \mathcal{R}(\bar\rho, \bar s)$.
Any such instability, however, is confined to a narrow band of modes with ${\tau(q)>0}$ and requires ${\delta_2>0}$.

As a consequence, oscillatory instabilities are excluded for the generalized mass-redistribution mechanism [Eq.\@~\eqref{app:instabilityCon2}], which occurs when ${\delta_2<0}$.
Unstable oscillatory modes can therefore arise only in the type-I unstable regime.

For the reaction terms and parameter ranges considered here, numerical evaluation confirms the absence of linearly unstable modes with a nonzero imaginary part.
Linearly stable oscillatory modes may nevertheless occur [Fig.\@~\ref{figAPP:LSA2}], leading to transient oscillatory relaxation towards the homogeneous steady state.

\section{Multi-phase coexistence}
\label{sec:Coexistence}
\begin{figure}
    \centering
    \includegraphics[]{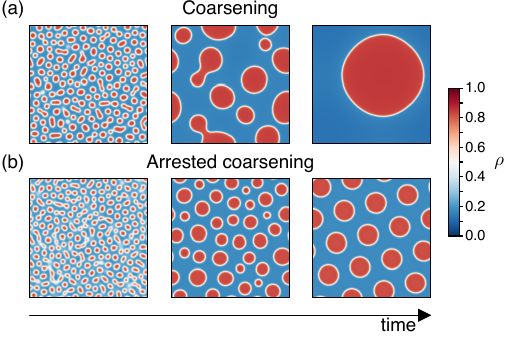}
    \caption{\textbf{Non-equilibrium phase separation}. Snapshots of the solute volume fraction $\rho$ from numerical simulations of a ternary mixture [Eq.\@~\eqref{eq:RescaledDynamics}] in a square domain of size ${L=200}$ with periodic boundary conditions for ${D=-0.3}$\@~(top) and ${D=0.3}$\@~(bottom). The system is initialized with small perturbations around the homogeneous steady state at average solute volume fraction\@~${\bar \rho=0.4}$. 
    Color indicates the local solute volume fraction $\rho$ (colorbar); in grayscale renderings, interfaces appear white.
    The remaining parameters are ${\ell=20}$, and ${\chi=2.4}$.}
    \label{fig:overview}
\end{figure}
In the previous section, we showed that spatial patterns can emerge from a linear instability of the homogeneous steady state.
This is further confirmed by numerical finite-element simulations presented in Ref.\@~\cite{prl} and Fig.\@~\ref{fig:overview}.

Within the linearly unstable regime, the system separates into solute-rich and solute-poor domains.
Depending on the chosen parameters, the late-time evolution of these domains either proceeds via continuous coarsening until only a single macroscopic domain remains\@~[Fig.\@~\ref{fig:overview}(a)], or coarsening is arrested, leading to the coexistence of multiple, roughly equally sized droplets\@~[Fig.\@~\ref{fig:overview}(b)].

Specifically, in Ref.\@~\cite{prl} we found that outside the equilibrium spinodal domain, i.e., when patterns arise via the generalized mass-redistribution instability [Eq.\@~\eqref{eq:MassRedistribution}], the phase-separated domains always coarsen.
Within the spinodal domain, by contrast, the late-time dynamics depend crucially on the relative solute diffusivities.
For ${D<0}$, the system invariably coarsens, whereas for ${D>0}$ coarsening is typically arrested\@~(Fig.\@~\ref{fig:overview})\@~\cite{prl}.

These observations naturally lead to two central questions, which we address in the following sections:
First, how does chemical activity modify multiphase coexistence?
Specifically, which physical principles determine the solute density and composition of the coexisting phases in the non-equilibrium steady state?
Second, under which conditions is coarsening arrested, and what mechanisms govern the selection of a finite steady-state domain size?

The first question is addressed in this section, where we analyze limiting cases of the general $N$-component model defined by Eqs.\@~\eqref{app:FreeEnergy2}, \eqref{eq:SoluteDynamics} and\@~\eqref{eq:PhiDynamicsWithReactions}.
From this analysis, we derive exact coexistence criteria for the chemically active mixture.
Specifically, we show that in the limit of fast chemical conversion, phase coexistence can be described by an effective free-energy functional that differs from the equilibrium free energy in Eq.\@~\eqref{app:FreeEnergy2}.
\subsection{Equilibrium limit}
In the absence of chemical activity, the steady state is characterized by vanishing solute fluxes.
From Eq.~\eqref{eq:PhiDynamics}, this implies
\begin{align}
    \boldsymbol{\nabla}\mu = 0\, , \quad
    \boldsymbol{\nabla} s_i = 0 \, , \qquad \forall i \in \{1,\dots, N-1\} \, .
\end{align}
That is, the steady-state composition is spatially uniform, and the total solute volume fractions in the coexisting phases are determined by the condition that the binary chemical potential $\mu$ is homogeneous across the system.

For a spherical droplet of radius $R$, with a dense solute-rich interior (phase 1) and a dilute exterior (phase 2), these conditions imply equality of the bulk chemical potentials\@~\cite{Weber.2019,Zwicker.2025},
\begin{align}
    \mu_1 = \mu_2 \, ,
    \label{eq:ChemicalEquilibrium}
\end{align}
and a curvature-dependent pressure imbalance between the two phases\@~(App.\@~\ref{sec:phaseCoexistence})~\cite{Weber.2019,Zwicker.2025},
\begin{align}
    \Pi_1 = \Pi_2 + \frac{(d-1)\gamma}{R} \, ,
    \label{eq:MechanicalEquilibrium}
\end{align}
where ${\Pi=\mu\rho-f(\rho)}$ denotes the local solute osmotic pressure, and the second term corresponds to the surface tension-induced Laplace pressure of the curved interface.
The (non-dimensional) surface tension~$\gamma$ is defined as the integral of the stationary solute profile $\rho(r)$ along the radial direction\@~(App.\@~\ref{sec:phaseCoexistence})\@~\cite{Bray.2002,Weber.2019}
\begin{align}
    \gamma
    =
    \kappa \int \mathrm{d}r \, (\partial_r \rho)^2 \, ,
    \label{eq:SurfaceTension}
\end{align}
with $r$ denoting the radial coordinate.
In the limit of weak curvature, i.e., for large droplets, $\gamma$ can be approximated by the corresponding expression for a planar interface.

\begin{figure*}
    \centering
    \includegraphics[]{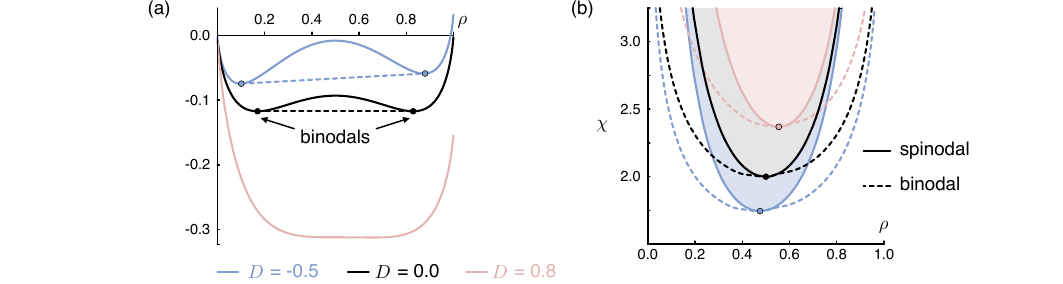}
    \caption{\textbf{Effective free energy} in the limit of fast chemical turnover.
    (a) Local free energy density of the effective free energy, Eq.\@\eqref{eq:NeqFreeEnergy}, for a ternary mixture with ${\chi=2.4}$, shown for different relative diffusivities: ${D=-0.5}$ (blue/gray), ${D=0}$ (black), ${D=0.8}$ (red/light gray). 
    The coexisting binodal solute volume fractions (bullets) are obtained from a common tangent construction (dashed lines).
    (b) Corresponding phase diagram in the ${(\rho,\chi)}$ plane.
    Solid lines indicate the spinodal boundaries [Eq.\@~\eqref{eq:Spinodal}], while dashed lines denote the binodals for a flat interface. Colored dots indicate the critical onset of phase separation.}
    \label{fig:freeEnergy}
\end{figure*}
\subsection{Fast chemical turnover}
\label{sec:FastTurnover}
In the presence of driven chemical reactions, the chemical
[Eq.~\eqref{eq:ChemicalEquilibrium}] and mechanical
[Eq.~\eqref{eq:MechanicalEquilibrium}] equilibrium conditions derived above no longer apply.
To illustrate this, we consider the limit of fast chemical turnover, in which local chemical relaxation is much faster than diffusive transport.
In this regime, the composition variables rapidly relax to their local reactive fixed points and are thus enslaved by the local solute volume fraction,
${s_i(\boldsymbol{r},t)=\bar s_i[\rho(\boldsymbol{r},t)]}$.
As a consequence, the ${N-1}$ composition fields can be adiabatically eliminated.

Using the non-dimensional variables introduced above, the total solute volume fraction $\rho$ then obeys an effective dynamical equation of the form [Eq.\@~\eqref{eq:SoluteDynamics}]
\begin{align}
    \partial_t \rho(\boldsymbol{r},t) =
    \boldsymbol{\nabla}\cdot\left[
    \left( \sum_{i=1}^{N-1} \frac{D_i \bar s_i}{\bar D} \right)
    \rho (1-\rho)\, \boldsymbol{\nabla} \eta
    \right] \, ,
    \label{eq:SoluteDynamics2}
\end{align}
where the mass-redistribution potential
\begin{align}
    \eta(\boldsymbol{r},t)
    =
    \mu(\rho)
    +
    \log \left(
        \sum_{i=1}^{N-1} \frac{D_i \bar s_i(\rho)}{\bar D}
    \right)
    \label{eq:MassRedistribution2}
\end{align}
only depends on the local solute volume fraction $\rho$.
It can therefore be written as the functional derivative of an effective free-energy functional
\begin{align}
    \eta(\boldsymbol{r},t)
    =
    \frac{\delta \mathcal{F}_\mathrm{neq}}{\delta \rho(\boldsymbol{r},t)} \, .
\end{align}
The resulting non-equilibrium free energy functional is given by
\begin{align}
    \hspace{-0.5mm} \mathcal{F}_\mathrm{neq}[\rho] = 
    \int_V \mathrm{d} \boldsymbol{r} \,\Bigg[
    &\frac{\kappa}{2}\left(\boldsymbol{\nabla}\rho\right)^2 + f(\rho) \; +\notag \\
     &+ \int^\rho \mathrm{d} \rho^\prime  \log \left( \sum_{i=1}^{N-1} \frac{D_i \bar s_i(\rho^\prime)}{\bar D} \right) \Bigg]\, .
     \label{eq:NeqFreeEnergy}
\end{align}
This functional has the form of a Flory-Huggins free energy of a symmetric binary mixture, augmented by an activity-induced contribution that explicitly depends on the reactive fixed-point composition $\{\bar s_i(\rho)\}$ and the solute diffusivities $\{D_i\}$.
As illustrated in Fig.\@~\ref{fig:freeEnergy}(a) for a ternary mixture, the last term in Eq.\@~\eqref{eq:NeqFreeEnergy} generally breaks the ${\rho\to1-\rho}$ symmetry of the equilibrium local free-energy density $f(\rho)$ [Eq.\@~\eqref{app:FHdensity}].
As a result, the equilibrium spinodal domain [${f^{\prime\prime}(\bar\rho)<0}$], where phase separation occurs spontaneously, is altered to 
\begin{align}
    f^{\prime\prime}(\rho)+ \frac{\sum_{i=1}^{N-1} D_i \partial_\rho\bar s_i(\rho)}{\sum_{i=1}^{N-1} D_i \bar s_i(\rho)} < 0 \, .
    \label{eq:Spinodal}
\end{align}
The emergent spinodal boundaries are shown in Fig.\@~\ref{fig:freeEnergy}(b) for several values of~$D$.
One finds that driven chemical turnover increases the system's tendency to phase-separate when
\begin{align}
    \sum_{i=1}^{N-1} D_i \partial_\rho\bar s_i(\rho) < 0 \,.
\end{align}

In the steady state, the absence of solute fluxes implies ${\boldsymbol{\nabla}\eta=0}$.
As a consequence, the coexistence conditions, Eqs.\@~\eqref{eq:ChemicalEquilibrium}, \eqref{eq:MechanicalEquilibrium}, are modified to
\begin{subequations}
\label{eq:ModifiedCoexistence}
\begin{align}
    \mu_\mathrm{1}-\mu_2  =&  \log\left( \frac{\sum_{i=1}^{N-1} D_i \bar s_i(\rho_2)}{\sum_{i=1}^{N-1} D_i \bar s_i(\rho_1)}\right) 
    \label{eq:ModifiedChemical}\, ,\\[2mm]
    \Pi_1 -\Pi_2 =& \int_{\rho_1}^{\rho_2}\mathrm{d} \rho^\prime \left( \frac{\sum_{i=1}^{N-1} D_i \rho^\prime \partial_\rho \bar s_i(\rho^\prime)}{\sum_{i=1}^{N-1} D_i \bar s_i(\rho^\prime)}\right) \notag \\
    &+ \frac{(d-1) \gamma_\mathrm{neq}}{R} \, . \label{eq:ModifiedMechanical}
\end{align}
\end{subequations}

Here, the non-equilibrium surface tension $\gamma_{\mathrm{neq}}$ is defined analogously to its equilibrium counterpart [Eq.\@~\eqref{eq:SurfaceTension}], but evaluated using the stationary non-equilibrium density profile.

\begin{figure}
    \centering
    \includegraphics[]{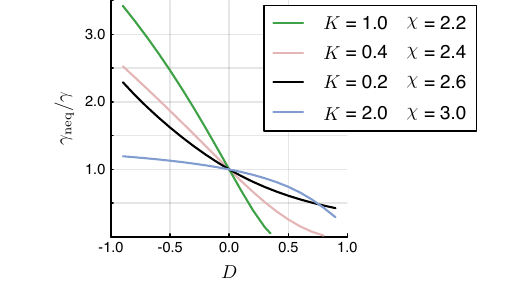}
    \caption{\textbf{Non-equilibrium surface tension}. The effective surface tension\@~$\gamma_\mathrm{neq}$ [Eq.\@~\eqref{eq:ModifiedMechanical}] of a ternary mixture in the limit of rapid chemical turnover\@~(${\ell \to 0}$) is estimated by numerically evaluating Eq.\@~\eqref{eq:SurfaceTension}, using the exact stationary solute profile\@~$\rho(r)$ for several values of the Flory-Huggins parameter $\chi$ and the reaction parameter $\rho_0$. The stationary profile is inferred from finite element simulations of the dynamical equations, Eq.\@~\eqref{eq:RescaledDynamics}, in a one-dimensional system of size ${L=1000}$, subject to no-flux boundary conditions and with an average solute volume fraction ${\bar \rho=0.5}$. The results are normalized by the corresponding equilibrium surface tension $\gamma$, which is recovered for ${D=0}$.}
    \label{fig:surfaceTension}
\end{figure}

We estimate $\gamma_{\mathrm{neq}}$ by numerically evaluating Eq.\@~\eqref{eq:SurfaceTension} for stationary flat-interface profiles obtained from finite-element simulations.
For a ternary mixture, the resulting surface tension as a function of the relative solute diffusivity $D$ is shown in Fig.\@~\ref{fig:surfaceTension} for several values of the Flory-Huggins parameter $\chi$ and the reaction parameter~$\rho_0$.
Depending on the relative solute diffusivity $D$, one finds that chemical activity can either enhance or reduce the effective surface tension compared to the equilibrium case~[Fig.\@~\ref{fig:surfaceTension}].
In particular, when the faster-diffusing species is locally enriched within the high-density phase (${D>0}$), the effective surface tension can be strongly reduced, suggesting the presence of long-lived capillary fluctuations.

Taken together, our analysis shows that chemical activity modifies phase coexistence in three distinct ways:
\begin{itemize}
    \item[(i)] It alters the chemical-potential balance through a composition-dependent contribution, which reflects the coupling between transport and local composition\@~[Eq.\@~\eqref{eq:ModifiedChemical}].
    \item[(ii)] It introduces an additional contribution to the pressure balance that depends on how the reactive fixed-point composition varies with density\@~[Eq.\@~\eqref{eq:ModifiedMechanical}].
    \item[(iii)] Lastly, chemical activity modifies the Laplace pressure through a renormalization of the interfacial tension, thereby affecting curvature-induced corrections to coexistence.
\end{itemize}

\begin{figure}[]
    \centering
    \includegraphics[]{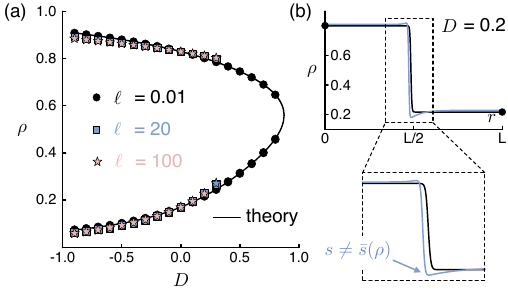}
    \caption{\textbf{Non-equilibrium coexistence}. (a) Coexisting plateau densities (markers) obtained from numerical simulations of a ternary mixture governed by Eq.\@~\eqref{eq:RescaledDynamics}.
    Simulations were performed for a one-dimensional systems of size ${L=500}$ (black circles, red squares) and ${L=1000}$ (blue stars), with no-flux boundary conditions, an average solute volume fraction ${\bar\rho=0.5}$, and ${\ell=\{0.01,20,100\}}$.
    Plateau values were extracted from the stationary solute density profiles [panel (b)] at the system boundaries, provided the system separates into only two macroscopic phases.
    The solid black line shows the theoretical coexistence densities in the limit of rapid chemical turnover, obtained by numerically solving Eq.\@~\eqref{eq:ModifiedCoexistence}.
    (b) Stationary solute profile $\rho(r)$ for ${D=0.2}$, and ${\ell = \{0.01, 20\}}$ (black/red).
    The remaining parameters used for both panels are ${\chi=2.4}$ and ${K=0.5}$.}
    \label{fig:binodals}
\end{figure}

As for an ordinary binary mixture, determining the coexisting solute volume fractions $\rho_{1,2}$ (binodals) corresponds to a common-tangent construction of the homogeneous part of the effective free energy, Eq.\@~\eqref{eq:NeqFreeEnergy}; see Fig.\@~\ref{fig:freeEnergy}.
However, since the effective free energy is no longer symmetric, this construction has to be carried out numerically, by numerically solving the modified coexistence conditions, Eqs.\@~\eqref{eq:ModifiedCoexistence}.
As shown in Fig.\@~\ref{fig:binodals}(a), for a ternary mixture, the coexistence densities obtained in this way agree with the exact plateau values extracted from direct numerical solutions of Eq.\@~\eqref{eq:RescaledDynamics}.

For slower chemical turnover, corresponding to finite values of the parameter $\ell$ [Eq.\@~\eqref{app:EffectiveParameters}], one observes small deviations between the common-tangent prediction and the simulation results~[Fig.\@~\ref{fig:binodals}(a)].
These originate from the diffusive coupling between the solute volume fraction~$\rho$ and the local composition~$s$ [Eq.\@~\eqref{eq:RescaledDynamics}]:

As shown in Fig.\@~\ref{fig:binodals}(b), far from the interface the system approaches a homogeneous plateau with ${s\approx\bar s(\rho)}$.
However, near the interface, the system is no longer in local chemical equilibrium.
As a result, the diffusive coupling between density and composition fields introduces an intrinsically nonlocal contribution to the dynamics, which is not captured by adiabatic elimination of the composition variable $s$.
Nevertheless, Fig.\@~\ref{fig:binodals}(a) shows that for macroscopically large phases the effective free energy in Eq.\@~\eqref{eq:NeqFreeEnergy} still provides a reasonable approximation of the coexisting plateau densities, even for ${\ell\gg1}$.

\section{Arrested coarsening}
\label{sec:singleDroplet}
A description based on the local effective free energy, Eq.\@~\eqref{eq:NeqFreeEnergy}, implies continued coarsening, just as for an equilibrium mixture.
It, thus, cannot account for the emergence of arrested states.
Instead, to identify the physical mechanism underlying arrested coarsening, we develop a genuinely non-equilibrium description based on a sharp-interface approximation.

For simplicity, we focus on a ternary mixture in a dilute regime, where individual droplets are well separated and interact only through a common background density.
In this limit, many-body effects are negligible, and the late-time dynamics can be reduced to the behavior of a single isolated droplet.
Accordingly, we consider a spherical droplet of radius $R$ embedded in an infinite two- or three-dimensional medium.
Using spherical coordinates with the droplet centered at the origin, we construct approximate steady-state solutions of the dynamical equations [Eq.\@~\eqref{eq:RescaledDynamics}].

A necessary condition for arrested coarsening is the existence of a stationary solution at finite $R$.
In the absence of such a fixed point, droplets either shrink and disappear (${R\to0}$) or grow without bound (${R\to\infty}$), corresponding to suppressed phase separation or unbounded coarsening, respectively.
Depending on the system parameters, we find that the model admits zero, one, or two stationary solutions, which we associate with the regimes of suppressed phase separation, unbounded coarsening, and arrested coarsening, respectively.
Finally, we compare the theoretical predictions obtained from this analysis with direct numerical simulations.
\subsection{Non-equilibrium steady-state conditions}
\label{app:steady_state} 

Far from the droplet, the total solute volume fraction approaches a homogeneous background value $\rho_\infty$.
The absence of gradients in the far field implies that both the local composition and the chemical potential relax to their local chemical equilibrium values,
\begin{align}
    s_\infty 
    =
    \bar s(\rho_\infty)
    \,, \qquad 
    \mu_\infty
    =
    f'(\rho_\infty) \, .
    \label{app:FarFieldEquilibrium}
\end{align}

In the steady state, the absence of net solute mass fluxes further requires the mass-redistribution potential to be spatially uniform, ${\boldsymbol{\nabla}\eta = 0}$\@~[Eq.\@~\eqref{eq:RescaledRhoDynamics}].
Using the explicit form of $\eta$ [Eq.\@~\eqref{eq:TernaryMassRedistribution}], this yields
\begin{align}
    (1 + D s)\,\boldsymbol{\nabla}\mu 
    + D\,\boldsymbol{\nabla}s 
    = 
    0 \, .
    \label{app:stationarity_eta}
\end{align}
For $D=0$, this condition reduces to the familiar equilibrium condition ${\boldsymbol{\nabla}\mu=0}$, whereas for ${D\neq 0}$, composition gradients contribute an additional non-equilibrium drive that modifies the coexistence conditions.
Evaluating Eq.\@~\eqref{app:stationarity_eta} at the droplet interface (${r=R}$) and in the far field (${r\to\infty}$) yields the exact \emph{steady-state condition}
\begin{align}
    \mu_\infty - \mu(R) 
    =
    \log
    \left[
        \frac{1 + D\, s(R)}{1 + D\, s_\infty}
    \right] \, ,
    \label{eq:steady-state-condition}
\end{align}
which shows that a finite droplet size can only be maintained if the chemical-potential difference between the droplet and the far field is exactly balanced by a corresponding change in local composition.

To determine the full steady-state profiles, we substitute the stationarity condition Eq.~\eqref{app:stationarity_eta} into the steady-state equation for the composition field, Eq.~\eqref{eq:RescaledSDynamics}.
This yields the profile equation
\begin{subequations}
\label{app:steadyStateCondition}
\begin{align}
    \boldsymbol{\nabla}\left[K(\rho,s)\,
    \boldsymbol{\nabla} s \right]
    + 
    \ell^{-2} \, \mathcal{R}(\rho,s) 
    = 0 \, ,
    \label{eq:ProfileEquation}
\end{align}
with the effective mobility
\begin{align}
    K(\rho,s) \equiv M(\rho)\,
    \frac{1 - D^2}{1 + D\,s} \, .
\end{align}
\end{subequations}

Equation\@~\eqref{app:steadyStateCondition} highlights a key non-equilibrium feature absent in equilibrium systems:
Composition gradients\@~$\boldsymbol{\nabla}s$ can be sustained in the steady state provided that spatially inhomogeneous reactions (${\mathcal{R}\neq 0}$) continuously drive the system out of equilibrium.
In the equilibrium limit ${\mathcal{R}=0}$, stationarity enforces ${\boldsymbol{\nabla}s=0}$, and the composition becomes spatially uniform.

\subsection{Sharp-interface approximation}
\label{sec:sharpInterface}
\begin{figure*}
\centering
\includegraphics[width=\linewidth]{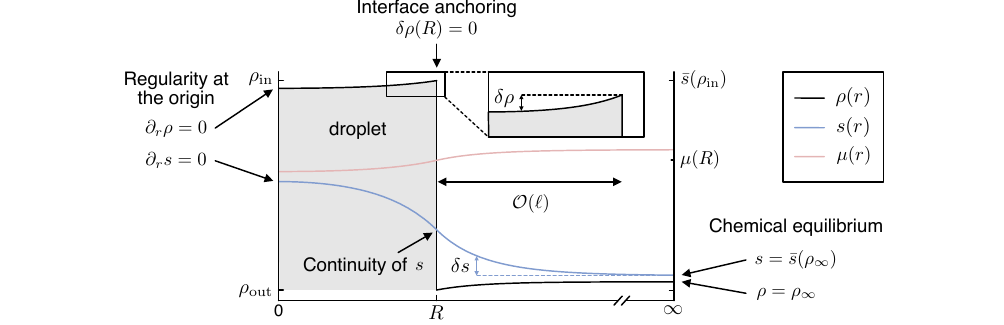}
\caption{\textbf{Sharp-interface construction}. An infinitely sharp interface at ${r=R}$ separates the high-density phase (gray) from the low-density phase (white). The solute volume fraction $\rho$ (black) remains close to the equilibrium coexistence densities\@~$\rho_\text{in/out}$\@~[Eq.\@~\eqref{app:linearGibbsThomson}], and exhibits a discontinuous jump across the interface.
In contrast, the chemical potential (red) and composition (blue) profiles vary smoothly over the characteristic length scale\@~$\ell$ and are continuous across the interface.
Regularity of the solutions enforces vanishing gradients of the density and composition profiles at the origin, while in the far field $\rho$ relaxes towards the fixed background density $\rho_\infty$, and the composition $s$ reaches local chemical equilibrium ${s=\bar s(\rho_\infty)}$. 
Interface anchoring ensures that density perturbations $\delta\rho$ vanish at\@~${r=R}$.}
\label{figAPP:sharpInterface}
\end{figure*}

The quasilinear elliptic equation for the stationary composition, Eq.\@~\eqref{app:steadyStateCondition}, does not admit a closed-form solution in general. 
To make analytical progress, we therefore adopt a sharp-interface approximation\@~(Fig.\@~\ref{figAPP:sharpInterface}) that assumes a separation of length scales.
Specifically, we assume that the (equilibrium) interfacial width $w$ between the high- and low-density regimes is small compared to both the droplet radius $R$ and the reactive screening length $\ell$.

The first condition ${w \ll R}$ is satisfied for macroscopically large droplets, while ${w \ll \ell}$ corresponds to slow reactive turnover (weak chemical activity), since\@~$\ell$ increases with the reaction timescale\@~$\tau$\@~[Eq.\@~\eqref{app:EffectiveParameters}]. 

Additionally, we restrict our analysis to systems sufficiently far from criticality.
For an equilibrium binary mixture, the interfacial width scales as
${w^2 \sim \kappa / f''(\rho_c)}$\@~\cite{Weber.2019}, such that
${w \to \infty}$ as ${\chi \downarrow 2}$ and ${f''(\rho_c) \to 0}$ [Eq.\@~\eqref{app:FHdensity}],
rendering the sharp-interface approximation invalid near the critical point.

Within the sharp interface limit, the composition $s$ remains essentially constant across the interface region, whereas the solute profile\@~$\rho$ exhibits a discontinuous jump (Fig.\@~\ref{figAPP:sharpInterface}). 
This allows us to simplify the non-linear profile equation, Eq.\@~\eqref{app:steadyStateCondition}, by treating the droplet interior and exterior as two quasi-homogeneous regions that can be analyzed independently.
The two regions are then coupled solely through boundary conditions at the droplet interface, enforcing continuity of fluxes and the appropriate matching of density and composition profiles\@~[Fig.\@~\ref{figAPP:sharpInterface}].

\subsection{Piecewise linearization of the solute profile} 
In the sharp-interface limit, the composition $s$ varies only weakly across the narrow interface, such that locally ${\boldsymbol{\nabla}s \approx 0}$ (Fig.\@~\ref{figAPP:sharpInterface}).
As a result, the stationarity condition Eq.\@~\eqref{app:stationarity_eta} reduces to ${\boldsymbol{\nabla}\mu \approx 0}$ across the interface.
This implies that the interface is in \emph{local thermodynamic equilibrium}, even though the system as a whole is driven out of equilibrium.
As a result, the chemical potential at the interface (${r=R}$) is fixed by the Gibbs-Thomson relation \cite{Eyre.1993,Garcke.2000} (App.\@~\ref{sec:phaseCoexistence})
\begin{align}
    \mu(R) &\approx \frac{(d - 1) \, \gamma}{(\rho_+ - \rho_-) R} \, .
\end{align}
Similarly, the solute volume fractions adjacent to the interface are given by the curvature-corrected coexistence solute volume fractions $\rho_{\mathrm{in}}(R)$ and $\rho_{\mathrm{out}}(R)$ of an equilibrium droplet.
This allows us to approximate the stationary density profile $\rho (r)$ in the inner and outer domains by expanding
\begin{subequations}\label{eq:rho_linearization_radial}
\begin{align}
  \rho(r) &\equiv \rho_\text{in} + \delta\rho_\text{in}(r), && 0 \le r < R \, ,\\
  \rho(r) &\equiv \rho_\text{out} + \delta\rho_\text{out}(r), && r > R \, ,
\end{align}
\end{subequations}
where we enforce \emph{interface anchoring}
\begin{align}
    \delta\rho_\text{in}(R)
    =
    \delta\rho_\text{out}(R)
    =
    0 \, .
\label{eq:anchoring_rho_radial}
\end{align}

Since gradients in the density are subleading away from the interface, the chemical potential to leading order obeys the \emph{local} bulk relation
\begin{align}
\label{eq:bulk_local_relation_radial}          
    \mu(r) \;\simeq\; f'(\rho) \, ,
    \qquad \text{for } r\neq R
    \, .
\end{align}
This implies a proportionality relation between local shifts in the chemical potentials and the densities,
\begin{align}
    \delta\mu_{\alpha}(r)
    &\equiv
    \mu(r) - f'(\rho_{\alpha}) \notag \\
    &\approx
    f''(\rho_{\alpha})\;
    \delta\rho_{\alpha}(r) \, ,
\label{eq:bulk_closure_radial}
\end{align}
where ${\alpha \in \{\text{in, out}\}}$, and ${f'(\rho_\alpha) = \mu(R)}$ due to local equilibrium at the interface.
A detailed analysis of the conditions under which this linearization is justified, i.e., when the deviations $\delta\rho_\alpha(r)$ can be treated as small, is provided in App.\@~\ref{sec:validity}.

Since in steady state the mass-redistribution potential is uniform, ${\eta = \eta_\text{stat}}$, there is also a local relation between the binary chemical potential and composition [Eq.\@~\eqref{eq:TernaryMassRedistribution}], 
\begin{align}
    \mu(r) \;=\; \eta_\text{stat} -\log\big[1+D\,s(r)\big]
    \, ,
\label{eq:mu_from_eta_radial}
\end{align}
which, when combined with Eq.~\eqref{eq:bulk_closure_radial}, yields the relations between the density shifts $\delta \rho_\alpha (r)$ (${\alpha\in\{\text{in},\text{out}\}}$) and the local composition $s(r)$
\begin{subequations}\label{eq:rho_profiles_LO_radial}
\begin{align}
    \delta \rho_\alpha(r)
    &\approx
    \frac{\eta_\text{stat} - \log\!\big[1+D\,s(r)\big] - f'(\rho_\alpha)}
           {f''(\rho_\alpha)}\,.
\end{align}
\end{subequations}

Far-field matching to the prescribed background [Eq.\@~\eqref{app:FarFieldEquilibrium}] fixes the value of the stationary mass-redistribution potential via
\begin{align}
    \eta_\text{stat} = \mu_\infty + \log(1+Ds_\infty)\, .
    \label{eq:EtaStat}
\end{align}
Alternatively, the constant value of the mass-redistribution potential can be fixed at the interface such that [Eq.\@~\eqref{eq:steady-state-condition}] $${\eta_\text{stat} = \mu(R) + \log[1 + D \, s(R)]}\, .$$ 
Then, one gets
\begin{align}
\label{eq:clousre-relation-full}
    \delta \rho_\alpha(r)
    \approx
    \frac{\mu(r) - \mu (R)}{f''(\rho_\alpha)} 
    \approx
    \frac{1}{f''(\rho_\alpha)}
    \log\frac{1+D\,s(R)}{1+D\,s(r)} \, .
\end{align}

From the latter equation we find that the sign of the density shift $\delta\rho_\alpha(r)$ is fixed by the monotonicity of the logarithm:
For ${D>0}$ (A-state diffuses faster) and $s$ larger inside than outside (A enriched inside the droplet), one has $${1+D\,s(R) > 1+D\,s(r)} \, \, \quad \text{for }\quad r>R \, .$$
Hence, ${\delta\rho_\text{out}(r)>0}$, corresponding to a slight oversaturation compared to the equilibrium case, while for ${r<R}$, ${\delta\rho_\text{in}(r)<0}$, corresponding to a slight undersaturation inside (Fig.\@~\ref{figAPP:sharpInterface}).

\subsection{Composition anchoring}
\label{sec:CompositionAnchoring}
Concerning the stationary composition profile $s(r)$, there are two related questions: 
Can it be systematically expanded around suitable reference values, and what is a good choice for these anchoring values? 
Unlike the density field, which exhibits a sharp discontinuity across the interface, the composition field is continuous and varies smoothly on the reactive length scale\@~$\ell$.
As a result, the total variation of $s$ within either bulk domain is not necessarily small\@~(Fig.\@~\ref{figAPP:sharpInterface}).

For \emph{large droplets} (${R \gg \ell}$), the stationary composition profile $s(r)$ interpolates between an interior plateau value\@~$\bar s_\mathrm{in}$ and the far-field value $s_\infty=\bar s(\rho_\infty)$ (Fig.\@~\ref{figAPP:sharpInterface}).
A natural choice in this regime is therefore a \emph{plateau anchoring},
\begin{subequations}
\begin{alignat}{3}
  s_\mathrm{in}(r)
  &\equiv
  \bar s_\mathrm{in}+\delta s_\mathrm{in}(r)\,, &\qquad& 0\le r<R\, ,\\
  s_\mathrm{out}(r)
  &\equiv
  s_{\infty}+\delta s_\mathrm{out}(r)\,, &\qquad& r>R \, .
\end{alignat}
\end{subequations}
Note however, that the interior plateau value $\bar s_\mathrm{in}$ is generally not known and must be determined self-consistently, although a first estimate may be obtained from the effective free-energy description discussed in Sec.\@~\ref{sec:FastTurnover}.

Alternatively, if the system is only weakly super- or undersaturated, such that ${\rho_\infty \approx \rho_\mathrm{out}}$, and the interior plateau differs only weakly from the local reactive equilibrium value $\bar s(\rho_\mathrm{in})$, a more convenient anchoring is obtained by expanding about the reactive equilibria associated with the coexistence densities,
\begin{align}
      s_\alpha(r)
      \equiv
      \bar s_\alpha+\delta s_\alpha(r),\qquad 
      \bar s_\alpha\equiv\bar s(\rho_\alpha)\, .
\end{align}

In both anchoring schemes, however, the deviations\@~$\delta s_\alpha(r)$ cannot be assumed to be small everywhere.
In particular, near the interface, the deviation from the respective bulk anchoring value is of order
$\mathcal{O}(s_\infty-\bar s_\mathrm{in})$ or $\mathcal{O}(\bar s(\rho_\mathrm{in})-\bar s(\rho_\mathrm{out}))$, which, depending on the reaction scheme, can be of order unity.

Conversely, for \emph{small droplets} (${R\ll \ell}$), the composition cannot relax to its inner reactive equilibrium within the droplet. 
Starting from its far-field value $s_\infty$, the composition typically varies only weakly across the droplet and remains close to this reference value throughout the system.
In this regime, plateau anchoring inside the droplet is no longer appropriate.
Instead, a more suitable choice is to expand around a common reference value\@~$s^*$ in both bulk domains
\begin{align}
    s_\alpha (r) 
    \equiv
    s^* + \delta s_\alpha (r) \, .
\end{align}
Possible choices for $s^*$ include the far-field composition\@~$s_\infty$ or, alternatively, the reactive equilibrium values\@~$\bar s(\rho_\alpha)$ evaluated at the interfacial coexistence densities.

Lastly, one may choose to anchor the expansion at the interfacial composition value $s(R)$ itself, corresponding to \emph{interface anchoring},
\begin{align}
    s_\alpha (r) 
    \equiv
    s(R) + \delta s_\alpha (r) \, ,
\end{align}
with the advantage that\@~${\delta s_\alpha(R) = 0}$ on both sides of the interface.
This mirrors the choice made for the density field, and ensures that both density and composition fluctuations vanish at the interface.
However, one again faces the problem that\@~$s(R)$ needs to be determined self-consistently.

Since the optimal anchoring strategy depends on the droplet size, the reaction kinetics, and the degree of supersaturation, no single choice is universally appropriate.
We therefore adopt the general ansatz
\begin{align}
    s_\alpha (r) 
    \equiv
    \tilde s_\alpha + \delta s_\alpha (r) \, ,
    \label{app:compositionExpansion}
\end{align}
which allows the reference values $\tilde s_\alpha$ to be specified a posteriori.
This flexible formulation enables a unified treatment of both small- and large-droplet limits, as well as different reaction schemes.

\subsection{Linearization of the composition profile}

Linearizing the equation for the stationary composition profile,  Eq.\@~\eqref{app:steadyStateCondition}, to leading order in $\delta s_\alpha$ and $\delta\rho_\alpha$ gives
\begin{align}
    K_\alpha\,
    \boldsymbol{\nabla}^2
    \delta s_\alpha
    +
    \ell^{-2}
    \big[ \mathcal R|_\alpha + 
    \partial_s\mathcal R|_\alpha \,\delta s_\alpha
    +
    \partial_\rho\mathcal R|_\alpha \, \delta\rho_\alpha
    \big]
    = 0 
    \,,
\label{app:s-profile-start}
\end{align}
where $|_\alpha$ indicates evaluation at $(\rho_\alpha,\tilde s_\alpha)$, and we have defined kinetic coefficients
\begin{align}
    K_\alpha
    \equiv
    M(\rho_\alpha)\;
    \frac{1-D^2}{1+D\,\tilde s_\alpha}
    \, .
\label{app:Kalpha}
\end{align} 

To close Eq.\@~\eqref{app:s-profile-start}, we need a relation between the variations in density and composition.
Using  Eq.\@~\eqref{eq:mu_from_eta_radial}, the chemical potential $\mu$ and local composition are related via
\begin{align}
    \mu(r)-f'(\rho_\alpha) 
    +
    \log 
    \left(
    \frac{1+D \, s_\alpha(r)}{1+D\,\tilde s_\alpha}
    \right)
    =
    C_\alpha \, ,
    \label{app:constPotential2}
\end{align}
where the right-hand side is the \emph{region-wise constant}
\begin{align}
    C_\alpha 
    \equiv
    \eta_{\text{stat}}
    -
    f'(\rho_\alpha)
    -
    \log
    \big(
    1+D\,\tilde s_\alpha
    \big) 
    \, .
    \label{eq:region-wise-constant-C}    
\end{align}
Linearizing the logarithm around $\tilde s_\alpha$ gives
\begin{align}
    \log\left(
    \frac{1+D \, s_\alpha(r)}{1+D\,\tilde s_\alpha}
    \right)
    \simeq
    \frac{D\, \delta s_\alpha(r)}{1+D\,\tilde s_\alpha}\;
    \, ,
\end{align}
and using Eq.\@~\eqref{eq:bulk_closure_radial} yields the \emph{closure relation}
\begin{align}
    f''(\rho_\alpha)\,
    \delta\rho_\alpha(r)
    +
    \frac{D}{1+D\,\tilde s_\alpha}\;
    \delta s_\alpha(r)
    =
    C_\alpha
    \, ,
\label{eq:closure-relation-rho-s}
\end{align}
valid to leading order in $(\delta\rho_\alpha,\delta s_\alpha)$.
Introducing the abbreviations
\begin{align}
    \beta_\alpha
    \equiv
    \frac{D}{f''(\rho_\alpha)\,[\,1+D\,\tilde s_\alpha\,]}
    \, ,
    \qquad
    c_\alpha
    \equiv
    \frac{C_\alpha}{f''(\rho_\alpha)} 
    \, ,
    \label{app:Beta-and-small-c}
\end{align}
the closure relation, Eq.\@~\eqref{eq:closure-relation-rho-s}, can be recast in compact form as
\begin{align}
    \delta\rho_\alpha(r)
    \;=\; 
    -\,\beta_\alpha\,
    \delta s_\alpha (r)
    \;+\; 
    c_\alpha
    \,.
\label{eq:closure-relation-rho-s-short}
\end{align}

Here, ${\beta_\alpha}$ quantifies the linear susceptibility of the density to composition changes within phase ${\alpha}$, while ${c_\alpha}$ denotes phase-dependent integration constants, which are related via [Eqs.\@~\eqref{eq:region-wise-constant-C},\@~\eqref{app:Beta-and-small-c}]
\begin{align}
    &c_\text{in} f''(\rho_\text{in})+\log(1+D \, \tilde s_\text{in}) \notag \\  = \;
    &c_\text{out} f''(\rho_\text{out})+\log(1+D \, \tilde s_\text{out}) \,.
    \label{app:small-c-restriction}
\end{align}
Substituting the closure relation, Eq.\@~\eqref{eq:closure-relation-rho-s-short}, into the linearized profile equation for the composition, Eq.\@~\eqref{app:s-profile-start}, gives a closed equation for the composition shifts\@~$\delta s_\alpha (\boldsymbol{r})$:
\begin{align}
    0 = \;
    &K_\alpha\,
    \boldsymbol{\nabla}^2
    \delta s_\alpha (\boldsymbol{r})
    +
    \ell^{-2}
    \Big[ 
    \mathcal R|_\alpha
    +
    \partial_\rho\mathcal R|_\alpha
    \,c_\alpha
    \notag \\ 
    &+\big(
    \partial_s \mathcal R|_\alpha
    -
    \beta_\alpha\,
    \partial_\rho\mathcal R|_\alpha
    \big) \,
    \delta s_\alpha (\boldsymbol{r})
    \Big]\, .
\label{app:s-profile-inhom}
\end{align}
This yields the \emph{inhomogeneous modified Helmholtz} (screened Poisson) equation
\begin{align}
    \boldsymbol{\nabla}^2 \, 
    \delta s_\alpha (\boldsymbol{r})
    - \ell_\alpha^{-2}\,
    \delta s_\alpha (\boldsymbol{r})
    =
    -\,S_\alpha
    \, ,
\label{app:s-profile-Helmholtz}
\end{align}
where we have defined the \emph{source term} 
\begin{align}
    S_\alpha
    \equiv
    \frac{\mathcal R|_\alpha+ \partial_\rho\mathcal R|_\alpha \, c_\alpha }{K_\alpha\, \ell^2}
    \, ,
\end{align}
and the \emph{reactive composition screening lengths}
\begin{align}
    \ell_\alpha^{2}
    \equiv
    \frac{K_\alpha\,\ell^2}{-\,\partial_s\mathcal R|_\alpha+\beta_\alpha\,\partial_\rho\mathcal R|_\alpha}
    \, .
\label{eq:screening_length}
\end{align}
We refer to $\ell_\alpha$ as reactive composition screening lengths, since they set the exponential relaxation scale of $s(r)$ inside and outside the droplet.
Absorbing the constant source into a shifted target yields the offset form
\begin{align}
    \boldsymbol{\nabla}^2 \, \delta s_\alpha 
    + \ell_\alpha^{-2}\,
    \big(
    \delta s_\alpha^0-\delta s_\alpha
    \big)
    = 0
    \,,
\label{app:s-profile-offset-pde}
\end{align}
where the offsets are given by
\begin{align}
    \delta s_\alpha^0
    \;=\;
    S_\alpha \, \ell_\alpha^2
    \;=\; 
    \frac{\mathcal R|_\alpha+\partial_\rho\mathcal R|_\alpha\,c_\alpha}{-\,\partial_s\mathcal R|_\alpha+\beta_\alpha\,\partial_\rho\mathcal R|_\alpha} 
    \, .
    \label{app:offsets}
\end{align}

To solve these piecewise linear Helmholtz problems, Eq.\@~\eqref{app:s-profile-offset-pde}, they have to be supplemented by additional boundary conditions at the origin (${r=0}$), and at the interface\@~(${r=R}$).
Specifically, regularity at the origin enforces a non-singularity condition:
\begin{align}
    \partial_r s\big|_{r=0}
    =
    0 \, .
    \label{app:regularitcCondition}
\end{align}

Additionally, our sharp interface approximation implies that the composition field $s(r)$ is continuous across the interface 
\begin{align}
    s\big|_{r \to R^-} 
    = 
    s\big|_{r \to R^+}  \, .
\label{app:ContinuityCondition}
\end{align}

Lastly, integrating the modified Helmholtz equation, Eq.\@~\eqref{app:steadyStateCondition}, over a thin spherical shell ${\mathcal{S}_\varepsilon=\{\,R-\varepsilon < r < R+\varepsilon\,\}}$ and applying the divergence theorem yields
\begin{align*}
        0
        &=
        \int_{\mathcal S_\varepsilon}
        \boldsymbol{\nabla}
        \cdot
        \big[K(\rho,s)\,\boldsymbol{\nabla}s\big]
        \, {\mathrm d} V
       + 
       \frac{1}{\ell^2}   
        \int_{\mathcal S_\varepsilon} \mathcal R(\rho,s) \, {\mathrm d}V
        \notag \\
        &= \int_{\partial\mathcal S_\varepsilon}
        K(\rho,s) \,\partial_r s \, 
        {\mathrm d}A
        + 
        {\mathcal O} (\varepsilon) \, ,
\end{align*}
where $\partial_r$ is the outward normal derivative. 
Taking ${\varepsilon \to 0}$ yields the condition
\begin{align}
    -K_\text{in} \,\partial_r s\big|_{r \to R^-} 
    =
    -K_\text{out} \,\partial_r s\big|_{r \to R^+} \, .
    \label{app:FluxContinuity}
\end{align}
corresponding to the continuity of diffusive composition fluxes across the interface.
\subsection{Radially symmetric composition profiles}

In radial symmetry, the Laplace operator is given by
\begin{align}
    \boldsymbol{\nabla}^2
    =
    \partial_r^2
    +
    \frac{d-1}{r}\, \partial_r \, .
\end{align}
Defining ${u_\alpha(r) \equiv \delta s_\alpha(r)-\delta s_\alpha^0}$, the modified Helmholtz equation, Eq.\@~\eqref{app:s-profile-offset-pde}, thus, becomes
\begin{align}
    \partial_r^2u_\alpha
    +
    \frac{d-1}{r} \, \partial_r u_\alpha 
    - 
    \frac{1}{\ell_\alpha^2}\,u_\alpha
    =
    0 \, .
\end{align}
To proceed, we define the auxiliary variable $${u_\alpha(r) = r^{-\nu}\,y_\alpha(z)}\, ,$$ with ${z=r/\ell_\alpha}$. 
Choosing ${2\nu=d-2}$, the differential equation simplifies to the \emph{modified Bessel equation}
\begin{align}
    z^2 \, \partial_z^2 y + z \, \partial_zy - (z^2 + \nu^2)\,y 
    = 0
    \,,
\end{align}
whose two independent solutions are the \emph{modified Bessel functions} of order\@~$\nu$:
\begin{subequations}
\begin{alignat}{2}
    &I_\nu(z)
    \quad
    &&\text{(regular at $z=0$)} \, , \\
    &K_\nu(z)
    \quad
    &&\text{(decays for $z\to\infty$)}
    \, .
\end{alignat}
\end{subequations}
Enforcing regularity at the origin and decay at infinity yields the bulk solutions
\begin{subequations}
\label{app:FirstSolution}
\begin{align}
    s_\text{in}(r)  
    &=  \tilde s_\text{in} + \delta s_\text{in}^0 
    + B_\text{in}\,H_\text{in}(r) \, ,\\
    s_\text{out}(r) 
    &=  \tilde s_\text{out} + \delta s_\text{out}^0 
    + B_\text{out}\,H_\text{out}(r) \, , 
\end{align}
\end{subequations}
with 
\begin{subequations}
\begin{align}
    H_\text{in}(r) 
    &\equiv
    r^{-\nu} 
    I_\nu \big(r/\ell_\text{in}\big) \, , 
    \\
    H_\text{out}(r)
    &\equiv 
    r^{-\nu} K_\nu \big(r/\ell_\text{out}\big)\, .
\end{align}
\end{subequations}

This leaves us with \textit{four unknown parameters}: 
The amplitudes\@~$B_\alpha$, the stationary value of the mass-redistribution potential $\eta_\text{stat}$, and the stationary droplet radius $R$.
Since\@~$\eta_\text{stat}$ enters the offsets\@~$\delta s^0_\alpha$ [Eqs.\@~\eqref{eq:region-wise-constant-C},\@~\eqref{app:Beta-and-small-c},\@~\eqref{app:offsets}], they should also be considered undetermined, although only one of them is truly independent, as they are related by the constraint in Eq.\@~\eqref{app:small-c-restriction}.

These unknown parameters need to be determined from the \textit{five remaining boundary conditions}: 
Two from interface anchoring of the density shifts\@~$\delta \rho(r)$\@~[Eq.\@~\eqref{eq:anchoring_rho_radial}], one each from continuity of the composition profile $s(r)$ [Eq.\@~\eqref{app:ContinuityCondition}] and the diffusive flux ${J(r)=-K_\alpha\partial_rs}$\@~[Eq.\@~\eqref{app:FluxContinuity}], and one from the far-field boundary condition ${\lim_{r\to\infty}\rho(r)=\rho_\infty}$, which fixes the value of the stationary potential $\eta_\text{stat}$ [Eq.\@~\eqref{eq:EtaStat}].
As we will show, this apparent over-determination is resolved by an appropriate choice of the reference state\@~$\tilde s_\alpha$.

\subsubsection*{Interface anchoring of the density}

Interface anchoring of the inner and outer densities\@~[$\delta\rho_\alpha(R)=0$] implies [Eq.\@~\eqref{eq:closure-relation-rho-s-short}]
\begin{align}
    \beta_\alpha \, 
    \delta s_\alpha(R)
    =
    c_\alpha \, ,
\end{align}
or equivalently [Eq.\@~\eqref{app:FirstSolution}]
\begin{align}
    \beta_\alpha \left[\delta s_\alpha^0 + B_\alpha H_\alpha(R)\right]
    =
    c_\alpha 
    \, ,
\end{align}
from which one infers
\begin{subequations}
\label{app:SecondSolutionStep}
\begin{align}
    s_\text{in}(r) 
    &= \tilde s_\text{in} + \delta s_\text{in}^0 + \left(\frac{c_\text{in}}{\beta_\text{in}}-\delta s^0_\text{in}\right)\, \frac{ H_\mathrm{in}(r)}{H_\mathrm{in}(R)} \, , \\
    s_\text{out}(r) 
    &\;=\; \tilde s_\text{out} + \delta s_\text{out}^0 + \left(\frac{c_\text{out}}{\beta_\text{out}}-\delta s^0_\text{out}\right)\, \frac{H_\mathrm{out}(r)}{H_\mathrm{out}(R)} \, .    
\end{align}
\end{subequations}
\subsubsection*{Far field relaxation}
\label{sec:farFieldRelaxation}
Evaluating the closure relation [Eq.\@~\eqref{eq:closure-relation-rho-s-short}] in the far-field\@~${(r\to\infty)}$ yields
\begin{align}
    c_\text{out}
    = 
    \rho_\infty - \rho_\text{out}
    + 
    \beta_\text{out} \, \delta s_\text{out}^0 
    \, ,
    \label{app:closure-at-infinity}
\end{align}
where we imposed ${\delta\rho(\infty)=\rho_\infty-\rho_\text{out}}$.
This ensures that the density profile converges to the correct value $\rho_\infty$.
Assuming ${c_\mathrm{out}\neq0}$, and inserting the definition of the offset\@~$\delta s^0_\text{out}$, Eq.\@~\eqref{app:offsets}, we find
\begin{align}
    c_\text{out} 
    =
    (\rho_\infty-\rho_\text{out})\left[1+\beta_\text{out} \bar s^\prime(\rho_\text{out})\right]
    - \beta_\text{out} \frac{\mathcal R|_\alpha }{\partial_s\mathcal R|_\alpha} \, .
    \label{app:General-cout}
\end{align}
Using Eq.\@~\eqref{app:offsets}, the composition offset in the outer domain is, thus, given by 
\begin{align}
    \delta s^0_\text{out} 
    =
    -\frac{\mathcal R|_\text{out} }{\partial_s\mathcal R|_\text{out}} + \bar s^\prime(\rho_\text{out}) (\rho_\infty-\rho_\text{out}) \,.
    \label{app:generalOuterOffset}
\end{align}
When expanding around ${\tilde s_\text{out}=\bar s(\rho_\text{out})}$, one has ${\mathcal R |_\text{out}=0}$ and the above expression reduces to 
\begin{align}
    \delta s^0_\text{out} 
    =
    \bar s^\prime(\rho_\text{out}) (\rho_\infty-\rho_\text{out}) \, ,
\end{align}
whereas for ${\tilde s_\text{out}}$ close to ${\bar s(\rho_\text{out})}$, we linearize in the difference ${\tilde s_\text{out}-\bar s(\rho_\text{out})}$ to obtain
\begin{align}
    \delta s^0_\text{out} 
    \approx
    \tilde s_\text{out}- \bar s(\rho_\text{out}) + \bar s^\prime(\rho_\text{out}) (\rho_\infty-\rho_\text{out}) \, .
\end{align}
For the special choice ${\tilde s_\text{out}=s_\infty}$, the latter gives
\begin{align}
    \delta s^0_\text{out} 
    =
    0 \, .
    \label{app:farFieldExpansionDelta}
\end{align}
That is, when expanding around the far-field composition, one recovers ${\lim_{r\to\infty}\delta s(r)=0}$ as expected.

The only exception to Eq.\@~\eqref{app:generalOuterOffset} occurs if one applies interface anchoring of the composition shift $\delta s_\text{out}(r)$, i.e., expands around ${\tilde s_\text{out}=s(R)}$.
In this case, ${c_\text{out}=0}$ [Eq.\@~\eqref{eq:closure-relation-rho-s-short}], and Eq.\@~\eqref{app:closure-at-infinity} reduces to
\begin{align}
    \big[
    1+\beta_\text{out} \, \bar s '(\rho_\text{out})
    \big] \;
    ( \rho_\infty - \rho_\text{out})
    \;=\; 
    \beta_\text{out} \;
    \frac{\mathcal R|_\text{out}}{\,\partial_s\mathcal R|_\text{out}}  \, .
\end{align}
Since $\beta_\text{out}$ and the reaction terms $\mathcal R|_\text{out} \, ,\partial_s\mathcal R|_\text{out}$ all depend on the reference value ${\tilde s_\text{out}=s(R)}$, this yields an implicit equation for $s(R)$, which needs to be solved self-consistently at fixed $R$.
The constant offset $\delta s^0_\text{out}$ then follows from its definition in Eq.\@~\eqref{app:offsets}.

In summary, far-field relaxation of the density profile towards $\rho_\infty$ fixes the value of $\delta s_\text{out}^0$ (or $c_\text{out}$), and thus also the stationary mass-redistribution potential $\eta_\text{stat}$.

\subsubsection*{Continuity of the composition profile}
\label{sec:continuity}
Continuity of the composition profile at the interface [${s_\text{in}(R)=s_\text{out}(R)}$] demands  [Eq.\@~\eqref{app:SecondSolutionStep}]
\begin{align}
    \tilde s_\text{in} + \frac{c_\text{in}}{\beta_\text{in}} 
    =
    \tilde s_\text{out} + \frac{c_\text{out}}{\beta_\text{out}}
    \, .
    \label{app:continuityComposition2}
\end{align}

This seems to contradict Eq.\@~\eqref{app:small-c-restriction}, which also relates $c_\text{in/out}$ and was derived by imposing a constant stationary potential $\eta_\text{stat}$ across both domains.
This apparent contradiction can be resolved as follows:

Using that, within the sharp interface limit, the chemical potential is continuous across the interface, a constant stationary mass-redistribution potential [Eq.\@~\eqref{eq:TernaryMassRedistribution}] implies
\begin{align}
    \log\left[1+D\,s_\text{in}(R)\right] 
    =
    \log\left[1+D\,s_\text{out}(R)\right] \, .
\end{align}
Since the logarithm is strictly monotonous, this is equivalent to the continuity condition ${s_\text{in}(R)=s_\text{out}(R)}$ when considering the exact stationary profile. 
Expanding the above equation to leading order in the perturbations ${\delta s_\text{in/out}(R)=c_\text{in/out}/\beta_\text{in/out}}$  [Eq.\@~\eqref{app:SecondSolutionStep}] reproduces\@~Eq.\@~\eqref{app:small-c-restriction}:
\begin{align}
    &\log(1+ D\,\tilde s_\text{in}) + \frac{D\,c_\text{in}}{(1+ D\,\tilde s_\text{in})\,\beta_\text{in}} \notag\\
    =\;
    &\log(1+ D\,\tilde s_\text{out}) + \frac{D\,c_\text{out}}{(1+ D\,\tilde s_\text{out})\,\beta_\text{out}} \, .
    \label{app:small-c-restriction2}
\end{align}

The two continuity conditions, Eqs.\@~\eqref{app:continuityComposition2} and \eqref{app:small-c-restriction2}, are, thus, only identical, when expanding around the same reference composition in both domains (${\tilde s_\text{in}=\tilde s_\text{out}}$), which is the preferred choice for small droplets\@~(${R \ll l}$).
However, this does not imply that there is no consistent linearization scheme for large droplets (${R \gg l}$) as the difference between Eqs.\@~\eqref{app:continuityComposition2} and \eqref{app:small-c-restriction2} only appears at subleading order $\mathcal{O}(\delta s_\text{in/out}^2)$.
In this case, it is reasonable to work with Eq.\@~\eqref{app:continuityComposition2} as it implies continuity of the linearized composition profile itself, rather than its logarithm.

Thus, the last remaining unknown is the stationary radius $R$, which needs to be determined self-consistently by imposing continuity of the radial diffusive flux\@~[Eq.\@~\eqref{app:FluxContinuity}].

\subsubsection*{Current matching}
Flux-continuity across the interface [Eq.\@~\eqref{app:FluxContinuity}] implies
\begin{align}
    &K_\text{in} \, \left(\frac{c_\text{in}}{\beta_\text{in}}-\delta s^0_\text{in}\right)\, \frac{H^\prime_\mathrm{in}(R)}{H_\mathrm{in}(R)} \notag \\
    =\;
    &K_\text{out} \, \left(\frac{c_\text{out}}{\beta_\text{out}}-\delta s^0_\text{out}\right)\, \frac{H^\prime_\mathrm{out}(R)}{H_\mathrm{out}(R)}   
    \, ,
\end{align}
where the radial derivatives of the auxiliary functions $H_\text{in/out}(r)$ are given by
\begin{subequations}
\begin{align}
    H'_\mathrm{in}(r)
    &= -\nu \, r^{-\nu-1}I_\nu(z)+r^{-\nu}\frac{1}{\ell_\mathrm{in}}\,I_\nu'(z) \, , \\
    H'_\mathrm{out}(r)&= -\nu \, r^{-\nu-1}K_\nu(z)+r^{-\nu}\frac{1}{\ell_\mathrm{out}}\,K_\nu'(z)\,.
\end{align}
\end{subequations}
Using the identities
\begin{subequations}
\begin{align}
    I_\nu'(z) 
    &=
    I_{\nu+1}(z) + \frac{\nu}{z}I_\nu(z) \, , \quad  \\
    K_\nu'(z)
    &=
    -\,K_{\nu+1}(z)+\frac{\nu}{z}K_\nu(z) \, ,
\end{align}
\end{subequations}
one finds
\begin{subequations}
\begin{align}
\frac{H_\mathrm{in}'(R)}{H_\mathrm{in}(R)}
    &
   = \frac{1}{\ell_\mathrm{in}}\frac{I_{\nu+1}(R/\ell_\mathrm{in})}{I_\nu(R/\ell_\mathrm{in})} \, , 
   \\
    \frac{H_\mathrm{out}'(R)}{H_\mathrm{out}(R)}
    &= -\,\frac{1}{\ell_\mathrm{out}}\frac{K_{\nu+1}(R/\ell_\mathrm{out})}{K_\nu(R/\ell_\mathrm{out})}
    \,.
\end{align}
\end{subequations}
The \emph{matching condition} for the interface fluxes, thus, reads
\begin{align}
    -&\frac{K_\text{out}}{\ell_\text{out}} \, \left(\frac{c_\text{out}}{\beta_\text{out}}-\delta s^0_\text{out}\right)\,\frac{K_{\nu+1}(R/\ell_\mathrm{out})}{K_\nu(R/\ell_\mathrm{out})} \notag\\
    &=\frac{K_\text{in}}{\ell_\text{in}} \, \left(\frac{c_\text{in}}{\beta_\text{in}}-\delta s^0_\text{in}\right)\, 
    \frac{I_{\nu+1}(R/\ell_\mathrm{in})}{I_\nu(R/\ell_\mathrm{in})} 
    \label{app:MatchingCondition}
\end{align}
which yields a highly non-linear equation for the stationary radius $R$, that generally needs to be solved numerically.

\subsection{Numerical solutions to the self-consistency equation}
\label{sec:NumericalSolutions}
To explore the number of stationary solutions and their physical interpretation, we solve the self-consistency equation, Eq.\@~\eqref{app:MatchingCondition}, for several different anchoring choices\@~$\tilde s_\alpha$\@~[Eq.\@~\eqref{app:compositionExpansion}], all of which yield qualitatively similar results\@~(Fig.\@~\ref{figAPP:Comparison}).

Specifically, we apply a Newton root-finding algorithm, with a range of different primers~(App.\@~\ref{sec:Numerics}). 
To analyze the influence of varying degrees of super-/undersaturation and diffusivity contrasts, we vary $\rho_\infty$ and $D$ for a three dimensional droplet (${d=3}$), while keeping the other parameters fixed at
${\chi=2.4}$, ${K=0.5}$, ${\kappa=0.8}$, and ${\ell=200}$.

\subsubsection*{Supersaturated background}
\begin{figure*}[]
    \centering
    \includegraphics[]{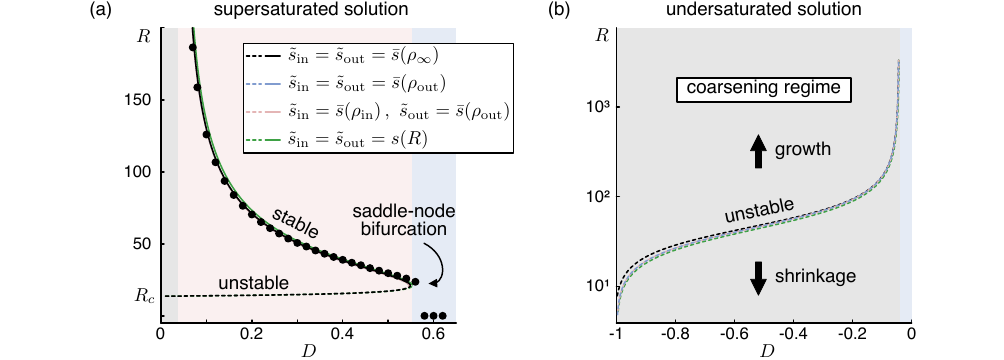}
    \caption{\textbf{Steady state radius} $R$ of a three-dimensional droplet, obtained from sharp-interface theory with different anchoring choices for the composition field $s$ (boxed inset). The stationary radii (lines) were obtained by numerically solving Eq.\@~\eqref{app:MatchingCondition} using a Newton root-finding algorithm for a supersaturated [(a), ${\rho_\infty=0.18>\rho_-=0.17}$] and an undersaturated [(b), ${\rho_\infty=0.16<\rho_-=0.17}$] solution. 
    Solid lines correspond to a stable droplet size, whereas dashed lines indicate an unstable fixed point. 
    Bullets mark the stationary droplet sizes observed in finite-element simulations of an isolated droplet in a spherically symmetric system of size ${L=15\ell}$, with no-flux boundary conditions at the origin and Dirichlet conditions enforcing far-field matching of the density\@~[Eq.\@~\eqref{app:FarFieldEquilibrium}] and local chemical equilibrium\@~[${s=\bar s(\rho_\infty)}$] at the outer system boundary.
    Depending on the value of\@~$D$, the droplet either coarsens (gray shaded area), approaches a stationary size (red shaded area), or dissolves, indicating that phase separation is fully suppressed (blue shaded area). Remaining parameters: ${\chi=2.4}$ (corresponding to ${\rho_-=0.17}$) and ${\ell=200}$.}
    \label{figAPP:Comparison}
\end{figure*}
We first consider the case of a supersaturated mixture, ${\rho_\infty>\rho_-}$, with $\rho_-$ denoting the minimum of the local free energy density $f(\rho)$ of a binary mixture [Eq.\@~\eqref{app:FHdensity}].
Under such conditions, droplet formation and growth can occur in a thermodynamic equilibrium system\@~\cite{Bray.2002,Weber.2019}.

The steady state radii obtained from numerically solving Eq.\@~\eqref{app:MatchingCondition}, are shown in Fig.\@~\ref{figAPP:Comparison}(a).
For vanishing diffusivity contrast\@~(${D=0}$), Eq.\@~\eqref{app:MatchingCondition} admits a single stationary solution.
In this limit, the dynamics reduce to those of an ordinary binary mixture, allowing us to identify this solution with the equilibrium nucleation radius $R_c$\@~\cite{Bray.2002,Lifshitz.1961}.
It therefore represents an unstable fixed point.

For increasing diffusivity contrast\@~$D$, this branch of the solution persists with only minor quantitative changes, indicating that the nucleation barrier is only weakly affected by the non-equilibrium drive.
However, above a critical threshold $D_c$, a second stationary solution appears.
Since a one-dimensional dynamical system cannot support two adjacent unstable fixed points, this additional solution must be stable, which is further confirmed by direct numerical simulations [Fig.\@~\ref{figAPP:Comparison}(a)].
This stable fixed point corresponds to a droplet of finite, stationary radius and therefore signals the onset of arrested coarsening.

When increasing $D$ even further, both fixed points vanish in a saddle–node bifurcation, beyond which phase separation is fully suppressed\@~[Fig.\@~\ref{figAPP:Comparison}(a)].

To test our theoretical predictions, we performed finite-element simulations of an isolated spherical droplet in a three-dimensional system of radial extent\@~${L=15 \, \ell}$.
The system is initialized with a step profile between densities $\rho_\pm$ and an initial droplet radius\@~${R=30}$.
No-flux conditions were imposed at the origin (${r=0}$), whereas at the outer system-boundary\@~(${r=L}$) we enforce far-field matching\@~(${\rho=\rho_\infty}$) and chemical equilibrium [${s=\bar s(\rho_\infty)}$].
After a sufficiently large runtime\@~[App.\@~\ref{sec:Numerics}], we extract the droplet radius from the position of the ${\rho=0.5}$ level-set.

As shown in Fig.\@~\ref{figAPP:Comparison}, the data obtained from numerical simulations is in good agreement with the predictions of the sharp-interface theory.
Within the arrested coarsening regime, the observed droplet radii agree with the predicted stationary solutions.
Upon approaching the transition to the coarsening regime, the observed droplet radius diverges, signaling the loss of a finite-size steady state.
Consistently, within the coarsening regime droplets grow without bound until their size reaches the system scale.
Conversely, beyond the saddle–node bifurcation, no finite-radius solution exists to exist and all droplets shrink until they dissolve.
These results demonstrate that the sharp-interface theory correctly captures both the existence of arrested coarsening and the transition to regular coarsening dynamics, as well as the suppression of phase-separation due to chemical activity.

We further observe that all anchoring choices perform comparably well, showing only minor quantitative differences. 
While interface anchoring [${\tilde s_\text{in/out} = s(R)}$] exhibits slightly larger deviations for bigger droplets, it yields improved agreement with the simulation results in the vicinity of the saddle–node bifurcation.
For the remaining anchoring choices, the sharp-interface theory tends to slightly underestimate the stationary radii as\@~$D$ increases.
This trend is consistent with the expectation that, for large diffusivity contrasts, the density profile departs more strongly from its equilibrium plateau values, indicating the gradual breakdown of the linearization scheme.

\subsubsection*{Undersaturated background}
We next consider the case of an undersaturated solution, ${\rho_\infty < \rho_-}$, where under equilibrium conditions, droplets of all sizes shrink, and phase separation is fully suppressed\@~\cite{Bray.2002}.

In the presence of chemical conversion, FEM simulations presented in Ref.\@~\cite{prl} show that droplets can form even under conditions where phase separation is thermodynamically excluded, provided the slower-diffusing species is enriched within dense regions. 
This observation is supported by our sharp-interface theory, which predicts that for sufficiently negative $D$, Eq.\@~\eqref{app:MatchingCondition} admits a single stationary solution\@~[Fig.\@~\ref{figAPP:Comparison}(b)], even for an undersaturated solution. 

While for the supersaturated case, the stability of this solution could be inferred by connecting the solution branch to the well-known ${D=0}$ limiting case, for undersaturated solutions, this argument is not available.
For strongly negative\@~$D$, the solution extends to small radii $R$\@~[Fig.\@~\ref{figAPP:Comparison}(b)], which suggests that it is unstable, as small droplets are always disfavored due to the diverging Laplace pressure. 
This is further confirmed by finite element simulations, which show that for\@~${D<0}$ no finite-size solutions exist, and sufficiently large droplets always grow.

In contrast to the supersaturated case, the non-equilibrium nucleation radius in undersaturated solutions depends sensitively on the diffusivity contrast\@~$D$ and can vary over several orders of magnitude\@~[Fig.\@~\ref{figAPP:Comparison}(b)].
As before, different anchoring choices show minimal impact on the predicted solution.

In summary, our sharp-interface theory accurately captures the steady states of an isolated spherical droplet. 
For supersaturated solutions, chemical activity only weakly affects droplet nucleation, but the long-term behavior depends sensitively on the diffusivity contrast\@~$D$: 
At small\@~$D$, droplets behave essentially as in thermodynamic systems and grow indefinitely; above a threshold\@~$D_c$, coarsening is arrested and droplets reach a stable finite size; and at even larger\@~$D$, droplet formation is fully suppressed. 
In contrast, for undersaturated solutions, where droplet formation is precluded in thermodynamic systems, chemical activity can enable droplet nucleation, provided the slower diffusing species is enriched in dense regions.

\subsection{Analytical solutions to the self-consistency equation}
Having established the structure of the solution space, we now derive approximate analytical solutions by expanding the matching condition Eq.\@~\eqref{app:MatchingCondition} in the limits of small and large droplets.

For simplicity, we focus on a spherical droplet in three spatial dimensions\@~(${d=3}$); a similar procedure can be applied analogously in arbitrary dimensions.
The matching condition, Eq.\@~\eqref{app:MatchingCondition}, for the stationary state involves ratios of modified Bessel functions. 
In three spatial dimensions one has ${\nu=1/2}$, and the half-integer identities
\begin{align}
\frac{I_{3/2}(z)}{I_{1/2}(z)} = \coth(z) - \frac{1}{z},
\quad
\frac{K_{3/2}(z)}{K_{1/2}(z)} = 1 + \frac{1}{z},
\label{eq:halfinteger-ratios}
\end{align}
hold for all ${z>0}$.
To rewrite the matching condition in a more compact form, we define the material combinations
\begin{align}
    \alpha_\text{in/out}=
    K_\text{in/out}\left(\frac{c_\text{in/out}}{\beta_\text{in/out}}-\delta s^0_\text{in/out}\right) \, .
\end{align}
Using Eq.\@~\eqref{eq:halfinteger-ratios} in the flux matching condition then yields
\begin{align}
    \frac{\alpha_\text{in}}{\ell_\text{in}}
    \left(
    \coth \frac{R}{\ell_\text{in}}
    -
    \frac{\ell_\text{in}}{R}
    \right)
    \;=\;
    -\,\frac{\alpha_\text{out}}{\ell_\text{out}}
    \left(
    1
    +
    \frac{\ell_\text{out}}{R}
    \right).
    \label{eq:matching-d3}
\end{align}
or, equivalently, the implicit expression
\begin{align}
R
\;=\;
\frac{
\alpha_\text{in}
-
\alpha_\text{out}
}{
\displaystyle
\frac{\alpha_\text{in}}{\ell_\text{in}}\,
\coth \left(\frac{R}{\ell_\text{in}}\right)
+
\frac{\alpha_\text{out}}{\ell_\text{out}}
}.
\label{eq:R-implicit-d3}
\end{align}

Equations \eqref{eq:matching-d3}–\eqref{eq:R-implicit-d3} provide the \emph{exact} condition for the stationary droplet radius in three dimensions.
It is important to recall, however, that for a curved interface, the coexistence densities satisfy ${\rho_\text{in/out}\approx \rho_\pm+ \mathcal{O}(R^{-1})}$, as detailed in App.\@~\ref{sec:phaseCoexistence}. 
Consequently, the prefactors\@~$\alpha_\text{in/out}$ and\@~$\ell_\text{in/out}$, inherit a weak $R$ dependence.

Crucially, this dependence vanishes for large enough droplets:
For ${R\gg1}$ the coexistence densities converge to their planar interface value, ${\rho_\text{in/out}\to \rho_\pm}$, with\@~$\rho_\pm$ denoting the minima of the binary free energy density\@~[Eq.\@~\eqref{app:FHdensity}].
In this regime, all prefactors become effectively constant, provided that the expansion of the composition field is not anchored at ${\tilde s=s(R)}$.
With these simplifications in mind, we now analyze two asymptotic regimes:
\begin{itemize}
    \item[(i)] Small to medium-sized droplets whose radius obeys ${1\ll R\ll \ell_\text{in/out}}$, and
    \item[(ii)] large droplets, where ${R \gg \ell_\text{in/out}\gg 1}$.
\end{itemize}
\subsubsection*{Small droplet limit}
\begin{figure}[]
    \centering
    \includegraphics[]{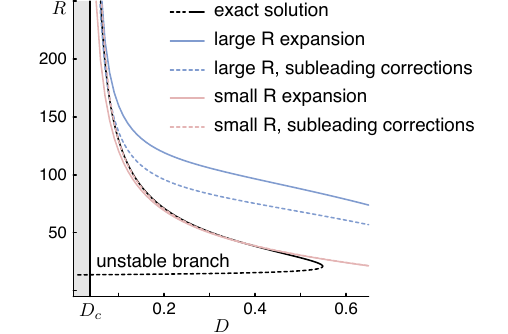}
    \caption{\textbf{Approximate solutions} for the stationary radius\@~$R$, obtained by expanding Eq.\@~\eqref{app:MatchingCondition} in the limit of small (${1\ll R\ll \ell_\text{in/out}}$, red) and large (${1 \ll \ell_\text{in/out} \ll R}$, blue) droplets for a supersaturated solution (${\rho_\infty >\rho_-}$). Solid and dashed colored lines correspond to the leading and next-to-leading order expressions in Eqs.\@~\eqref{app:SmallDropletSolution}, \eqref{eq:R-large-leading-no-bang}, and \eqref{eq:R-large-corrected-no-bang}, respectively. The exact solution (black), is determined through a Newton root-finding procedure, with solid (dashed) lines indicating (un-)stable solutions. All parameters were chosen as in Fig.\@~\ref{figAPP:Comparison}(a), with ${\tilde s_\text{in}=\tilde s_\text{out}=s_\infty}$.}
    \label{figAPP:Approximations}
\end{figure}
For ${1\ll R\ll \ell_\text{in/out}}$, we use the expansion
\begin{align}
    \coth(z) = \frac{1}{z} + \frac{z}{3} - \frac{z^3}{45} +\mathcal{O}(z^5) \, ,
    \qquad z=\frac{R}{\ell_\text{in}} \,.
\end{align}
Only retaining the leading nontrivial order, Eq.\@~\eqref{eq:matching-d3} then reduces to
\begin{align}
    \frac{\alpha_\text{in}R}{3\,\ell_\text{in}^2} = -\frac{\alpha_\text{out}}{\ell_\mathrm{out}}-\frac{\alpha_\text{out}}{R} \, .
\end{align}
The physically admissible positive root yields the small-radius approximation for the stationary droplet radius:
\begin{align}
    R_\text{small}= -\frac{3}{2}\ell_\text{in} \frac{\alpha_\text{out}\ell_\text{in}}{\alpha_\text{in}\ell_\text{out}}\left(1+\sqrt{1-\frac{4}{3}\frac{\alpha_\text{in}\ell_\text{out}^2}{\alpha_\text{out}\ell_\text{in}^2}}\right)\,,
    \label{app:SmallDropletSolution}
\end{align}
where all quantities depending on $\rho_\text{in/out}$ are evaluated at the minima\@~$\rho_\pm$ of the free energy density $f(\rho)$ [Eq.\@~\eqref{app:FHdensity}].
The above expression shows that the stationary radius scales with the characteristic length scale of the inner domain $\ell_\text{in}$.

Comparison with the exact numerical solution of Eq.\@~\eqref{app:MatchingCondition} demonstrates that our approximation agrees surprisingly well (Fig.\@~\ref{figAPP:Approximations}), even for large radii\@~$R$.
Including the next-to-leading order term in the expansion of $\coth(z)$ further improves quantitative accuracy.
However, because we neglect surface-tension–induced corrections to the coexistence densities\@~$\rho_\text{in/out}$, the approximation cannot capture the unstable nucleation radius and applies only to the stable branch.
As a result, it does not predict the bifurcation point.

\subsubsection*{Large droplet limit}
For large $z$, we use the asymptotic expansion
\begin{align}
    \coth(z) = 1 + 2 e^{-2z} + \mathcal{O}(e^{-4z}) \, , \qquad
    z=R/\ell_\text{in} \, .
\end{align}
At leading order, Eq.\@~\eqref{eq:R-implicit-d3} then yields the large-radius estimate
\begin{align}
    R_{0}
    =
    \frac{\xi_\text{in}\ell_\text{in}-\xi_\text{out}\ell_\text{out}}{\xi_\text{in}+\xi_\text{out}}\,,
    \label{eq:R-large-leading-no-bang}
\end{align}
where we introduced
\begin{align}
    \xi_\text{in} \equiv \frac{\alpha_\text{in}}{\ell_\text{in}},
    \qquad
    \xi_\text{out} \equiv \frac{\alpha_\text{out}}{\ell_\text{out}}.
\end{align}
Including the first exponentially small correction from ${\coth(z)}$ gives
\begin{align}
    R_{\mathrm{large}}
    &\approx
    \frac{\xi_\text{in}\ell_\text{in}-\xi_\text{out}\ell_\text{out}}
    {\xi_\text{in}+\xi_\text{out}+2\,\xi_\text{in}\exp\left(-\frac{2 R_{0}}{\ell_\text{in}}\right)}
    \nonumber \\
    &=
    R_{0}\left[
    1
    -
    \frac{2\,\xi_\text{in}}{\xi_\text{in}+\xi_\text{out}}
    \exp\left(-\frac{2 R_{0}}{\ell_\text{in}}\right)
    \right].
    \label{eq:R-large-corrected-no-bang}
\end{align}
As shown in Fig.\@~\ref{figAPP:Approximations}, Eq.\@~\eqref{eq:R-large-leading-no-bang} accurately predicts the stationary droplet size in the large-$R$ limit, and its range of validity extends further when subleading corrections\@~[Eq.\@~\eqref{eq:R-large-corrected-no-bang}] are included.

Equation\@~\eqref{eq:R-large-leading-no-bang} further allows us to estimate the onset of arrested coarsening.
Specifically, the stationary radius\@~$R_0$ diverges when $${\xi_\text{in}+\xi_\text{out}= 0\, .}$$
This marks the transition between the coarsening regime, where no (stable) stationary radius exists, and the arrested regime, where droplets converge to a unique stationary size.

Expanding this condition to leading orders in $D$ and the supersaturation\@~${\varepsilon=\rho_\infty-\rho_-}$ yields an estimate for the critical relative diffusivity 
\begin{align}
    D_c 
    =
    &\left(\tilde s_\text{out}-\tilde s_\text{in}+\frac{\mathcal R_\text{out}}{\partial_s\mathcal R_\text{out}}-\frac{\mathcal R_\text{in}}{\partial_s\mathcal R_\text{in}}\right)^{-1} \times \notag \\
    &\left(1+\sqrt{\frac{\partial_s\mathcal R_\text{out}}{\partial_s \mathcal R_\text{in}}}\,\right)f''(\rho_-)\, \varepsilon
    \, ,
    \label{app:criticalD}
\end{align}
above which the system exhibits a stationary solution of finite size.
If the expansion is performed around the same reference composition ${\tilde s_\text{in}=\tilde s_\text{out}=\bar s(\rho_\text{out})}$ in both domains, one has ${\mathcal R_\text{out}=0}$, and the above equation simplifies to the more compact form
\begin{align}
    D_c= -\frac{\partial_s\mathcal R_\text{in}}{\mathcal R_\text{in}}\left(1+\sqrt{\frac{\partial_s\mathcal R_\text{out}}{\partial_s \mathcal R_\text{in}}}\,\right)f''(\rho_-)\,\varepsilon \, ,
    \label{app:criticalD2}
\end{align}

This analysis leads to two main conclusions:
First, achieving arrested coarsening at higher supersaturation requires a larger diffusivity contrast $D$.
Second, the critical threshold $D_c$ decreases with increasing reactive turnover inside the droplet:
When\@~$\mathcal R_\text{in}$ is large, even a modest diffusivity difference can suffice to arrest coarsening.
\section{Summary \& Outlook}
\label{sec:Summary}
In this work, we studied pattern formation and phase coexistence in incompressible chemically active mixtures composed of an inert solvent and multiple solute species.
The solutes are thermodynamically identical, sharing the same molecular volume, internal energy, and interparticle interactions, and differ only in their diffusivities.
Chemical reactions interconvert the solutes, thereby modifying local transport properties without altering interparticle interactions.
As a result, chemical activity enters the dynamics purely through kinetic asymmetries, while the short-range interactions remain unchanged.
Our model, thus, yields a generalization of classical mass-conserving reaction-diffusion systems\@~\cite{Brauns.2020,Frey.2022} to dense, interacting mixtures.

Using linear stability analysis and numerical finite-element simulations, we demonstrated that such chemically active mixtures can form patterns of coexisting solute-rich and solute-poor domains.
Depending on the reaction scheme and parameter regime, these patterns form via two distinct routes.
When short-range interactions alone are insufficient to drive phase separation, patterns can nevertheless emerge through a generalized mass-redistribution instability, provided the chemical dynamics locally enriches slower-diffusing solutes within dense regions.
As shown in Ref.\@~\cite{prl}, this mechanism invariably leads to uninterrupted coarsening.
Conversely, when short-range interactions favor phase separation, chemical activity can interrupt the coarsening process if it locally enriches the faster-diffusing species within solute-rich domains.

In the coarsening regime, we demonstrated that the solute volume fractions of the coexisting phases can be approximated by an effective free-energy description obtained by adiabatically eliminating the local composition variables in favor of their reactive fixed points.
This description becomes exact in the limit of rapid chemical turnover but generally neglects nonlocal effects arising from diffusive coupling between solute density and composition.

To characterize the arrested coarsening regime, we employed a sharp-interface theory, constructing approximate stationary solutions for an isolated spherical droplet. 
From this, we derived analytical expressions for the stationary droplet radius and for the critical conditions marking the onset of arrested coarsening, in good agreement with numerical simulations.
Our analysis shows that steady finite droplet sizes arise only when chemical activity locally enriches the faster-diffusing species within dense domains, allowing non-equilibrium composition-driven fluxes to balance surface-tension–driven coarsening.

The same sharp-interface approach also provides insight into the nucleation of individual droplets.
In particular, we obtained an estimate for the non-equilibrium nucleation radius that separates shrinking from growing droplets and thus controls the onset of phase separation outside the linearly unstable regime.
We showed that, even in parameter regimes where phase separation is precluded in thermodynamic equilibrium, chemical activity can generate a finite nucleation barrier, provided the slower-diffusing species is locally enriched within high-density domains.

While our sharp-interface analysis focused on isolated droplets, it naturally suggests a route toward understanding dense regimes where many droplets interact [Fig.\@~\ref{fig:overview}(b)].
As shown by the single-droplet theory, due to the finite relaxation length of the composition field, the internal composition of a droplet depends on its size (Fig.\@~\ref{fig:model}).
In a multi-droplet setting, this implies systematic composition differences between droplets of different radii, which in turn generate non-equilibrium fluxes that redistribute solutes between droplets.
Such size-dependent, composition-mediated mass exchange is expected to bias transport from larger to smaller droplets, thereby arresting classical Ostwald ripening.
A quantitative description of this regime could be developed by extending our single droplet theory along the lines of Refs.\@~\cite{Zwicker.2015,Wurtz.2018}.

\subsection{Chemically active mixtures}
A wide range of studies on chemically active mixtures have shown that chemical reactions can strongly affect phase separation and coarsening, often leading to finite-size droplets and non-equilibrium steady states\@~\cite{Glotzer.1994,Carati.1997,Zwicker.2015,Wurtz.2018,Osmanovic.2023,Li.2020,Chen.2024,Bauermann.2025}.

In most of these models, it is assumed that chemical reactions convert molecular states with distinct short-range interactions\@~\cite{Zwicker.2015,Wurtz.2018,Bauermann.2025r8,Bauermann.2025,Fries.2025,Alston.2022}, such that chemical activity directly modifies the thermodynamic driving forces for phase separation.
Here, we instead focus on a limit in which chemical reactions leave intermolecular interactions unchanged and affect the system solely through local transport properties.
Our results demonstrate that such kinetic regulation is sufficient to induce arrested coarsening, underscoring that non-equilibrium transport effects must be treated on equal footing with short-range interactions when analyzing phase separation in chemically active systems.

Recent work\@~\cite{Bauermann.2025} has shown that arrested coarsening in chemically active mixtures with interaction-switching solutes can be understood in terms of coupled dynamics between a conserved, phase-separating field and an additional non-conserved field. 
We find that the present system admits an analogous description, with the total solute volume fraction acting as the conserved, phase-separating variable and the local composition serving as the non-conserved variable mediating chemical activity. 
Closely related ideas were explored in our own work\@~\cite{Rasshofer.2025}, where we introduced a minimal continuum model to isolate the generic consequences of such coupled-field dynamics. 
From this perspective, it seems that arrested coarsening emerges generically when non-equilibrium driving of the non-conserved field sustains spatial gradients that feed back onto transport of the conserved quantity, thereby altering coarsening dynamics and phase coexistence.

\subsection{Fluctuations, coalescence, and interfacial dynamics}
The model considered here is purely deterministic and therefore only addresses classical Ostwald ripening~\cite{Ostwald.1897,Lifshitz.1961} as the dominant coarsening mechanism.
In many phase-separating systems, however, coarsening is also driven by droplet coalescence, whereby thermal fluctuations lead to droplet motion, collisions, and subsequent merging\@~\cite{Siggia.1979,Weber.2019}. 
In our model, chemical conversion directly controls the effective mobility of droplets through their internal composition, which itself depends on droplet size. 
As a result, larger droplets may acquire higher effective mobilities, in stark contrast to classical Brownian diffusion, where mobility decreases with size. 
Such a size-dependent, activity-controlled mobility is expected to strongly modify collision rates and coalescence dynamics, potentially giving rise to qualitatively new coarsening behavior.

Stochastic effects are likewise expected to be important for our understanding of interfacial properties.
Activity is known to alter the statistics of capillary fluctuations in phase-separated systems\@~\cite{Fausti.2021,Maire.2025,Cho.2025}.
In the limit of fast chemical turnover, our theory reduces to ordinary relaxational dynamics governed by an effective free energy with a well-defined non-equilibrium surface tension.
This suggests that, in this regime, interfacial fluctuations may still obey equilibrium-like scaling\@~\cite{Bray.2001}.
Beyond this limit, however, the coupling between composition, transport, and interface motion may lead to qualitatively different fluctuation statistics.

A quantitative understanding of coalescence dynamics and interfacial fluctuations requires a stochastic generalization of the coupled density–composition dynamics studied here.
Developing such a description constitutes an important direction for future work and may reveal fundamental differences between kinetically regulated and interaction-regulated active mixtures.
\subsection{Active matter}
Beyond their direct implications for chemically active mixtures, our results reveal a close analogy to phase separation in active matter.
In coarse-grained descriptions of MIPS of non-interacting particles, activity enters through an effective free-energy functional containing a kinetic contribution\@~${\int^\rho \mathrm{d}\rho'\,\log [v(\rho')]}$\@~\cite{Tailleur.2008,Cates.2014,Solon.2018}, where $v(\rho)$ denotes the effective density-dependent self-propulsion speed.
In the limit of fast chemical turnover, we obtain a formally analogous structure:
The effective free energy governing the solute density acquires an activity-induced term ${\int^\rho \mathrm{d}\rho'\log[ D_{\mathrm{eff}}(\bar s(\rho^\prime))]}$, where the effective mobility $D_{\mathrm{eff}}(\bar s(\rho))$ is set by the local, density-dependent composition $\bar s(\rho)$.

While the microscopic origin of non-equilibrium driving is different, both classes of systems realize similar coarse-grained phase-separation dynamics.
This connection can be pushed further:
Extending the adiabatic elimination scheme discussed in Sec.\@~\ref{sec:Coexistence} by a gradient expansion of the form
${s = \bar s(\rho) + \alpha\nabla^2\rho + \mathcal{O}(\boldsymbol{\nabla}^4\rho)}$ allows to incorporate weak deviations from local reactive equilibrium.
In close analogy to Ref.\@~\cite{Robinson.2025}, such an expansion is expected to generate non-variational gradient terms, enabling a controlled mapping onto Active Model B\@~\cite{Wittkowski.2014} dynamics.

\begin{acknowledgments} 
    We thank Henrik Weyer and Tobias Roth for stimulating discussions.
    This research was funded by the Deutsche Forschungsgemeinschaft (DFG, German Research Foundation) under Germany's Excellence Strategy – EXC3092 – 533751719.
    We further acknowledge support from the John Templeton Foundation, and the U.S. National Science Foundation (NSF) under Grant No.\@~PHY-2309135 through the Kavli Institute for Theoretical Physics (KITP).
\end{acknowledgments}

\cleardoublepage
\appendix
\section{Relation between the interaction and stiffness parameter}
\label{sec:coarseGraining}
The Flory-Huggins parameter $\chi$ and the stiffness parameter $\tilde \kappa$ appearing in the free energy of a binary mixture, Eq.\@~\eqref{app:FreeEnergy2} ca be related by direct coarse graining of a (square) lattice-based model with equally sized particles and lattice spacing $\nu^{1/d}$.
For a given lattice configuration, the total interaction energy is given by
\begin{align*}
    E = \frac{k_\mathrm{B} T}{2}\sum_{j\sim i} \, \Big\{
    &{\tilde \varepsilon}_\text{PS} \, \left[\sigma_i(1-\sigma_j)+\sigma_j(1-\sigma_i)\right] \notag \\ + \;
    &{\tilde \varepsilon}_\text{PP} \, \sigma_i \sigma_j  + 
    {\tilde \varepsilon}_\text{SS} \, (1-\sigma_i)(1-\sigma_j)\Big\} \, , 
\end{align*}
where ${\tilde \varepsilon}_\text{SP}$, ${\tilde \varepsilon}_\text{PP}$, and ${\tilde \varepsilon}_\text{SS}$ denote the dimensionless pairwise solute-solvent, solute-solute, and solvent-solvent interaction energies, respectively.
The variables $\sigma_i$ specify whether a lattice site is occupied by a solute (${\sigma_i=1}$) or solvent (${\sigma_i=0}$) particle.
Coarse-graining this expression via 
\begin{align}
    \sum_i \ &\to \, \nu^{-1} \int\mathrm{d}\boldsymbol{r} \, , \\
    \langle \sigma_i\rangle \, &\to\ \rho(r) \, , \\
    \langle\sigma_{i+1}-\sigma_i\rangle \, &\to \ \nu^{1/d} \,\partial_i\rho \, ,
\end{align}
one obtains 
\begin{align*}
    E = \frac{k_B T}{\nu} \int \mathrm{d}\boldsymbol{r}  \left[ \frac{\chi \nu^{2/d}}{z} (\boldsymbol{\nabla} \rho)^2+\chi\rho(1-\rho) + \mathcal{O}(\rho) \right] \, ,
\end{align*}
where $${\chi= z\,(2\tilde \varepsilon_\text{SP}-\tilde \varepsilon_\text{PP}-\tilde \varepsilon_\text{SS})/2} \, ,$$
Using ${\varepsilon_{\alpha\beta}=z{\tilde\varepsilon}_{\alpha\beta}}$, this agrees with our previous definition of the Flory-Huggins parameter [Eq.\@~\eqref{app:FloryHugginsParameter}].
Comparing with Eq.\@~\eqref{app:FreeEnergy2}, we read off the relation ${\tilde \kappa=2\chi\nu^{2/d}/z}$, which relates the stiffness parameter $\tilde\kappa$ to the short-range interactions encoded by $\chi$.
To ensure thermodynamic stability against small-scale fluctuations, we employ the more general form
\begin{align}
    \tilde\kappa = \max \left(\frac{2\chi\nu^{2/d}}{z},0 \right) \, .
\end{align}
This choice reflects that in the thermodynamically stable regime ${\chi<0}$,  the local free-energy correctly describes the system's tendency to mix, whereas a negative prefactor for the stiffness term would induce unphysical large gradients.

\section{Equilibrium phase coexistence}
\label{sec:phaseCoexistence}
For equal diffusivities, the dynamics of the mixture [Eq.\@~\eqref{eq:SoluteDynamics}] are governed solely by the (non-dimensionalized) chemical potential of a binary mixture
\begin{align}
    \mu 
    =
    -\kappa  \, 
    \boldsymbol{\nabla}^2 \rho 
    + f'(\rho) \, ,
    \label{app:ChemicalPotential}
\end{align}
and the steady state is characterized by the condition ${\boldsymbol{\nabla}\mu = 0}$.
The same steady-state condition also arises for thermodynamically consistent or linear conversion rates, as in both cases the solute composition becomes homogeneous throughout the system, and the mass-redistribution potential $\eta$ [Eq.\@~\eqref{eq:MassRedistribution}] reduces to the chemical potential $\mu$ plus a constant shift.

In this section, for the convenience of the reader, we briefly recapitulate the resulting phase-coexistence conditions.
To obtain the coexistence conditions, we follow the standard projection method, see e.g.~Ref.~\cite{Bray.2002}. 
Specifically, we consider a \emph{spherical droplet} of radius ${R}$ 
embedded in a surrounding phase, and adopt spherical coordinates with radial coordinate ${r}$. 
Due to spherical symmetry, the Laplacian in $d$ spatial dimensions reduces to
\begin{align}
    \nabla^2 = \partial_{r}^2 + \frac{d-1}{r}\,\partial_{r} \, .
\end{align}
For constant $\mu$, multiplying Eq.\@~\eqref{app:ChemicalPotential} by ${\partial_{r}\rho}$ and integrating across the interfacial region yields
\begin{align}
    \mu \, (\rho_\text{in} &- \rho_\text{out}) - f(\rho_\text{in}) + f(\rho_\text{out})
    = \notag \\[2mm]
    &\kappa \int_{R-\varepsilon}^{R+\varepsilon} \mathrm{d}r
    \left[ \frac{\partial_{r} (\partial_r \rho)^2}{2} + \frac{d - 1}{r} \, (\partial_r \rho)^2 \right] \, ,
    \label{app:projectionINtegral}
\end{align}
where the integration range $r \in [R-\varepsilon, R+\varepsilon]$ is taken large enough such that the profile\@~$\rho(r)$ approaches the bulk values $\rho_\text{in/out}$ at the boundaries.
Assuming constant plateau values $\rho_\text{in/out}$, the first term on the right-hand side of Eq.~\eqref{app:projectionINtegral} integrates to zero, whereas the second term remains finite for curved interfaces (${d \geq 2}$). 
In the \emph{sharp–interface limit}, where the interfacial width ${w \sim \sqrt{\kappa}}$ \cite{Weber.2019,Bray.2002} is much smaller than the droplet radius ${R}$, the term ${(\partial_{r}\rho)^2}$ is sharply localized at the interface.
We, thus, approximate ${1/r \simeq 1/R}$ in the last term of Eq.\@~\eqref{app:projectionINtegral}, and define the surface tension
\begin{align}
    \gamma 
    \equiv
    \kappa \int_{R-\varepsilon}^{R+\varepsilon} \! \mathrm{d}r \,  (\partial_r \rho)^2 \, .
    \label{app:SurfaceTension}
\end{align}
Substituting into Eq.~\eqref{app:projectionINtegral} then yields the \textit{Gibbs-Thomson} relation
\begin{align}
    \mu (R) 
    = \frac{f(\rho_\text{in}) - f(\rho_\text{out})}{\rho_\text{in} - \rho_\text{out}} + \frac{(d - 1) \, \gamma}{(\rho_\text{in} - \rho_\text{out}) R} \, .
    \label{app:GibbsThomson}
\end{align}

The above equation shows that the uniform chemical potential is set by two contributions: 
The first corresponds to the common-tangent construction for a flat interface, while the second encodes the curvature-induced Laplace pressure correction.
In the bulk phases, where gradients are negligible, one has ${\mu \approx f'(\rho_\text{in/out})}$\@~[Eq.\@~\ref{app:ChemicalPotential}].
The coexisting bulk densities are therefore determined by a common--tangent construction that acquires a curvature correction:
\begin{align}
    f(\rho_\text{out}) + f'(\rho_\text{out})\, (\rho_\text{in} - \rho_\text{out}) = f(\rho_\text{in}) + \frac{(d - 1) \, \gamma}{R} \, .
\end{align}
The additional term proportional to $\gamma/R$ represents the curvature–induced shift of the coexistence condition.
Introducing the local osmotic pressure ${\Pi = \mu \rho-f(\rho)}$, the above equation can be reformulated as
\begin{align}
    \Pi_\text{in} = \Pi_\text{out} + \frac{(d-1) \gamma}{R} \, .
\end{align}
That is, the osmotic pressures inside and outside the droplet differ by a curvature-dependent \emph{Laplace pressure}.
This pressure imbalance arises from interfacial tension and reflects the energetic cost associated with maintaining a curved interface.

For a \emph{flat interface}, the fact that we are considering a symmetric binary mixture with ${f(\rho) = f(1 - \rho)}$ implies 
\begin{align}
    \rho_\text{in/out} = \rho_\pm \, , \qquad \mu=0 \, .
\end{align}
Assuming that for a spherical droplet the coexisting densities only deviate weakly, we write $\rho_\text{in/out} = \rho_\pm + \delta \rho_\pm$.
Expanding Eq.\@~\eqref{app:GibbsThomson} to leading order in~$R^{-1}$ yields
\begin{subequations}
\begin{align}
    \rho_\text{in/out} &= \rho_\pm +  \frac{(d - 1) \, \gamma}{(\rho_+ - \rho_-) f''(\rho_\pm) R} \, , \qquad \\
    \mu &\approx \frac{(d - 1) \, \gamma}{(\rho_+ - \rho_-) R} \, .
    \label{app:linearGibbsThomson}
\end{align}
\end{subequations}

Note that, in principle, the surface tension $\gamma$ [Eq.\@~\eqref{app:SurfaceTension}] also weakly depends on droplet radius $R$.
To leading order in $R^{-1}$, it can be approximated by its flat-interface value, which is obtained by integrating Eq.\@~\eqref{app:SurfaceTension} across the stationary profile of a one-dimensional system.
In this work, we exclusively use ${\chi=2.4}$ (${\kappa=0.8}$ for ${d=3}$), which gives ${\gamma \approx 0.088}$ when integrating along the numerically obtained profile for a system of size ${L=1000}$\@~(${\bar \rho =0.5}$).
\section{Validity of linearization in the density shifts} 
\label{sec:validity}
In this section, we specify under which conditions the linearization, Eq.\@~\eqref{eq:rho_linearization_radial}, of the dynamical equations in the density perturbations $\delta\rho_\alpha$ (${\alpha\in\{\text{in, out}\}}$) is justified.
That is, we ask under which circumstances the density profile $\rho(r)$ only weakly deviates from the coexisting equilibrium plateau densities $\rho_\text{in/out}$.
Inspecting Eq.\@~\eqref{eq:clousre-relation-full}, this is the case whenever
\begin{align}
  \bigg|\log\frac{1 + D\,s(R)}{1 + D\,s(r)}\bigg| \ll 1 .
\end{align}
Assume that ${1 + D\,s(r)}$ is uniformly bounded from below by ${\delta > 0}$.
Then, by the mean value theorem for ${g(x)=\log(1+Dx)}$, there exists ${\xi}$ between ${s(R)}$ and ${s(r)}$ such that
\begin{align}
  \bigg|\log\frac{1 + D\,s(R)}{1 + D\,s(r)}\bigg|
  &= \frac{|D|}{\,|1 + D\,\xi|\,}\,|s(R)-s(r)| \notag \\
  &\le \frac{|D|}{\delta}\,|s(r)-s(R)| \, .
\end{align}
This implies that 
\begin{align}
  |\delta\rho_\alpha(r)|
  \le \frac{1}{|f''(\rho_\alpha)|}\,\frac{|D|}{\delta}\,|s(r)-s(R)| \, .
\end{align}
In short, if ${|1+D\,s|\ge\delta>0}$ and ${|D| \, |s(r)-s(R)|\ll \delta}$, then ${|\delta\rho_\alpha(r)|\ll 1}$, validating the linearized bulk closure.

One can also ask the reverse question: 
Given a weak level of super- or under-saturation ${\varepsilon_{\mathrm{sup}}\equiv|\rho_\infty-\rho_{\text{out}}|\ll 1}$, what does this imply for the non-equilibrium drive ${|D|\,\Delta s}$, in particular for the steady–state contrast ${|D|\,|s_\infty-s(R)|}$? 
By the mean–value theorem for $f'(\rho)$,
\begin{align}
    \mu_\infty-\mu(R)
    =
    f''(\rho_*)\,[\rho_\infty-\rho_{\text{out}}],\qquad \rho_*\in[\rho_{\text{out}},\rho_\infty]
    \, .
\end{align}
Assuming local convexity of the free-energy density along the outer segment so that ${0<c_{\text{out}}\le f''(\rho)\le C_{\text{out}}}$ for ${\rho\in[\rho_{\text{out}},\rho_\infty]}$, this implies
\begin{align}
    c_{\text{out}}\,\varepsilon_{\mathrm{sup}}\le|\mu_\infty-\mu(R)|\le C_{\text{out}}\,\varepsilon_{\mathrm{sup}}
\end{align}
Constancy of ${\eta=\mu+\log(1+D s)}$ implies (again using the mean-value theorem)
\begin{align}
    \mu_\infty-\mu(R)
    =
    -\log\!\frac{1+D \, s(R)}{1+D \, s_\infty}
    =
    \frac{D}{1+D\,\xi}\;
    \Delta s \, ,
\end{align}
for some ${\xi\in[s(R),s_\infty]}$. Since ${\delta\le|1+D\,\xi|\le 2}$ we obtain
\begin{align}
    \delta\,c_{\text{out}}\,
    \varepsilon_{\mathrm{sup}}
    \le
    {|D|} \, |\Delta s|
    \le
    2\,C_{\text{out}} \,
    \varepsilon_{\mathrm{sup}}
    \,.
\end{align}
Thus, provided the system remains uniformly away from ${1+D \, s=0}$ and the outer segment lies in a convex (stable) basin of $f$, weak super/under-saturation enforces a small composition drive, ${|D|\,|\Delta s| = \mathcal{O}(\varepsilon_{\mathrm{sup}})}$, with constants set by the curvature bounds of $f$ and the gap $\delta$.

\section{Numerical methods}
\label{sec:Numerics}
\begin{table*}
    \centering
    \caption{Simulation parameters applied for finite element simulations.}
    \begin{ruledtabular}
    \begin{tabular}{lccccc}
        & Mesh size & Rel.\@~tolerance & System size $L$ & Sim.\@~Time $T\,(1+D)$ & Noise $\xi$\\[1mm]
        \hline\\[-2mm]
        Fig.\@~\ref{fig:overview}
        & 0.5
        & $10^{-5}$ 
        & $200$ 
        & $10^8$ 
        & $0.01$ \\[2mm]
        
        Fig.\@~\ref{fig:surfaceTension}  
        & 0.05 
        & $10^{-6}$ 
        & $400$ 
        & $10^7$ 
        & 0 \\[2mm]
        
        Fig.\@~\ref{fig:binodals}
        & 0.2 
        & $10^{-6}$ 
        & $500,1000$ 
        & $10^{10}$ 
        & 0 \\[2mm]

        Fig.\@~\ref{figAPP:Comparison}
        & 0.2 
        & $10^{-6}$ 
        & $15\ell$ 
        & $10^9$ 
        & 0 
        \end{tabular}
    \end{ruledtabular}
    \label{tab:FigureParams}
\end{table*}

Finite element simulations were performed using Comsol Multiphysics Version 6.1\@~\cite{AB.2023}, employing the Pardiso solver with adaptive time stepping, an implicit 5th-order backward differentiation formula, and relative tolerances as specified in Table\@~\ref{tab:FigureParams}. 
All simulation files and Mathematica~\cite{Inc..2024} notebooks are made available in a dedicated Zenodo repository\@~\cite{ZN2}.

Two-dimensional simulations used a triangular Delaunay tessellation with linear Lagrangian shape functions, while one-dimensional simulations used an equidistant grid. 
The maximum element size applied in both cases is detailed in Table\@~\ref{tab:FigureParams}.
The remaining solver settings were kept at Comsol’s default values.
Simulation runtimes $T$ are also provided in Table\@~\ref{tab:FigureParams}.

All simulations were performed at the level of the individual volume fractions\@~$\phi_\text{A/B}$, governed by the dynamical equations [Eq.\@~\eqref{eq:PhiDynamics}] for a ternary mixture with ${D_\text{AB} = 0}$ and including the reaction term $\tilde{\mathcal R}$.

\textit{Surface tension.}
To compute the non-equilibrium surface tension shown in Fig.\@~\ref{fig:surfaceTension}, we initialized a one-dimensional system of length ${L=400}$ with a step profile interpolating between the equilibrium coexistence densities $\rho_\pm$.
The jump is located at ${L=200}$, such that the average solute volume fraction equals ${\bar\rho=0.5}$. 
The stationary solute density profile\@~$\rho(r)$ was obtained by evolving the initial state according to Eq.\@~\eqref{eq:SoluteDynamics2} up to a runtime $T$ and using no-flux boundary conditions.
Finally, the non-equilibrium surface tension was evaluated using Eq.\@~\eqref{eq:SurfaceTension} by numerically integrating the stationary density profile with Comsol’s built-in integration routine.

\textit{2d Bulk simulations.}
The two-dimensional simulations presented in Fig.\@~\ref{fig:overview} were carried out in a square domain of size $L$ with periodic boundary conditions. 
Simulations were initialized using the average density ${\bar \rho}$ as well as the corresponding stationary composition\@~$\bar s(\bar \rho)$.
For both fields\@~$\phi_\text{A/B}$, we applied a weak random perturbation of the form
\begin{align}
    \phi_\text{A/B} \big\rvert_{t=0}= \bar \rho\frac{1\pm\bar s(\bar \rho)}{2}[1+\xi\,\eta(\boldsymbol{r})] \, ,
\end{align}
with $\eta$ denoting a uniformly distributed random variable ${\eta \in [-1,1]}$ and a noise amplitude $\xi$ as specified in Table\@~\ref{tab:FigureParams}.

\textit{1d Bulk simulations.}
The one-dimensional simulations presented in Fig.\@~\ref{fig:binodals} were conducted in a domain of size ${L=500}$ for ${\ell = \{0.01,20\}}$ and ${L=1000}$ for ${\ell=100}$, using no-flux boundary conditions.
The system was initialized with a step profile interpolating between the equilibrium coexistence densities of a flat interface $\rho_\pm$, enforcing local chemical equilibrium via ${s(r)=\bar s(\rho)}$.
The coexisting plateau values were obtained by evaluating the stationary solute profile at the system boundaries after a runtime $T$ (Tab.\@~\ref{tab:FigureParams}), provided the system separated into exactly two macroscopic domains.
If multiple high-density domains persisted, the corresponding data were omitted, as the system is then in the arrested-coarsening regime.

\textit{Single droplet simulations.}
The single droplet simulations presented in Fig.\@~\ref{figAPP:Comparison} were conducted in a one-dimensional spherically symmetric system of size ${L=15\ell}$, initialized with a step function profile between densities $\rho_\pm$ and an initial droplet radius ${R=30}$.
At the outer system-boundary, Dirichlet boundary conditions were applied to both fields $\phi_\text{A/B}$, enforcing
\begin{align}
    \phi_\text{A/B}\big\rvert_{r=L}= \rho_\infty\frac{1\pm\bar s(\rho_\infty)}{2} \, ,
\end{align}
while for the chemical potential, we employ no-flux boundary conditions.
This setup ensures local chemical equilibrium at the system boundary.


\begin{thebibliography}{94}%
\makeatletter
\providecommand \@ifxundefined [1]{%
 \@ifx{#1\undefined}
}%
\providecommand \@ifnum [1]{%
 \ifnum #1\expandafter \@firstoftwo
 \else \expandafter \@secondoftwo
 \fi
}%
\providecommand \@ifx [1]{%
 \ifx #1\expandafter \@firstoftwo
 \else \expandafter \@secondoftwo
 \fi
}%
\providecommand \natexlab [1]{#1}%
\providecommand \enquote  [1]{``#1''}%
\providecommand \bibnamefont  [1]{#1}%
\providecommand \bibfnamefont [1]{#1}%
\providecommand \citenamefont [1]{#1}%
\providecommand \href@noop [0]{\@secondoftwo}%
\providecommand \href [0]{\begingroup \@sanitize@url \@href}%
\providecommand \@href[1]{\@@startlink{#1}\@@href}%
\providecommand \@@href[1]{\endgroup#1\@@endlink}%
\providecommand \@sanitize@url [0]{\catcode `\\12\catcode `\$12\catcode `\&12\catcode `\#12\catcode `\^12\catcode `\_12\catcode `\%12\relax}%
\providecommand \@@startlink[1]{}%
\providecommand \@@endlink[0]{}%
\providecommand \url  [0]{\begingroup\@sanitize@url \@url }%
\providecommand \@url [1]{\endgroup\@href {#1}{\urlprefix }}%
\providecommand \urlprefix  [0]{URL }%
\providecommand \Eprint [0]{\href }%
\providecommand \doibase [0]{http://dx.doi.org/}%
\providecommand \selectlanguage [0]{\@gobble}%
\providecommand \bibinfo  [0]{\@secondoftwo}%
\providecommand \bibfield  [0]{\@secondoftwo}%
\providecommand \translation [1]{[#1]}%
\providecommand \BibitemOpen [0]{}%
\providecommand \bibitemStop [0]{}%
\providecommand \bibitemNoStop [0]{.\EOS\space}%
\providecommand \EOS [0]{\spacefactor3000\relax}%
\providecommand \BibitemShut  [1]{\csname bibitem#1\endcsname}%
\let\auto@bib@innerbib\@empty
\bibitem [{\citenamefont {Chen}(2002)}]{Chen.2002}%
  \BibitemOpen
  \bibfield  {author} {\bibinfo {author} {\bibfnamefont {Long-Qing}\ \bibnamefont {Chen}},\ }\bibfield  {title} {\enquote {\bibinfo {title} {{Phase-field models for microstructure evolution}},}\ }\href {\doibase 10.1146/annurev.matsci.32.112001.132041} {\bibfield  {journal} {\bibinfo  {journal} {Annual Review of Materials Research}\ }\textbf {\bibinfo {volume} {32}},\ \bibinfo {pages} {113--140} (\bibinfo {year} {2002})}\BibitemShut {NoStop}%
\bibitem [{\citenamefont {Hyman}\ \emph {et~al.}(2014)\citenamefont {Hyman}, \citenamefont {Weber},\ and\ \citenamefont {Jülicher}}]{Hyman.2014}%
  \BibitemOpen
  \bibfield  {author} {\bibinfo {author} {\bibfnamefont {Anthony~A.}\ \bibnamefont {Hyman}}, \bibinfo {author} {\bibfnamefont {Christoph~A.}\ \bibnamefont {Weber}}, \ and\ \bibinfo {author} {\bibfnamefont {Frank}\ \bibnamefont {Jülicher}},\ }\bibfield  {title} {\enquote {\bibinfo {title} {{Liquid-Liquid Phase Separation in Biology}},}\ }\href {\doibase 10.1146/annurev-cellbio-100913-013325} {\bibfield  {journal} {\bibinfo  {journal} {Annual Review of Cell and Developmental Biology}\ }\textbf {\bibinfo {volume} {30}},\ \bibinfo {pages} {39--58} (\bibinfo {year} {2014})}\BibitemShut {NoStop}%
\bibitem [{\citenamefont {Banani}\ \emph {et~al.}(2017)\citenamefont {Banani}, \citenamefont {Lee}, \citenamefont {Hyman},\ and\ \citenamefont {Rosen}}]{Banani.2017}%
  \BibitemOpen
  \bibfield  {author} {\bibinfo {author} {\bibfnamefont {Salman~F.}\ \bibnamefont {Banani}}, \bibinfo {author} {\bibfnamefont {Hyun~O.}\ \bibnamefont {Lee}}, \bibinfo {author} {\bibfnamefont {Anthony~A.}\ \bibnamefont {Hyman}}, \ and\ \bibinfo {author} {\bibfnamefont {Michael~K.}\ \bibnamefont {Rosen}},\ }\bibfield  {title} {\enquote {\bibinfo {title} {{Biomolecular condensates: organizers of cellular biochemistry}},}\ }\href {\doibase 10.1038/nrm.2017.7} {\bibfield  {journal} {\bibinfo  {journal} {Nature Reviews Molecular Cell Biology}\ }\textbf {\bibinfo {volume} {18}},\ \bibinfo {pages} {285--298} (\bibinfo {year} {2017})}\BibitemShut {NoStop}%
\bibitem [{\citenamefont {Siteur}\ \emph {et~al.}(2023)\citenamefont {Siteur}, \citenamefont {Liu}, \citenamefont {Rottschäfer}, \citenamefont {Heide}, \citenamefont {Rietkerk}, \citenamefont {Doelman}, \citenamefont {Boström},\ and\ \citenamefont {Koppel}}]{Siteur.2023}%
  \BibitemOpen
  \bibfield  {author} {\bibinfo {author} {\bibfnamefont {Koen}\ \bibnamefont {Siteur}}, \bibinfo {author} {\bibfnamefont {Quan-Xing}\ \bibnamefont {Liu}}, \bibinfo {author} {\bibfnamefont {Vivi}\ \bibnamefont {Rottschäfer}}, \bibinfo {author} {\bibfnamefont {Tjisse van~der}\ \bibnamefont {Heide}}, \bibinfo {author} {\bibfnamefont {Max}\ \bibnamefont {Rietkerk}}, \bibinfo {author} {\bibfnamefont {Arjen}\ \bibnamefont {Doelman}}, \bibinfo {author} {\bibfnamefont {Christoffer}\ \bibnamefont {Boström}}, \ and\ \bibinfo {author} {\bibfnamefont {Johan van~de}\ \bibnamefont {Koppel}},\ }\bibfield  {title} {\enquote {\bibinfo {title} {{Phase-separation physics underlies new theory for the resilience of patchy ecosystems}},}\ }\href {\doibase 10.1073/pnas.2202683120} {\bibfield  {journal} {\bibinfo  {journal} {Proceedings of the National Academy of Sciences}\ }\textbf {\bibinfo {volume} {120}},\ \bibinfo {pages} {e2202683120} (\bibinfo {year} {2023})}\BibitemShut {NoStop}%
\bibitem [{\citenamefont {Koppel}\ \emph {et~al.}(2008)\citenamefont {Koppel}, \citenamefont {Gascoigne}, \citenamefont {Theraulaz}, \citenamefont {Rietkerk}, \citenamefont {Mooij},\ and\ \citenamefont {Herman}}]{Koppel.2008}%
  \BibitemOpen
  \bibfield  {author} {\bibinfo {author} {\bibfnamefont {Johan van~de}\ \bibnamefont {Koppel}}, \bibinfo {author} {\bibfnamefont {Joanna~C.}\ \bibnamefont {Gascoigne}}, \bibinfo {author} {\bibfnamefont {Guy}\ \bibnamefont {Theraulaz}}, \bibinfo {author} {\bibfnamefont {Max}\ \bibnamefont {Rietkerk}}, \bibinfo {author} {\bibfnamefont {Wolf~M.}\ \bibnamefont {Mooij}}, \ and\ \bibinfo {author} {\bibfnamefont {Peter M.~J.}\ \bibnamefont {Herman}},\ }\bibfield  {title} {\enquote {\bibinfo {title} {{Experimental Evidence for Spatial Self-Organization and Its Emergent Effects in Mussel Bed Ecosystems}},}\ }\href {\doibase 10.1126/science.1163952} {\bibfield  {journal} {\bibinfo  {journal} {Science}\ }\textbf {\bibinfo {volume} {322}},\ \bibinfo {pages} {739--742} (\bibinfo {year} {2008})}\BibitemShut {NoStop}%
\bibitem [{\citenamefont {Brangwynne}\ \emph {et~al.}(2009)\citenamefont {Brangwynne}, \citenamefont {Eckmann}, \citenamefont {Courson}, \citenamefont {Rybarska}, \citenamefont {Hoege}, \citenamefont {Gharakhani}, \citenamefont {Jülicher},\ and\ \citenamefont {Hyman}}]{Brangwynne.2009}%
  \BibitemOpen
  \bibfield  {author} {\bibinfo {author} {\bibfnamefont {Clifford~P.}\ \bibnamefont {Brangwynne}}, \bibinfo {author} {\bibfnamefont {Christian~R.}\ \bibnamefont {Eckmann}}, \bibinfo {author} {\bibfnamefont {David~S.}\ \bibnamefont {Courson}}, \bibinfo {author} {\bibfnamefont {Agata}\ \bibnamefont {Rybarska}}, \bibinfo {author} {\bibfnamefont {Carsten}\ \bibnamefont {Hoege}}, \bibinfo {author} {\bibfnamefont {Jöbin}\ \bibnamefont {Gharakhani}}, \bibinfo {author} {\bibfnamefont {Frank}\ \bibnamefont {Jülicher}}, \ and\ \bibinfo {author} {\bibfnamefont {Anthony~A.}\ \bibnamefont {Hyman}},\ }\bibfield  {title} {\enquote {\bibinfo {title} {{Germline P Granules Are Liquid Droplets That Localize by Controlled Dissolution/Condensation}},}\ }\href {\doibase 10.1126/science.1172046} {\bibfield  {journal} {\bibinfo  {journal} {Science}\ }\textbf {\bibinfo {volume} {324}},\ \bibinfo {pages} {1729--1732} (\bibinfo {year} {2009})}\BibitemShut {NoStop}%
\bibitem [{\citenamefont {Feric}\ \emph {et~al.}(2016)\citenamefont {Feric}, \citenamefont {Vaidya}, \citenamefont {Harmon}, \citenamefont {Mitrea}, \citenamefont {Zhu}, \citenamefont {Richardson}, \citenamefont {Kriwacki}, \citenamefont {Pappu},\ and\ \citenamefont {Brangwynne}}]{Feric.2016}%
  \BibitemOpen
  \bibfield  {author} {\bibinfo {author} {\bibfnamefont {Marina}\ \bibnamefont {Feric}}, \bibinfo {author} {\bibfnamefont {Nilesh}\ \bibnamefont {Vaidya}}, \bibinfo {author} {\bibfnamefont {Tyler~S.}\ \bibnamefont {Harmon}}, \bibinfo {author} {\bibfnamefont {Diana~M.}\ \bibnamefont {Mitrea}}, \bibinfo {author} {\bibfnamefont {Lian}\ \bibnamefont {Zhu}}, \bibinfo {author} {\bibfnamefont {Tiffany~M.}\ \bibnamefont {Richardson}}, \bibinfo {author} {\bibfnamefont {Richard~W.}\ \bibnamefont {Kriwacki}}, \bibinfo {author} {\bibfnamefont {Rohit~V.}\ \bibnamefont {Pappu}}, \ and\ \bibinfo {author} {\bibfnamefont {Clifford~P.}\ \bibnamefont {Brangwynne}},\ }\bibfield  {title} {\enquote {\bibinfo {title} {{Coexisting Liquid Phases Underlie Nucleolar Subcompartments}},}\ }\href {\doibase 10.1016/j.cell.2016.04.047} {\bibfield  {journal} {\bibinfo  {journal} {Cell}\ }\textbf {\bibinfo {volume} {165}},\ \bibinfo {pages} {1686--1697} (\bibinfo {year} {2016})}\BibitemShut {NoStop}%
\bibitem [{\citenamefont {Ben-Jacob}\ \emph {et~al.}(2000)\citenamefont {Ben-Jacob}, \citenamefont {Cohen},\ and\ \citenamefont {Levine}}]{Ben-Jacob.2000}%
  \BibitemOpen
  \bibfield  {author} {\bibinfo {author} {\bibfnamefont {Eshel}\ \bibnamefont {Ben-Jacob}}, \bibinfo {author} {\bibfnamefont {Inon}\ \bibnamefont {Cohen}}, \ and\ \bibinfo {author} {\bibfnamefont {Herbert}\ \bibnamefont {Levine}},\ }\bibfield  {title} {\enquote {\bibinfo {title} {{Cooperative self-organization of microorganisms}},}\ }\href {\doibase 10.1080/000187300405228} {\bibfield  {journal} {\bibinfo  {journal} {Advances in Physics}\ }\textbf {\bibinfo {volume} {49}},\ \bibinfo {pages} {395--554} (\bibinfo {year} {2000})}\BibitemShut {NoStop}%
\bibitem [{\citenamefont {Liebchen}\ and\ \citenamefont {Löwen}(2018)}]{Liebchen.2018}%
  \BibitemOpen
  \bibfield  {author} {\bibinfo {author} {\bibfnamefont {Benno}\ \bibnamefont {Liebchen}}\ and\ \bibinfo {author} {\bibfnamefont {Hartmut}\ \bibnamefont {Löwen}},\ }\bibfield  {title} {\enquote {\bibinfo {title} {{Synthetic Chemotaxis and Collective Behavior in Active Matter}},}\ }\href {\doibase 10.1021/acs.accounts.8b00215} {\bibfield  {journal} {\bibinfo  {journal} {Accounts of Chemical Research}\ }\textbf {\bibinfo {volume} {51}},\ \bibinfo {pages} {2982--2990} (\bibinfo {year} {2018})}\BibitemShut {NoStop}%
\bibitem [{\citenamefont {Palacci}\ \emph {et~al.}(2013)\citenamefont {Palacci}, \citenamefont {Sacanna}, \citenamefont {Steinberg}, \citenamefont {Pine},\ and\ \citenamefont {Chaikin}}]{Palacci.2013}%
  \BibitemOpen
  \bibfield  {author} {\bibinfo {author} {\bibfnamefont {Jeremie}\ \bibnamefont {Palacci}}, \bibinfo {author} {\bibfnamefont {Stefano}\ \bibnamefont {Sacanna}}, \bibinfo {author} {\bibfnamefont {Asher~Preska}\ \bibnamefont {Steinberg}}, \bibinfo {author} {\bibfnamefont {David~J.}\ \bibnamefont {Pine}}, \ and\ \bibinfo {author} {\bibfnamefont {Paul~M.}\ \bibnamefont {Chaikin}},\ }\bibfield  {title} {\enquote {\bibinfo {title} {{Living Crystals of Light-Activated Colloidal Surfers}},}\ }\href {\doibase 10.1126/science.1230020} {\bibfield  {journal} {\bibinfo  {journal} {Science}\ }\textbf {\bibinfo {volume} {339}},\ \bibinfo {pages} {936--940} (\bibinfo {year} {2013})}\BibitemShut {NoStop}%
\bibitem [{\citenamefont {Doi}(2013)}]{Doi.2013}%
  \BibitemOpen
  \bibfield  {author} {\bibinfo {author} {\bibfnamefont {Masao}\ \bibnamefont {Doi}},\ }\href {\doibase 10.1093/acprof:oso/9780199652952.001.0001} {\emph {\bibinfo {title} {{Soft Matter Physics}}}}\ (\bibinfo  {publisher} {Oxford University Press},\ \bibinfo {year} {2013})\BibitemShut {NoStop}%
\bibitem [{\citenamefont {Liu}\ and\ \citenamefont {Goldenfeld}(1989)}]{Liu.1989}%
  \BibitemOpen
  \bibfield  {author} {\bibinfo {author} {\bibfnamefont {Fong}\ \bibnamefont {Liu}}\ and\ \bibinfo {author} {\bibfnamefont {Nigel}\ \bibnamefont {Goldenfeld}},\ }\bibfield  {title} {\enquote {\bibinfo {title} {{Dynamics of phase separation in block copolymer melts}},}\ }\href {\doibase 10.1103/physreva.39.4805} {\bibfield  {journal} {\bibinfo  {journal} {Physical Review A}\ }\textbf {\bibinfo {volume} {39}},\ \bibinfo {pages} {4805--4810} (\bibinfo {year} {1989})}\BibitemShut {NoStop}%
\bibitem [{\citenamefont {Muratov}(2002)}]{Muratov.2002}%
  \BibitemOpen
  \bibfield  {author} {\bibinfo {author} {\bibfnamefont {C.~B.}\ \bibnamefont {Muratov}},\ }\bibfield  {title} {\enquote {\bibinfo {title} {{Theory of domain patterns in systems with long-range interactions of Coulomb type}},}\ }\href {\doibase 10.1103/physreve.66.066108} {\bibfield  {journal} {\bibinfo  {journal} {Physical Review E}\ }\textbf {\bibinfo {volume} {66}},\ \bibinfo {pages} {066108} (\bibinfo {year} {2002})}\BibitemShut {NoStop}%
\bibitem [{\citenamefont {Kumar}\ and\ \citenamefont {Safran}(2023)}]{Kumar.2023}%
  \BibitemOpen
  \bibfield  {author} {\bibinfo {author} {\bibfnamefont {Amit}\ \bibnamefont {Kumar}}\ and\ \bibinfo {author} {\bibfnamefont {Samuel~A.}\ \bibnamefont {Safran}},\ }\bibfield  {title} {\enquote {\bibinfo {title} {{Fluctuations and Shape Dependence of Microphase Separation in Systems with Long-Range Interactions}},}\ }\href {\doibase 10.1103/physrevlett.131.258401} {\bibfield  {journal} {\bibinfo  {journal} {Physical Review Letters}\ }\textbf {\bibinfo {volume} {131}},\ \bibinfo {pages} {258401} (\bibinfo {year} {2023})}\BibitemShut {NoStop}%
\bibitem [{\citenamefont {Winter}\ \emph {et~al.}(2025)\citenamefont {Winter}, \citenamefont {Liu}, \citenamefont {Ziepke}, \citenamefont {Dadunashvili},\ and\ \citenamefont {Frey}}]{Winter.2025}%
  \BibitemOpen
  \bibfield  {author} {\bibinfo {author} {\bibfnamefont {Antonia}\ \bibnamefont {Winter}}, \bibinfo {author} {\bibfnamefont {Yuhao}\ \bibnamefont {Liu}}, \bibinfo {author} {\bibfnamefont {Alexander}\ \bibnamefont {Ziepke}}, \bibinfo {author} {\bibfnamefont {George}\ \bibnamefont {Dadunashvili}}, \ and\ \bibinfo {author} {\bibfnamefont {Erwin}\ \bibnamefont {Frey}},\ }\bibfield  {title} {\enquote {\bibinfo {title} {{Phase separation on deformable membranes: Interplay of mechanical coupling and dynamic surface geometry}},}\ }\href {\doibase 10.1103/physreve.111.044405} {\bibfield  {journal} {\bibinfo  {journal} {Physical Review E}\ }\textbf {\bibinfo {volume} {111}},\ \bibinfo {pages} {044405} (\bibinfo {year} {2025})}\BibitemShut {NoStop}%
\bibitem [{\citenamefont {Style}\ \emph {et~al.}(2018)\citenamefont {Style}, \citenamefont {Sai}, \citenamefont {Fanelli}, \citenamefont {Ijavi}, \citenamefont {Smith-Mannschott}, \citenamefont {Xu}, \citenamefont {Wilen},\ and\ \citenamefont {Dufresne}}]{Style.2018}%
  \BibitemOpen
  \bibfield  {author} {\bibinfo {author} {\bibfnamefont {Robert~W.}\ \bibnamefont {Style}}, \bibinfo {author} {\bibfnamefont {Tianqi}\ \bibnamefont {Sai}}, \bibinfo {author} {\bibfnamefont {Nicoló}\ \bibnamefont {Fanelli}}, \bibinfo {author} {\bibfnamefont {Mahdiye}\ \bibnamefont {Ijavi}}, \bibinfo {author} {\bibfnamefont {Katrina}\ \bibnamefont {Smith-Mannschott}}, \bibinfo {author} {\bibfnamefont {Qin}\ \bibnamefont {Xu}}, \bibinfo {author} {\bibfnamefont {Lawrence~A.}\ \bibnamefont {Wilen}}, \ and\ \bibinfo {author} {\bibfnamefont {Eric~R.}\ \bibnamefont {Dufresne}},\ }\bibfield  {title} {\enquote {\bibinfo {title} {{Liquid-Liquid Phase Separation in an Elastic Network}},}\ }\href {\doibase 10.1103/physrevx.8.011028} {\bibfield  {journal} {\bibinfo  {journal} {Physical Review X}\ }\textbf {\bibinfo {volume} {8}},\ \bibinfo {pages} {011028} (\bibinfo {year} {2018})},\ \Eprint {http://arxiv.org/abs/1709.00500} {1709.00500} \BibitemShut {NoStop}%
\bibitem [{\citenamefont {Zhang}\ \emph {et~al.}(2021)\citenamefont {Zhang}, \citenamefont {Lee}, \citenamefont {Meir}, \citenamefont {Brangwynne},\ and\ \citenamefont {Wingreen}}]{Zhang.2021}%
  \BibitemOpen
  \bibfield  {author} {\bibinfo {author} {\bibfnamefont {Yaojun}\ \bibnamefont {Zhang}}, \bibinfo {author} {\bibfnamefont {Daniel S.~W.}\ \bibnamefont {Lee}}, \bibinfo {author} {\bibfnamefont {Yigal}\ \bibnamefont {Meir}}, \bibinfo {author} {\bibfnamefont {Clifford~P.}\ \bibnamefont {Brangwynne}}, \ and\ \bibinfo {author} {\bibfnamefont {Ned~S.}\ \bibnamefont {Wingreen}},\ }\bibfield  {title} {\enquote {\bibinfo {title} {{Mechanical Frustration of Phase Separation in the Cell Nucleus by Chromatin}},}\ }\href {\doibase 10.1103/physrevlett.126.258102} {\bibfield  {journal} {\bibinfo  {journal} {Physical Review Letters}\ }\textbf {\bibinfo {volume} {126}},\ \bibinfo {pages} {258102} (\bibinfo {year} {2021})}\BibitemShut {NoStop}%
\bibitem [{\citenamefont {Vidal-Henriquez}\ and\ \citenamefont {Zwicker}(2021)}]{Vidal-Henriquez.2021}%
  \BibitemOpen
  \bibfield  {author} {\bibinfo {author} {\bibfnamefont {Estefania}\ \bibnamefont {Vidal-Henriquez}}\ and\ \bibinfo {author} {\bibfnamefont {David}\ \bibnamefont {Zwicker}},\ }\bibfield  {title} {\enquote {\bibinfo {title} {{Cavitation controls droplet sizes in elastic media}},}\ }\href {\doibase 10.1073/pnas.2102014118} {\bibfield  {journal} {\bibinfo  {journal} {Proceedings of the National Academy of Sciences}\ }\textbf {\bibinfo {volume} {118}},\ \bibinfo {pages} {e2102014118} (\bibinfo {year} {2021})},\ \Eprint {http://arxiv.org/abs/2102.02506} {2102.02506} \BibitemShut {NoStop}%
\bibitem [{\citenamefont {Ostwald}(1897)}]{Ostwald.1897}%
  \BibitemOpen
  \bibfield  {author} {\bibinfo {author} {\bibfnamefont {W.}~\bibnamefont {Ostwald}},\ }\bibfield  {title} {\enquote {\bibinfo {title} {{Studien über die Bildung und Umwandlung fester Körper}},}\ }\href {\doibase 10.1515/zpch-1897-2233} {\bibfield  {journal} {\bibinfo  {journal} {Zeitschrift für Physikalische Chemie}\ }\textbf {\bibinfo {volume} {22U}},\ \bibinfo {pages} {289--330} (\bibinfo {year} {1897})}\BibitemShut {NoStop}%
\bibitem [{\citenamefont {Wagner}(1961)}]{Wagner.1961}%
  \BibitemOpen
  \bibfield  {author} {\bibinfo {author} {\bibfnamefont {Carl}\ \bibnamefont {Wagner}},\ }\bibfield  {title} {\enquote {\bibinfo {title} {{Theorie der Alterung von Niederschlägen durch Umlösen (Ostwald‐Reifung)}},}\ }\href {\doibase 10.1002/bbpc.19610650704} {\bibfield  {journal} {\bibinfo  {journal} {Zeitschrift für Elektrochemie, Berichte der Bunsengesellschaft für physikalische Chemie}\ }\textbf {\bibinfo {volume} {65}},\ \bibinfo {pages} {581--591} (\bibinfo {year} {1961})}\BibitemShut {NoStop}%
\bibitem [{\citenamefont {Lifshitz}\ and\ \citenamefont {Slyozov}(1961)}]{Lifshitz.1961}%
  \BibitemOpen
  \bibfield  {author} {\bibinfo {author} {\bibfnamefont {I.M.}\ \bibnamefont {Lifshitz}}\ and\ \bibinfo {author} {\bibfnamefont {V.V.}\ \bibnamefont {Slyozov}},\ }\bibfield  {title} {\enquote {\bibinfo {title} {{The kinetics of precipitation from supersaturated solid solutions}},}\ }\href {\doibase 10.1016/0022-3697(61)90054-3} {\bibfield  {journal} {\bibinfo  {journal} {Journal of Physics and Chemistry of Solids}\ }\textbf {\bibinfo {volume} {19}},\ \bibinfo {pages} {35--50} (\bibinfo {year} {1961})}\BibitemShut {NoStop}%
\bibitem [{\citenamefont {Weber}\ \emph {et~al.}(2019)\citenamefont {Weber}, \citenamefont {Zwicker}, \citenamefont {Jülicher},\ and\ \citenamefont {Lee}}]{Weber.2019}%
  \BibitemOpen
  \bibfield  {author} {\bibinfo {author} {\bibfnamefont {Christoph~A}\ \bibnamefont {Weber}}, \bibinfo {author} {\bibfnamefont {David}\ \bibnamefont {Zwicker}}, \bibinfo {author} {\bibfnamefont {Frank}\ \bibnamefont {Jülicher}}, \ and\ \bibinfo {author} {\bibfnamefont {Chiu~Fan}\ \bibnamefont {Lee}},\ }\bibfield  {title} {\enquote {\bibinfo {title} {{Physics of active emulsions}},}\ }\href {\doibase 10.1088/1361-6633/ab052b} {\bibfield  {journal} {\bibinfo  {journal} {Reports on Progress in Physics}\ }\textbf {\bibinfo {volume} {82}},\ \bibinfo {pages} {064601} (\bibinfo {year} {2019})},\ \Eprint {http://arxiv.org/abs/1806.09552} {1806.09552} \BibitemShut {NoStop}%
\bibitem [{\citenamefont {Zwicker}\ \emph {et~al.}(2025)\citenamefont {Zwicker}, \citenamefont {Paulin},\ and\ \citenamefont {Burg}}]{Zwicker.2025}%
  \BibitemOpen
  \bibfield  {author} {\bibinfo {author} {\bibfnamefont {David}\ \bibnamefont {Zwicker}}, \bibinfo {author} {\bibfnamefont {Oliver~W}\ \bibnamefont {Paulin}}, \ and\ \bibinfo {author} {\bibfnamefont {Cathelijne~ter}\ \bibnamefont {Burg}},\ }\bibfield  {title} {\enquote {\bibinfo {title} {{Physics of droplet regulation in biological cells}},}\ }\href {\doibase 10.1088/1361-6633/ae12a7} {\bibfield  {journal} {\bibinfo  {journal} {Reports on Progress in Physics}\ }\textbf {\bibinfo {volume} {88}},\ \bibinfo {pages} {116601} (\bibinfo {year} {2025})}\BibitemShut {NoStop}%
\bibitem [{\citenamefont {Schnitzer}(1993)}]{Schnitzer.1993}%
  \BibitemOpen
  \bibfield  {author} {\bibinfo {author} {\bibfnamefont {Mark~J.}\ \bibnamefont {Schnitzer}},\ }\bibfield  {title} {\enquote {\bibinfo {title} {{Theory of continuum random walks and application to chemotaxis}},}\ }\href {\doibase 10.1103/physreve.48.2553} {\bibfield  {journal} {\bibinfo  {journal} {Physical Review E}\ }\textbf {\bibinfo {volume} {48}},\ \bibinfo {pages} {2553--2568} (\bibinfo {year} {1993})}\BibitemShut {NoStop}%
\bibitem [{\citenamefont {Tailleur}\ and\ \citenamefont {Cates}(2008)}]{Tailleur.2008}%
  \BibitemOpen
  \bibfield  {author} {\bibinfo {author} {\bibfnamefont {J.}~\bibnamefont {Tailleur}}\ and\ \bibinfo {author} {\bibfnamefont {M.~E.}\ \bibnamefont {Cates}},\ }\bibfield  {title} {\enquote {\bibinfo {title} {{Statistical Mechanics of Interacting Run-and-Tumble Bacteria}},}\ }\href {\doibase 10.1103/physrevlett.100.218103} {\bibfield  {journal} {\bibinfo  {journal} {Physical Review Letters}\ }\textbf {\bibinfo {volume} {100}},\ \bibinfo {pages} {218103} (\bibinfo {year} {2008})},\ \Eprint {http://arxiv.org/abs/0803.1069} {0803.1069} \BibitemShut {NoStop}%
\bibitem [{\citenamefont {Cates}\ and\ \citenamefont {Tailleur}(2014)}]{Cates.2014}%
  \BibitemOpen
  \bibfield  {author} {\bibinfo {author} {\bibfnamefont {Michael~E.}\ \bibnamefont {Cates}}\ and\ \bibinfo {author} {\bibfnamefont {Julien}\ \bibnamefont {Tailleur}},\ }\bibfield  {title} {\enquote {\bibinfo {title} {{Motility-Induced Phase Separation}},}\ }\href {\doibase 10.1146/annurev-conmatphys-031214-014710} {\bibfield  {journal} {\bibinfo  {journal} {Annual Review of Condensed Matter Physics}\ }\textbf {\bibinfo {volume} {6}},\ \bibinfo {pages} {1--26} (\bibinfo {year} {2014})},\ \Eprint {http://arxiv.org/abs/1406.3533} {1406.3533} \BibitemShut {NoStop}%
\bibitem [{\citenamefont {Murray}(2003)}]{Murray.2003}%
  \BibitemOpen
  \bibinfo {editor} {\bibfnamefont {J.D.}\ \bibnamefont {Murray}},\ ed.,\ \href {\doibase 10.1007/b98869} {\emph {\bibinfo {title} {{Mathematical Biology, II: Spatial Models and Biomedical Applications}}}},\ Interdisciplinary Applied Mathematics\ (\bibinfo  {publisher} {Springer New York},\ \bibinfo {address} {NY},\ \bibinfo {year} {2003})\BibitemShut {NoStop}%
\bibitem [{\citenamefont {Weyer}\ \emph {et~al.}(2025)\citenamefont {Weyer}, \citenamefont {Muramatsu},\ and\ \citenamefont {Frey}}]{Weyer.2025}%
  \BibitemOpen
  \bibfield  {author} {\bibinfo {author} {\bibfnamefont {Henrik}\ \bibnamefont {Weyer}}, \bibinfo {author} {\bibfnamefont {David}\ \bibnamefont {Muramatsu}}, \ and\ \bibinfo {author} {\bibfnamefont {Erwin}\ \bibnamefont {Frey}},\ }\bibfield  {title} {\enquote {\bibinfo {title} {{Chemotaxis-Induced Phase Separation}},}\ }\href {\doibase 10.1103/2933-45qc} {\bibfield  {journal} {\bibinfo  {journal} {Physical Review Letters}\ }\textbf {\bibinfo {volume} {135}},\ \bibinfo {pages} {208402} (\bibinfo {year} {2025})},\ \Eprint {http://arxiv.org/abs/2409.20090} {2409.20090} \BibitemShut {NoStop}%
\bibitem [{\citenamefont {Turing}(1952)}]{Turing.1952}%
  \BibitemOpen
  \bibfield  {author} {\bibinfo {author} {\bibfnamefont {Alan~Mathison}\ \bibnamefont {Turing}},\ }\bibfield  {title} {\enquote {\bibinfo {title} {{The chemical basis of morphogenesis}},}\ }\href {\doibase 10.1098/rstb.1952.0012} {\bibfield  {journal} {\bibinfo  {journal} {Philosophical Transactions of the Royal Society of London. Series B, Biological Sciences}\ }\textbf {\bibinfo {volume} {237}},\ \bibinfo {pages} {37--72} (\bibinfo {year} {1952})}\BibitemShut {NoStop}%
\bibitem [{\citenamefont {Halatek}\ \emph {et~al.}(2018)\citenamefont {Halatek}, \citenamefont {Brauns},\ and\ \citenamefont {Frey}}]{Halatek.2018b}%
  \BibitemOpen
  \bibfield  {author} {\bibinfo {author} {\bibfnamefont {J.}~\bibnamefont {Halatek}}, \bibinfo {author} {\bibfnamefont {F.}~\bibnamefont {Brauns}}, \ and\ \bibinfo {author} {\bibfnamefont {E.}~\bibnamefont {Frey}},\ }\bibfield  {title} {\enquote {\bibinfo {title} {{Self-organization principles of intracellular pattern formation}},}\ }\href {\doibase 10.1098/rstb.2017.0107} {\bibfield  {journal} {\bibinfo  {journal} {Philosophical Transactions of the Royal Society B: Biological Sciences}\ }\textbf {\bibinfo {volume} {373}},\ \bibinfo {pages} {20170107} (\bibinfo {year} {2018})},\ \Eprint {http://arxiv.org/abs/1802.07169} {1802.07169} \BibitemShut {NoStop}%
\bibitem [{\citenamefont {Frey}\ and\ \citenamefont {Brauns}(2022)}]{Frey.2022}%
  \BibitemOpen
  \bibfield  {author} {\bibinfo {author} {\bibfnamefont {Erwin}\ \bibnamefont {Frey}}\ and\ \bibinfo {author} {\bibfnamefont {Fridtjof}\ \bibnamefont {Brauns}},\ }\bibfield  {title} {\enquote {\bibinfo {title} {{Active Matter and Nonequilibrium Statistical Physics}},}\ }\href {\doibase 10.1093/oso/9780192858313.003.0011} {\ ,\ \bibinfo {pages} {347--445} (\bibinfo {year} {2022})}\BibitemShut {NoStop}%
\bibitem [{\citenamefont {Brauns}\ \emph {et~al.}(2020)\citenamefont {Brauns}, \citenamefont {Halatek},\ and\ \citenamefont {Frey}}]{Brauns.2020}%
  \BibitemOpen
  \bibfield  {author} {\bibinfo {author} {\bibfnamefont {Fridtjof}\ \bibnamefont {Brauns}}, \bibinfo {author} {\bibfnamefont {Jacob}\ \bibnamefont {Halatek}}, \ and\ \bibinfo {author} {\bibfnamefont {Erwin}\ \bibnamefont {Frey}},\ }\bibfield  {title} {\enquote {\bibinfo {title} {{Phase-Space Geometry of Mass-Conserving Reaction-Diffusion Dynamics}},}\ }\href {\doibase 10.1103/physrevx.10.041036} {\bibfield  {journal} {\bibinfo  {journal} {Physical Review X}\ }\textbf {\bibinfo {volume} {10}},\ \bibinfo {pages} {041036} (\bibinfo {year} {2020})},\ \Eprint {http://arxiv.org/abs/1812.08684} {1812.08684} \BibitemShut {NoStop}%
\bibitem [{\citenamefont {Brauns}\ \emph {et~al.}(2021)\citenamefont {Brauns}, \citenamefont {Weyer}, \citenamefont {Halatek}, \citenamefont {Yoon},\ and\ \citenamefont {Frey}}]{Brauns.2021}%
  \BibitemOpen
  \bibfield  {author} {\bibinfo {author} {\bibfnamefont {Fridtjof}\ \bibnamefont {Brauns}}, \bibinfo {author} {\bibfnamefont {Henrik}\ \bibnamefont {Weyer}}, \bibinfo {author} {\bibfnamefont {Jacob}\ \bibnamefont {Halatek}}, \bibinfo {author} {\bibfnamefont {Junghoon}\ \bibnamefont {Yoon}}, \ and\ \bibinfo {author} {\bibfnamefont {Erwin}\ \bibnamefont {Frey}},\ }\bibfield  {title} {\enquote {\bibinfo {title} {{Wavelength Selection by Interrupted Coarsening in Reaction-Diffusion Systems}},}\ }\href {\doibase 10.1103/physrevlett.126.104101} {\bibfield  {journal} {\bibinfo  {journal} {Physical Review Letters}\ }\textbf {\bibinfo {volume} {126}},\ \bibinfo {pages} {104101} (\bibinfo {year} {2021})},\ \Eprint {http://arxiv.org/abs/2005.01495} {2005.01495} \BibitemShut {NoStop}%
\bibitem [{\citenamefont {Glotzer}\ \emph {et~al.}(1994)\citenamefont {Glotzer}, \citenamefont {Stauffer},\ and\ \citenamefont {Jan}}]{Glotzer.1994}%
  \BibitemOpen
  \bibfield  {author} {\bibinfo {author} {\bibfnamefont {Sharon~C.}\ \bibnamefont {Glotzer}}, \bibinfo {author} {\bibfnamefont {Dietrich}\ \bibnamefont {Stauffer}}, \ and\ \bibinfo {author} {\bibfnamefont {Naeem}\ \bibnamefont {Jan}},\ }\bibfield  {title} {\enquote {\bibinfo {title} {{Monte Carlo simulations of phase separation in chemically reactive binary mixtures}},}\ }\href {\doibase 10.1103/physrevlett.72.4109} {\bibfield  {journal} {\bibinfo  {journal} {Physical Review Letters}\ }\textbf {\bibinfo {volume} {72}},\ \bibinfo {pages} {4109--4112} (\bibinfo {year} {1994})}\BibitemShut {NoStop}%
\bibitem [{\citenamefont {Carati}\ and\ \citenamefont {Lefever}(1997)}]{Carati.1997}%
  \BibitemOpen
  \bibfield  {author} {\bibinfo {author} {\bibfnamefont {Daniele}\ \bibnamefont {Carati}}\ and\ \bibinfo {author} {\bibfnamefont {René}\ \bibnamefont {Lefever}},\ }\bibfield  {title} {\enquote {\bibinfo {title} {{Chemical freezing of phase separation in immiscible binary mixtures}},}\ }\href {\doibase 10.1103/physreve.56.3127} {\bibfield  {journal} {\bibinfo  {journal} {Physical Review E}\ }\textbf {\bibinfo {volume} {56}},\ \bibinfo {pages} {3127--3136} (\bibinfo {year} {1997})}\BibitemShut {NoStop}%
\bibitem [{\citenamefont {Cates}\ \emph {et~al.}(2010)\citenamefont {Cates}, \citenamefont {Marenduzzo}, \citenamefont {Pagonabarraga},\ and\ \citenamefont {Tailleur}}]{Cates.2010}%
  \BibitemOpen
  \bibfield  {author} {\bibinfo {author} {\bibfnamefont {M.~E.}\ \bibnamefont {Cates}}, \bibinfo {author} {\bibfnamefont {D.}~\bibnamefont {Marenduzzo}}, \bibinfo {author} {\bibfnamefont {I.}~\bibnamefont {Pagonabarraga}}, \ and\ \bibinfo {author} {\bibfnamefont {J.}~\bibnamefont {Tailleur}},\ }\bibfield  {title} {\enquote {\bibinfo {title} {{Arrested phase separation in reproducing bacteria creates a generic route to pattern formation}},}\ }\href {\doibase 10.1073/pnas.1001994107} {\bibfield  {journal} {\bibinfo  {journal} {Proceedings of the National Academy of Sciences}\ }\textbf {\bibinfo {volume} {107}},\ \bibinfo {pages} {11715--11720} (\bibinfo {year} {2010})},\ \Eprint {http://arxiv.org/abs/1001.0057} {1001.0057} \BibitemShut {NoStop}%
\bibitem [{\citenamefont {Zwicker}\ \emph {et~al.}(2015)\citenamefont {Zwicker}, \citenamefont {Hyman},\ and\ \citenamefont {Jülicher}}]{Zwicker.2015}%
  \BibitemOpen
  \bibfield  {author} {\bibinfo {author} {\bibfnamefont {David}\ \bibnamefont {Zwicker}}, \bibinfo {author} {\bibfnamefont {Anthony~A.}\ \bibnamefont {Hyman}}, \ and\ \bibinfo {author} {\bibfnamefont {Frank}\ \bibnamefont {Jülicher}},\ }\bibfield  {title} {\enquote {\bibinfo {title} {{Suppression of Ostwald ripening in active emulsions}},}\ }\href {\doibase 10.1103/physreve.92.012317} {\bibfield  {journal} {\bibinfo  {journal} {Physical Review E}\ }\textbf {\bibinfo {volume} {92}},\ \bibinfo {pages} {012317} (\bibinfo {year} {2015})}\BibitemShut {NoStop}%
\bibitem [{\citenamefont {Wurtz}\ and\ \citenamefont {Lee}(2018{\natexlab{a}})}]{Wurtz.2018}%
  \BibitemOpen
  \bibfield  {author} {\bibinfo {author} {\bibfnamefont {Jean~David}\ \bibnamefont {Wurtz}}\ and\ \bibinfo {author} {\bibfnamefont {Chiu~Fan}\ \bibnamefont {Lee}},\ }\bibfield  {title} {\enquote {\bibinfo {title} {{Chemical-Reaction-Controlled Phase Separated Drops: Formation, Size Selection, and Coarsening}},}\ }\href {\doibase 10.1103/physrevlett.120.078102} {\bibfield  {journal} {\bibinfo  {journal} {Physical Review Letters}\ }\textbf {\bibinfo {volume} {120}},\ \bibinfo {pages} {078102} (\bibinfo {year} {2018}{\natexlab{a}})},\ \Eprint {http://arxiv.org/abs/1707.08433} {1707.08433} \BibitemShut {NoStop}%
\bibitem [{\citenamefont {Donau}\ and\ \citenamefont {Boekhoven}(2023)}]{Donau.2023}%
  \BibitemOpen
  \bibfield  {author} {\bibinfo {author} {\bibfnamefont {Carsten}\ \bibnamefont {Donau}}\ and\ \bibinfo {author} {\bibfnamefont {Job}\ \bibnamefont {Boekhoven}},\ }\bibfield  {title} {\enquote {\bibinfo {title} {{The chemistry of chemically fueled droplets}},}\ }\href {\doibase 10.1016/j.trechm.2022.11.003} {\bibfield  {journal} {\bibinfo  {journal} {Trends in Chemistry}\ }\textbf {\bibinfo {volume} {5}},\ \bibinfo {pages} {45--60} (\bibinfo {year} {2023})}\BibitemShut {NoStop}%
\bibitem [{\citenamefont {Tjhung}\ \emph {et~al.}(2018)\citenamefont {Tjhung}, \citenamefont {Nardini},\ and\ \citenamefont {Cates}}]{Tjhung.2018}%
  \BibitemOpen
  \bibfield  {author} {\bibinfo {author} {\bibfnamefont {Elsen}\ \bibnamefont {Tjhung}}, \bibinfo {author} {\bibfnamefont {Cesare}\ \bibnamefont {Nardini}}, \ and\ \bibinfo {author} {\bibfnamefont {Michael~E.}\ \bibnamefont {Cates}},\ }\bibfield  {title} {\enquote {\bibinfo {title} {{Cluster Phases and Bubbly Phase Separation in Active Fluids: Reversal of the Ostwald Process}},}\ }\href {\doibase 10.1103/physrevx.8.031080} {\bibfield  {journal} {\bibinfo  {journal} {Physical Review X}\ }\textbf {\bibinfo {volume} {8}},\ \bibinfo {pages} {031080} (\bibinfo {year} {2018})},\ \Eprint {http://arxiv.org/abs/1801.07687} {1801.07687} \BibitemShut {NoStop}%
\bibitem [{\citenamefont {Zwicker}\ \emph {et~al.}(2017)\citenamefont {Zwicker}, \citenamefont {Seyboldt}, \citenamefont {Weber}, \citenamefont {Hyman},\ and\ \citenamefont {Jülicher}}]{Zwicker.2017}%
  \BibitemOpen
  \bibfield  {author} {\bibinfo {author} {\bibfnamefont {David}\ \bibnamefont {Zwicker}}, \bibinfo {author} {\bibfnamefont {Rabea}\ \bibnamefont {Seyboldt}}, \bibinfo {author} {\bibfnamefont {Christoph~A.}\ \bibnamefont {Weber}}, \bibinfo {author} {\bibfnamefont {Anthony~A.}\ \bibnamefont {Hyman}}, \ and\ \bibinfo {author} {\bibfnamefont {Frank}\ \bibnamefont {Jülicher}},\ }\bibfield  {title} {\enquote {\bibinfo {title} {{Growth and division of active droplets provides a model for protocells}},}\ }\href {\doibase 10.1038/nphys3984} {\bibfield  {journal} {\bibinfo  {journal} {Nature Physics}\ }\textbf {\bibinfo {volume} {13}},\ \bibinfo {pages} {408--413} (\bibinfo {year} {2017})},\ \Eprint {http://arxiv.org/abs/1603.01571} {1603.01571} \BibitemShut {NoStop}%
\bibitem [{\citenamefont {Raßhofer}\ \emph {et~al.}(2025)\citenamefont {Raßhofer}, \citenamefont {Bauer}, \citenamefont {Ziepke}, \citenamefont {Maryshev},\ and\ \citenamefont {Frey}}]{Rasshofer.2025}%
  \BibitemOpen
  \bibfield  {author} {\bibinfo {author} {\bibfnamefont {Florian}\ \bibnamefont {Raßhofer}}, \bibinfo {author} {\bibfnamefont {Simon}\ \bibnamefont {Bauer}}, \bibinfo {author} {\bibfnamefont {Alexander}\ \bibnamefont {Ziepke}}, \bibinfo {author} {\bibfnamefont {Ivan}\ \bibnamefont {Maryshev}}, \ and\ \bibinfo {author} {\bibfnamefont {Erwin}\ \bibnamefont {Frey}},\ }\bibfield  {title} {\enquote {\bibinfo {title} {{Capillary wave formation in conserved active emulsions}},}\ }\href {\doibase 10.1103/lqws-dkmy} {\bibfield  {journal} {\bibinfo  {journal} {Physical Review Research}\ }\textbf {\bibinfo {volume} {7}},\ \bibinfo {pages} {043240} (\bibinfo {year} {2025})}\BibitemShut {NoStop}%
\bibitem [{\citenamefont {Häfner}\ and\ \citenamefont {Müller}(2024)}]{Haefner.2024}%
  \BibitemOpen
  \bibfield  {author} {\bibinfo {author} {\bibfnamefont {Gregor}\ \bibnamefont {Häfner}}\ and\ \bibinfo {author} {\bibfnamefont {Marcus}\ \bibnamefont {Müller}},\ }\bibfield  {title} {\enquote {\bibinfo {title} {{Reaction-Driven Diffusiophoresis of Liquid Condensates: Potential Mechanisms for Intracellular Organization}},}\ }\href {\doibase 10.1021/acsnano.3c12842} {\bibfield  {journal} {\bibinfo  {journal} {ACS Nano}\ }\textbf {\bibinfo {volume} {18}},\ \bibinfo {pages} {16530--16544} (\bibinfo {year} {2024})}\BibitemShut {NoStop}%
\bibitem [{\citenamefont {Demarchi}\ \emph {et~al.}(2023)\citenamefont {Demarchi}, \citenamefont {Goychuk}, \citenamefont {Maryshev},\ and\ \citenamefont {Frey}}]{Demarchi.2023}%
  \BibitemOpen
  \bibfield  {author} {\bibinfo {author} {\bibfnamefont {Leonardo}\ \bibnamefont {Demarchi}}, \bibinfo {author} {\bibfnamefont {Andriy}\ \bibnamefont {Goychuk}}, \bibinfo {author} {\bibfnamefont {Ivan}\ \bibnamefont {Maryshev}}, \ and\ \bibinfo {author} {\bibfnamefont {Erwin}\ \bibnamefont {Frey}},\ }\bibfield  {title} {\enquote {\bibinfo {title} {{Enzyme-Enriched Condensates Show Self-Propulsion, Positioning, and Coexistence}},}\ }\href {\doibase 10.1103/physrevlett.130.128401} {\bibfield  {journal} {\bibinfo  {journal} {Physical Review Letters}\ }\textbf {\bibinfo {volume} {130}},\ \bibinfo {pages} {128401} (\bibinfo {year} {2023})},\ \Eprint {http://arxiv.org/abs/2301.00392} {2301.00392} \BibitemShut {NoStop}%
\bibitem [{\citenamefont {Jambon-Puillet}\ \emph {et~al.}(2024)\citenamefont {Jambon-Puillet}, \citenamefont {Testa}, \citenamefont {Lorenz}, \citenamefont {Style}, \citenamefont {Rebane},\ and\ \citenamefont {Dufresne}}]{Jambon-Puillet.2024}%
  \BibitemOpen
  \bibfield  {author} {\bibinfo {author} {\bibfnamefont {Etienne}\ \bibnamefont {Jambon-Puillet}}, \bibinfo {author} {\bibfnamefont {Andrea}\ \bibnamefont {Testa}}, \bibinfo {author} {\bibfnamefont {Charlotta}\ \bibnamefont {Lorenz}}, \bibinfo {author} {\bibfnamefont {Robert~W.}\ \bibnamefont {Style}}, \bibinfo {author} {\bibfnamefont {Aleksander~A.}\ \bibnamefont {Rebane}}, \ and\ \bibinfo {author} {\bibfnamefont {Eric~R.}\ \bibnamefont {Dufresne}},\ }\bibfield  {title} {\enquote {\bibinfo {title} {{Phase-separated droplets swim to their dissolution}},}\ }\href {\doibase 10.1038/s41467-024-47889-y} {\bibfield  {journal} {\bibinfo  {journal} {Nature Communications}\ }\textbf {\bibinfo {volume} {15}},\ \bibinfo {pages} {3919} (\bibinfo {year} {2024})}\BibitemShut {NoStop}%
\bibitem [{\citenamefont {Agudo-Canalejo}\ and\ \citenamefont {Golestanian}(2019)}]{Agudo-Canalejo.2019}%
  \BibitemOpen
  \bibfield  {author} {\bibinfo {author} {\bibfnamefont {Jaime}\ \bibnamefont {Agudo-Canalejo}}\ and\ \bibinfo {author} {\bibfnamefont {Ramin}\ \bibnamefont {Golestanian}},\ }\bibfield  {title} {\enquote {\bibinfo {title} {{Active Phase Separation in Mixtures of Chemically Interacting Particles}},}\ }\href {\doibase 10.1103/physrevlett.123.018101} {\bibfield  {journal} {\bibinfo  {journal} {Physical Review Letters}\ }\textbf {\bibinfo {volume} {123}},\ \bibinfo {pages} {018101} (\bibinfo {year} {2019})},\ \Eprint {http://arxiv.org/abs/1901.09022} {1901.09022} \BibitemShut {NoStop}%
\bibitem [{\citenamefont {Saha}\ \emph {et~al.}(2020)\citenamefont {Saha}, \citenamefont {Agudo-Canalejo},\ and\ \citenamefont {Golestanian}}]{Saha.2020}%
  \BibitemOpen
  \bibfield  {author} {\bibinfo {author} {\bibfnamefont {Suropriya}\ \bibnamefont {Saha}}, \bibinfo {author} {\bibfnamefont {Jaime}\ \bibnamefont {Agudo-Canalejo}}, \ and\ \bibinfo {author} {\bibfnamefont {Ramin}\ \bibnamefont {Golestanian}},\ }\bibfield  {title} {\enquote {\bibinfo {title} {{Scalar Active Mixtures: The Nonreciprocal Cahn-Hilliard Model}},}\ }\href {\doibase 10.1103/physrevx.10.041009} {\bibfield  {journal} {\bibinfo  {journal} {Physical Review X}\ }\textbf {\bibinfo {volume} {10}},\ \bibinfo {pages} {041009} (\bibinfo {year} {2020})},\ \Eprint {http://arxiv.org/abs/2005.07101} {2005.07101} \BibitemShut {NoStop}%
\bibitem [{\citenamefont {Bauermann}\ \emph {et~al.}(2023)\citenamefont {Bauermann}, \citenamefont {Bartolucci}, \citenamefont {Boekhoven}, \citenamefont {Weber},\ and\ \citenamefont {Jülicher}}]{Bauermann.2023}%
  \BibitemOpen
  \bibfield  {author} {\bibinfo {author} {\bibfnamefont {Jonathan}\ \bibnamefont {Bauermann}}, \bibinfo {author} {\bibfnamefont {Giacomo}\ \bibnamefont {Bartolucci}}, \bibinfo {author} {\bibfnamefont {Job}\ \bibnamefont {Boekhoven}}, \bibinfo {author} {\bibfnamefont {Christoph~A.}\ \bibnamefont {Weber}}, \ and\ \bibinfo {author} {\bibfnamefont {Frank}\ \bibnamefont {Jülicher}},\ }\bibfield  {title} {\enquote {\bibinfo {title} {{Formation of liquid shells in active droplet systems}},}\ }\href {\doibase 10.1103/physrevresearch.5.043246} {\bibfield  {journal} {\bibinfo  {journal} {Physical Review Research}\ }\textbf {\bibinfo {volume} {5}},\ \bibinfo {pages} {043246} (\bibinfo {year} {2023})}\BibitemShut {NoStop}%
\bibitem [{\citenamefont {Boeynaems}\ \emph {et~al.}(2018)\citenamefont {Boeynaems}, \citenamefont {Alberti}, \citenamefont {Fawzi}, \citenamefont {Mittag}, \citenamefont {Polymenidou}, \citenamefont {Rousseau}, \citenamefont {Schymkowitz}, \citenamefont {Shorter}, \citenamefont {Wolozin}, \citenamefont {Bosch}, \citenamefont {Tompa},\ and\ \citenamefont {Fuxreiter}}]{Boeynaems.2018}%
  \BibitemOpen
  \bibfield  {author} {\bibinfo {author} {\bibfnamefont {Steven}\ \bibnamefont {Boeynaems}}, \bibinfo {author} {\bibfnamefont {Simon}\ \bibnamefont {Alberti}}, \bibinfo {author} {\bibfnamefont {Nicolas~L.}\ \bibnamefont {Fawzi}}, \bibinfo {author} {\bibfnamefont {Tanja}\ \bibnamefont {Mittag}}, \bibinfo {author} {\bibfnamefont {Magdalini}\ \bibnamefont {Polymenidou}}, \bibinfo {author} {\bibfnamefont {Frederic}\ \bibnamefont {Rousseau}}, \bibinfo {author} {\bibfnamefont {Joost}\ \bibnamefont {Schymkowitz}}, \bibinfo {author} {\bibfnamefont {James}\ \bibnamefont {Shorter}}, \bibinfo {author} {\bibfnamefont {Benjamin}\ \bibnamefont {Wolozin}}, \bibinfo {author} {\bibfnamefont {Ludo Van~Den}\ \bibnamefont {Bosch}}, \bibinfo {author} {\bibfnamefont {Peter}\ \bibnamefont {Tompa}}, \ and\ \bibinfo {author} {\bibfnamefont {Monika}\ \bibnamefont {Fuxreiter}},\ }\bibfield  {title} {\enquote {\bibinfo {title} {{Protein Phase Separation: A New Phase in Cell Biology}},}\ }\href {\doibase 10.1016/j.tcb.2018.02.004}
  {\bibfield  {journal} {\bibinfo  {journal} {Trends in Cell Biology}\ }\textbf {\bibinfo {volume} {28}},\ \bibinfo {pages} {420--435} (\bibinfo {year} {2018})}\BibitemShut {NoStop}%
\bibitem [{\citenamefont {Fritsch}\ \emph {et~al.}(2021)\citenamefont {Fritsch}, \citenamefont {Diaz-Delgadillo}, \citenamefont {Adame-Arana}, \citenamefont {Hoege}, \citenamefont {Mittasch}, \citenamefont {Kreysing}, \citenamefont {Leaver}, \citenamefont {Hyman}, \citenamefont {Jülicher},\ and\ \citenamefont {Weber}}]{Fritsch.2021}%
  \BibitemOpen
  \bibfield  {author} {\bibinfo {author} {\bibfnamefont {Anatol~W.}\ \bibnamefont {Fritsch}}, \bibinfo {author} {\bibfnamefont {Andrés~F.}\ \bibnamefont {Diaz-Delgadillo}}, \bibinfo {author} {\bibfnamefont {Omar}\ \bibnamefont {Adame-Arana}}, \bibinfo {author} {\bibfnamefont {Carsten}\ \bibnamefont {Hoege}}, \bibinfo {author} {\bibfnamefont {Matthäus}\ \bibnamefont {Mittasch}}, \bibinfo {author} {\bibfnamefont {Moritz}\ \bibnamefont {Kreysing}}, \bibinfo {author} {\bibfnamefont {Mark}\ \bibnamefont {Leaver}}, \bibinfo {author} {\bibfnamefont {Anthony~A.}\ \bibnamefont {Hyman}}, \bibinfo {author} {\bibfnamefont {Frank}\ \bibnamefont {Jülicher}}, \ and\ \bibinfo {author} {\bibfnamefont {Christoph~A.}\ \bibnamefont {Weber}},\ }\bibfield  {title} {\enquote {\bibinfo {title} {{Local thermodynamics govern formation and dissolution of Caenorhabditis elegans P granule condensates}},}\ }\href {\doibase 10.1073/pnas.2102772118} {\bibfield  {journal} {\bibinfo  {journal} {Proceedings of the National Academy of
  Sciences}\ }\textbf {\bibinfo {volume} {118}},\ \bibinfo {pages} {e2102772118} (\bibinfo {year} {2021})}\BibitemShut {NoStop}%
\bibitem [{\citenamefont {Alberti}\ \emph {et~al.}(2019)\citenamefont {Alberti}, \citenamefont {Gladfelter},\ and\ \citenamefont {Mittag}}]{Alberti.2019l6i}%
  \BibitemOpen
  \bibfield  {author} {\bibinfo {author} {\bibfnamefont {Simon}\ \bibnamefont {Alberti}}, \bibinfo {author} {\bibfnamefont {Amy}\ \bibnamefont {Gladfelter}}, \ and\ \bibinfo {author} {\bibfnamefont {Tanja}\ \bibnamefont {Mittag}},\ }\bibfield  {title} {\enquote {\bibinfo {title} {{Considerations and Challenges in Studying Liquid-Liquid Phase Separation and Biomolecular Condensates}},}\ }\href {\doibase 10.1016/j.cell.2018.12.035} {\bibfield  {journal} {\bibinfo  {journal} {Cell}\ }\textbf {\bibinfo {volume} {176}},\ \bibinfo {pages} {419--434} (\bibinfo {year} {2019})}\BibitemShut {NoStop}%
\bibitem [{\citenamefont {Hondele}\ \emph {et~al.}(2020)\citenamefont {Hondele}, \citenamefont {Heinrich}, \citenamefont {Rios},\ and\ \citenamefont {Weis}}]{Hondele.2020}%
  \BibitemOpen
  \bibfield  {author} {\bibinfo {author} {\bibfnamefont {Maria}\ \bibnamefont {Hondele}}, \bibinfo {author} {\bibfnamefont {Stephanie}\ \bibnamefont {Heinrich}}, \bibinfo {author} {\bibfnamefont {Paolo De~Los}\ \bibnamefont {Rios}}, \ and\ \bibinfo {author} {\bibfnamefont {Karsten}\ \bibnamefont {Weis}},\ }\bibfield  {title} {\enquote {\bibinfo {title} {{Membraneless organelles: phasing out of equilibrium}},}\ }\href {\doibase 10.1042/etls20190190} {\bibfield  {journal} {\bibinfo  {journal} {Emerging Topics in Life Sciences}\ }\textbf {\bibinfo {volume} {4}},\ \bibinfo {pages} {343--354} (\bibinfo {year} {2020})}\BibitemShut {NoStop}%
\bibitem [{\citenamefont {Rai}\ \emph {et~al.}(2018)\citenamefont {Rai}, \citenamefont {Chen}, \citenamefont {Selbach},\ and\ \citenamefont {Pelkmans}}]{Rai.2018}%
  \BibitemOpen
  \bibfield  {author} {\bibinfo {author} {\bibfnamefont {Arpan~Kumar}\ \bibnamefont {Rai}}, \bibinfo {author} {\bibfnamefont {Jia-Xuan}\ \bibnamefont {Chen}}, \bibinfo {author} {\bibfnamefont {Matthias}\ \bibnamefont {Selbach}}, \ and\ \bibinfo {author} {\bibfnamefont {Lucas}\ \bibnamefont {Pelkmans}},\ }\bibfield  {title} {\enquote {\bibinfo {title} {{Kinase-controlled phase transition of membraneless organelles in mitosis}},}\ }\href {\doibase 10.1038/s41586-018-0279-8} {\bibfield  {journal} {\bibinfo  {journal} {Nature}\ }\textbf {\bibinfo {volume} {559}},\ \bibinfo {pages} {211--216} (\bibinfo {year} {2018})}\BibitemShut {NoStop}%
\bibitem [{\citenamefont {Hofweber}\ and\ \citenamefont {Dormann}(2019)}]{Hofweber.2019}%
  \BibitemOpen
  \bibfield  {author} {\bibinfo {author} {\bibfnamefont {Mario}\ \bibnamefont {Hofweber}}\ and\ \bibinfo {author} {\bibfnamefont {Dorothee}\ \bibnamefont {Dormann}},\ }\bibfield  {title} {\enquote {\bibinfo {title} {{Friend or foe—Post-translational modifications as regulators of phase separation and RNP granule dynamics}},}\ }\href {\doibase 10.1074/jbc.tm118.001189} {\bibfield  {journal} {\bibinfo  {journal} {Journal of Biological Chemistry}\ }\textbf {\bibinfo {volume} {294}},\ \bibinfo {pages} {7137--7150} (\bibinfo {year} {2019})}\BibitemShut {NoStop}%
\bibitem [{\citenamefont {Owen}\ and\ \citenamefont {Shewmaker}(2019)}]{Owen.2019}%
  \BibitemOpen
  \bibfield  {author} {\bibinfo {author} {\bibfnamefont {Izzy}\ \bibnamefont {Owen}}\ and\ \bibinfo {author} {\bibfnamefont {Frank}\ \bibnamefont {Shewmaker}},\ }\bibfield  {title} {\enquote {\bibinfo {title} {{The Role of Post-Translational Modifications in the Phase Transitions of Intrinsically Disordered Proteins}},}\ }\href {\doibase 10.3390/ijms20215501} {\bibfield  {journal} {\bibinfo  {journal} {International Journal of Molecular Sciences}\ }\textbf {\bibinfo {volume} {20}},\ \bibinfo {pages} {5501} (\bibinfo {year} {2019})}\BibitemShut {NoStop}%
\bibitem [{\citenamefont {Snead}\ and\ \citenamefont {Gladfelter}(2019)}]{Snead.2019}%
  \BibitemOpen
  \bibfield  {author} {\bibinfo {author} {\bibfnamefont {Wilton~T.}\ \bibnamefont {Snead}}\ and\ \bibinfo {author} {\bibfnamefont {Amy~S.}\ \bibnamefont {Gladfelter}},\ }\bibfield  {title} {\enquote {\bibinfo {title} {{The Control Centers of Biomolecular Phase Separation: How Membrane Surfaces, PTMs, and Active Processes Regulate Condensation}},}\ }\href {\doibase 10.1016/j.molcel.2019.09.016} {\bibfield  {journal} {\bibinfo  {journal} {Molecular Cell}\ }\textbf {\bibinfo {volume} {76}},\ \bibinfo {pages} {295--305} (\bibinfo {year} {2019})}\BibitemShut {NoStop}%
\bibitem [{\citenamefont {Söding}\ \emph {et~al.}(2020)\citenamefont {Söding}, \citenamefont {Zwicker}, \citenamefont {Sohrabi-Jahromi}, \citenamefont {Boehning},\ and\ \citenamefont {Kirschbaum}}]{Soeding.2020}%
  \BibitemOpen
  \bibfield  {author} {\bibinfo {author} {\bibfnamefont {Johannes}\ \bibnamefont {Söding}}, \bibinfo {author} {\bibfnamefont {David}\ \bibnamefont {Zwicker}}, \bibinfo {author} {\bibfnamefont {Salma}\ \bibnamefont {Sohrabi-Jahromi}}, \bibinfo {author} {\bibfnamefont {Marc}\ \bibnamefont {Boehning}}, \ and\ \bibinfo {author} {\bibfnamefont {Jan}\ \bibnamefont {Kirschbaum}},\ }\bibfield  {title} {\enquote {\bibinfo {title} {{Mechanisms for Active Regulation of Biomolecular Condensates}},}\ }\href {\doibase 10.1016/j.tcb.2019.10.006} {\bibfield  {journal} {\bibinfo  {journal} {Trends in Cell Biology}\ }\textbf {\bibinfo {volume} {30}},\ \bibinfo {pages} {4--14} (\bibinfo {year} {2020})}\BibitemShut {NoStop}%
\bibitem [{\citenamefont {Wurtz}\ and\ \citenamefont {Lee}(2018{\natexlab{b}})}]{Wurtz.2018l4k}%
  \BibitemOpen
  \bibfield  {author} {\bibinfo {author} {\bibfnamefont {Jean~David}\ \bibnamefont {Wurtz}}\ and\ \bibinfo {author} {\bibfnamefont {Chiu~Fan}\ \bibnamefont {Lee}},\ }\bibfield  {title} {\enquote {\bibinfo {title} {{Stress granule formation via ATP depletion-triggered phase separation}},}\ }\href {\doibase 10.1088/1367-2630/aab549} {\bibfield  {journal} {\bibinfo  {journal} {New Journal of Physics}\ }\textbf {\bibinfo {volume} {20}},\ \bibinfo {pages} {045008} (\bibinfo {year} {2018}{\natexlab{b}})},\ \Eprint {http://arxiv.org/abs/1708.05697} {1708.05697} \BibitemShut {NoStop}%
\bibitem [{\citenamefont {Fu}\ \emph {et~al.}(2012)\citenamefont {Fu}, \citenamefont {Tang}, \citenamefont {Liu}, \citenamefont {Huang}, \citenamefont {Hwa},\ and\ \citenamefont {Lenz}}]{Fu.2012}%
  \BibitemOpen
  \bibfield  {author} {\bibinfo {author} {\bibfnamefont {Xiongfei}\ \bibnamefont {Fu}}, \bibinfo {author} {\bibfnamefont {Lei-Han}\ \bibnamefont {Tang}}, \bibinfo {author} {\bibfnamefont {Chenli}\ \bibnamefont {Liu}}, \bibinfo {author} {\bibfnamefont {Jian-Dong}\ \bibnamefont {Huang}}, \bibinfo {author} {\bibfnamefont {Terence}\ \bibnamefont {Hwa}}, \ and\ \bibinfo {author} {\bibfnamefont {Peter}\ \bibnamefont {Lenz}},\ }\bibfield  {title} {\enquote {\bibinfo {title} {{Stripe Formation in Bacterial Systems with Density-Suppressed Motility}},}\ }\href {\doibase 10.1103/physrevlett.108.198102} {\bibfield  {journal} {\bibinfo  {journal} {Physical Review Letters}\ }\textbf {\bibinfo {volume} {108}},\ \bibinfo {pages} {198102} (\bibinfo {year} {2012})},\ \Eprint {http://arxiv.org/abs/1204.0615} {1204.0615} \BibitemShut {NoStop}%
\bibitem [{\citenamefont {Keller}\ and\ \citenamefont {Segel}(1970)}]{Keller.1970}%
  \BibitemOpen
  \bibfield  {author} {\bibinfo {author} {\bibfnamefont {Evelyn~F.}\ \bibnamefont {Keller}}\ and\ \bibinfo {author} {\bibfnamefont {Lee~A.}\ \bibnamefont {Segel}},\ }\bibfield  {title} {\enquote {\bibinfo {title} {{Initiation of slime mold aggregation viewed as an instability}},}\ }\href {\doibase 10.1016/0022-5193(70)90092-5} {\bibfield  {journal} {\bibinfo  {journal} {Journal of Theoretical Biology}\ }\textbf {\bibinfo {volume} {26}},\ \bibinfo {pages} {399--415} (\bibinfo {year} {1970})}\BibitemShut {NoStop}%
\bibitem [{\citenamefont {Raßhofer}\ and\ \citenamefont {Frey}(2026)}]{prl}%
  \BibitemOpen
  \bibfield  {author} {\bibinfo {author} {\bibfnamefont {Florian}\ \bibnamefont {Raßhofer}}\ and\ \bibinfo {author} {\bibfnamefont {Erwin}\ \bibnamefont {Frey}},\ }\bibfield  {title} {\enquote {\bibinfo {title} {{Local composition controls pattern formation in conserved active emulsions}},}\ }\href@noop {} {\  (\bibinfo {year} {2026})},\ \bibinfo {note} {(submitted)}\BibitemShut {NoStop}%
\bibitem [{\citenamefont {Berry}\ \emph {et~al.}(2018)\citenamefont {Berry}, \citenamefont {Brangwynne},\ and\ \citenamefont {Haataja}}]{Berry.2018}%
  \BibitemOpen
  \bibfield  {author} {\bibinfo {author} {\bibfnamefont {Joel}\ \bibnamefont {Berry}}, \bibinfo {author} {\bibfnamefont {Clifford~P}\ \bibnamefont {Brangwynne}}, \ and\ \bibinfo {author} {\bibfnamefont {Mikko}\ \bibnamefont {Haataja}},\ }\bibfield  {title} {\enquote {\bibinfo {title} {{Physical principles of intracellular organization via active and passive phase transitions}},}\ }\href {\doibase 10.1088/1361-6633/aaa61e} {\bibfield  {journal} {\bibinfo  {journal} {Reports on Progress in Physics}\ }\textbf {\bibinfo {volume} {81}},\ \bibinfo {pages} {046601} (\bibinfo {year} {2018})}\BibitemShut {NoStop}%
\bibitem [{\citenamefont {Mao}\ \emph {et~al.}(2018)\citenamefont {Mao}, \citenamefont {Kuldinow}, \citenamefont {Haataja},\ and\ \citenamefont {Košmrlj}}]{Mao.2018}%
  \BibitemOpen
  \bibfield  {author} {\bibinfo {author} {\bibfnamefont {Sheng}\ \bibnamefont {Mao}}, \bibinfo {author} {\bibfnamefont {Derek}\ \bibnamefont {Kuldinow}}, \bibinfo {author} {\bibfnamefont {Mikko~P.}\ \bibnamefont {Haataja}}, \ and\ \bibinfo {author} {\bibfnamefont {Andrej}\ \bibnamefont {Košmrlj}},\ }\bibfield  {title} {\enquote {\bibinfo {title} {{Phase behavior and morphology of multicomponent liquid mixtures}},}\ }\href {\doibase 10.1039/c8sm02045k} {\bibfield  {journal} {\bibinfo  {journal} {Soft Matter}\ }\textbf {\bibinfo {volume} {15}},\ \bibinfo {pages} {1297--1311} (\bibinfo {year} {2018})},\ \Eprint {http://arxiv.org/abs/1810.03689} {1810.03689} \BibitemShut {NoStop}%
\bibitem [{\citenamefont {Cahn}\ and\ \citenamefont {Hilliard}(1958)}]{Cahn.1958}%
  \BibitemOpen
  \bibfield  {author} {\bibinfo {author} {\bibfnamefont {John~W}\ \bibnamefont {Cahn}}\ and\ \bibinfo {author} {\bibfnamefont {John~E}\ \bibnamefont {Hilliard}},\ }\bibfield  {title} {\enquote {\bibinfo {title} {{Free Energy of a Nonuniform System. I. Interfacial Free Energy}},}\ }\href {\doibase 10.1063/1.1744102} {\bibfield  {journal} {\bibinfo  {journal} {The Journal of Chemical Physics}\ }\textbf {\bibinfo {volume} {28}},\ \bibinfo {pages} {258--267} (\bibinfo {year} {1958})}\BibitemShut {NoStop}%
\bibitem [{\citenamefont {Rubinstein}\ and\ \citenamefont {Colby}(2003)}]{Rubinstein.2003}%
  \BibitemOpen
  \bibfield  {author} {\bibinfo {author} {\bibfnamefont {Michael}\ \bibnamefont {Rubinstein}}\ and\ \bibinfo {author} {\bibfnamefont {Ralph~H}\ \bibnamefont {Colby}},\ }\href {\doibase 10.1093/oso/9780198520597.001.0001} {\emph {\bibinfo {title} {{Polymer Physics}}}}\ (\bibinfo  {publisher} {Oxford University Press},\ \bibinfo {year} {2003})\BibitemShut {NoStop}%
\bibitem [{\citenamefont {Groot}\ and\ \citenamefont {Mazur}(1984)}]{Groot.1984}%
  \BibitemOpen
  \bibfield  {author} {\bibinfo {author} {\bibfnamefont {S.R.~de}\ \bibnamefont {Groot}}\ and\ \bibinfo {author} {\bibfnamefont {P.}~\bibnamefont {Mazur}},\ }\href {https://books.google.de/books?id=HFAIv43rlGkC} {\emph {\bibinfo {title} {{Non-equilibrium Thermodynamics}}}},\ Dover Books on Physics\ (\bibinfo  {publisher} {Dover Publications},\ \bibinfo {year} {1984})\BibitemShut {NoStop}%
\bibitem [{\citenamefont {Onuki}(2002)}]{Onuki.2002}%
  \BibitemOpen
  \bibfield  {author} {\bibinfo {author} {\bibfnamefont {Akira}\ \bibnamefont {Onuki}},\ }\href {\doibase 10.1017/cbo9780511534874} {\emph {\bibinfo {title} {{Phase Transition Dynamics}}}}\ (\bibinfo  {publisher} {Cambridge University Press},\ \bibinfo {address} {Oxford},\ \bibinfo {year} {2002})\BibitemShut {NoStop}%
\bibitem [{\citenamefont {Onsager}(1930)}]{Onsager.1930}%
  \BibitemOpen
  \bibfield  {author} {\bibinfo {author} {\bibfnamefont {Lars}\ \bibnamefont {Onsager}},\ }\bibfield  {title} {\enquote {\bibinfo {title} {{Reciprocal Relations in Irreversible Processes. I}},}\ }\href {\doibase 10.1103/physrev.37.405} {\bibfield  {journal} {\bibinfo  {journal} {Physical Review}\ }\textbf {\bibinfo {volume} {37}},\ \bibinfo {pages} {405--426} (\bibinfo {year} {1930})}\BibitemShut {NoStop}%
\bibitem [{\citenamefont {Onsager}(1931)}]{Onsager.1931}%
  \BibitemOpen
  \bibfield  {author} {\bibinfo {author} {\bibfnamefont {Lars}\ \bibnamefont {Onsager}},\ }\bibfield  {title} {\enquote {\bibinfo {title} {{Reciprocal Relations in Irreversible Processes. II.}}}\ }\href {\doibase 10.1103/physrev.38.2265} {\bibfield  {journal} {\bibinfo  {journal} {Physical Review}\ }\textbf {\bibinfo {volume} {38}},\ \bibinfo {pages} {2265--2279} (\bibinfo {year} {1931})}\BibitemShut {NoStop}%
\bibitem [{\citenamefont {Bo}\ \emph {et~al.}(2021)\citenamefont {Bo}, \citenamefont {Hubatsch}, \citenamefont {Bauermann}, \citenamefont {Weber},\ and\ \citenamefont {Jülicher}}]{Bo.2021}%
  \BibitemOpen
  \bibfield  {author} {\bibinfo {author} {\bibfnamefont {Stefano}\ \bibnamefont {Bo}}, \bibinfo {author} {\bibfnamefont {Lars}\ \bibnamefont {Hubatsch}}, \bibinfo {author} {\bibfnamefont {Jonathan}\ \bibnamefont {Bauermann}}, \bibinfo {author} {\bibfnamefont {Christoph~A.}\ \bibnamefont {Weber}}, \ and\ \bibinfo {author} {\bibfnamefont {Frank}\ \bibnamefont {Jülicher}},\ }\bibfield  {title} {\enquote {\bibinfo {title} {{Stochastic dynamics of single molecules across phase boundaries}},}\ }\href {\doibase 10.1103/physrevresearch.3.043150} {\bibfield  {journal} {\bibinfo  {journal} {Physical Review Research}\ }\textbf {\bibinfo {volume} {3}},\ \bibinfo {pages} {043150} (\bibinfo {year} {2021})}\BibitemShut {NoStop}%
\bibitem [{\citenamefont {Cotton}\ \emph {et~al.}(2022)\citenamefont {Cotton}, \citenamefont {Golestanian},\ and\ \citenamefont {Agudo-Canalejo}}]{Cotton.2022}%
  \BibitemOpen
  \bibfield  {author} {\bibinfo {author} {\bibfnamefont {Matthew~W.}\ \bibnamefont {Cotton}}, \bibinfo {author} {\bibfnamefont {Ramin}\ \bibnamefont {Golestanian}}, \ and\ \bibinfo {author} {\bibfnamefont {Jaime}\ \bibnamefont {Agudo-Canalejo}},\ }\bibfield  {title} {\enquote {\bibinfo {title} {{Catalysis-Induced Phase Separation and Autoregulation of Enzymatic Activity}},}\ }\href {\doibase 10.1103/physrevlett.129.158101} {\bibfield  {journal} {\bibinfo  {journal} {Physical Review Letters}\ }\textbf {\bibinfo {volume} {129}},\ \bibinfo {pages} {158101} (\bibinfo {year} {2022})},\ \Eprint {http://arxiv.org/abs/2205.12306} {2205.12306} \BibitemShut {NoStop}%
\bibitem [{\citenamefont {Kondepudi}\ and\ \citenamefont {Prigogine}(2023)}]{Kondepudi.2014}%
  \BibitemOpen
  \bibfield  {author} {\bibinfo {author} {\bibfnamefont {Dilip}\ \bibnamefont {Kondepudi}}\ and\ \bibinfo {author} {\bibfnamefont {Ilya}\ \bibnamefont {Prigogine}},\ }\bibfield  {title} {\enquote {\bibinfo {title} {{Modern Thermodynamics}},}\ }\href {\doibase 10.1002/9781118698723} {\bibfield  {journal} {\bibinfo  {journal} {Wiley}\ } (\bibinfo {year} {2023}),\ 10.1002/9781118698723}\BibitemShut {NoStop}%
\bibitem [{\citenamefont {Bray}(2002)}]{Bray.2002}%
  \BibitemOpen
  \bibfield  {author} {\bibinfo {author} {\bibfnamefont {A.~J.}\ \bibnamefont {Bray}},\ }\bibfield  {title} {\enquote {\bibinfo {title} {{Theory of phase-ordering kinetics}},}\ }\href {\doibase 10.1080/00018730110117433} {\bibfield  {journal} {\bibinfo  {journal} {Advances in Physics}\ }\textbf {\bibinfo {volume} {51}},\ \bibinfo {pages} {481--587} (\bibinfo {year} {2002})}\BibitemShut {NoStop}%
\bibitem [{\citenamefont {Cross}\ and\ \citenamefont {Hohenberg}(1993)}]{Cross.1993}%
  \BibitemOpen
  \bibfield  {author} {\bibinfo {author} {\bibfnamefont {M.~C.}\ \bibnamefont {Cross}}\ and\ \bibinfo {author} {\bibfnamefont {P.~C.}\ \bibnamefont {Hohenberg}},\ }\bibfield  {title} {\enquote {\bibinfo {title} {{Pattern formation outside of equilibrium}},}\ }\href {\doibase 10.1103/revmodphys.65.851} {\bibfield  {journal} {\bibinfo  {journal} {Reviews of Modern Physics}\ }\textbf {\bibinfo {volume} {65}},\ \bibinfo {pages} {851--1112} (\bibinfo {year} {1993})}\BibitemShut {NoStop}%
\bibitem [{\citenamefont {Eyre}(1993)}]{Eyre.1993}%
  \BibitemOpen
  \bibfield  {author} {\bibinfo {author} {\bibfnamefont {David~J.}\ \bibnamefont {Eyre}},\ }\bibfield  {title} {\enquote {\bibinfo {title} {{Systems of Cahn-Hilliard Equations}},}\ }\href {\doibase 10.1137/0153078} {\bibfield  {journal} {\bibinfo  {journal} {SIAM Journal on Applied Mathematics}\ }\textbf {\bibinfo {volume} {53}},\ \bibinfo {pages} {1686--1712} (\bibinfo {year} {1993})}\BibitemShut {NoStop}%
\bibitem [{\citenamefont {Garcke}\ and\ \citenamefont {Novick-Cohen}(2000)}]{Garcke.2000}%
  \BibitemOpen
  \bibfield  {author} {\bibinfo {author} {\bibfnamefont {Harald}\ \bibnamefont {Garcke}}\ and\ \bibinfo {author} {\bibfnamefont {Amy}\ \bibnamefont {Novick-Cohen}},\ }\bibfield  {title} {\enquote {\bibinfo {title} {{A singular limit for a system of degenerate Cahn-Hilliard equations}},}\ }\href {\doibase 10.57262/ade/1356651336} {\bibfield  {journal} {\bibinfo  {journal} {Advances in Differential Equations}\ }\textbf {\bibinfo {volume} {5}} (\bibinfo {year} {2000}),\ 10.57262/ade/1356651336}\BibitemShut {NoStop}%
\bibitem [{\citenamefont {Osmanović}\ and\ \citenamefont {Franco}(2023)}]{Osmanovic.2023}%
  \BibitemOpen
  \bibfield  {author} {\bibinfo {author} {\bibfnamefont {Dino}\ \bibnamefont {Osmanović}}\ and\ \bibinfo {author} {\bibfnamefont {Elisa}\ \bibnamefont {Franco}},\ }\bibfield  {title} {\enquote {\bibinfo {title} {{Chemical reaction motifs driving non-equilibrium behaviours in phase separating materials}},}\ }\href {\doibase 10.1098/rsif.2023.0117} {\bibfield  {journal} {\bibinfo  {journal} {Journal of the Royal Society Interface}\ }\textbf {\bibinfo {volume} {20}},\ \bibinfo {pages} {20230117} (\bibinfo {year} {2023})},\ \Eprint {http://arxiv.org/abs/2207.10135} {2207.10135} \BibitemShut {NoStop}%
\bibitem [{\citenamefont {Li}\ and\ \citenamefont {Cates}(2020)}]{Li.2020}%
  \BibitemOpen
  \bibfield  {author} {\bibinfo {author} {\bibfnamefont {Yuting~I}\ \bibnamefont {Li}}\ and\ \bibinfo {author} {\bibfnamefont {Michael~E}\ \bibnamefont {Cates}},\ }\bibfield  {title} {\enquote {\bibinfo {title} {{Non-equilibrium phase separation with reactions: a canonical model and its behaviour}},}\ }\href {\doibase 10.1088/1742-5468/ab7e2d} {\bibfield  {journal} {\bibinfo  {journal} {Journal of Statistical Mechanics: Theory and Experiment}\ }\textbf {\bibinfo {volume} {2020}},\ \bibinfo {pages} {053206} (\bibinfo {year} {2020})},\ \Eprint {http://arxiv.org/abs/2001.02563} {2001.02563} \BibitemShut {NoStop}%
\bibitem [{\citenamefont {Chen}\ \emph {et~al.}(2024)\citenamefont {Chen}, \citenamefont {Seyboldt}, \citenamefont {Sommer}, \citenamefont {Jülicher},\ and\ \citenamefont {Harmon}}]{Chen.2024}%
  \BibitemOpen
  \bibfield  {author} {\bibinfo {author} {\bibfnamefont {Xi}~\bibnamefont {Chen}}, \bibinfo {author} {\bibfnamefont {Rabea}\ \bibnamefont {Seyboldt}}, \bibinfo {author} {\bibfnamefont {Jens-Uwe}\ \bibnamefont {Sommer}}, \bibinfo {author} {\bibfnamefont {Frank}\ \bibnamefont {Jülicher}}, \ and\ \bibinfo {author} {\bibfnamefont {Tyler}\ \bibnamefont {Harmon}},\ }\bibfield  {title} {\enquote {\bibinfo {title} {{Droplet Differentiation by a Chemical Switch}},}\ }\href {\doibase 10.1103/physrevlett.133.028402} {\bibfield  {journal} {\bibinfo  {journal} {Physical Review Letters}\ }\textbf {\bibinfo {volume} {133}},\ \bibinfo {pages} {028402} (\bibinfo {year} {2024})}\BibitemShut {NoStop}%
\bibitem [{\citenamefont {Bauermann}\ \emph {et~al.}(2025{\natexlab{a}})\citenamefont {Bauermann}, \citenamefont {Bartolucci}, \citenamefont {Weber},\ and\ \citenamefont {Jülicher}}]{Bauermann.2025}%
  \BibitemOpen
  \bibfield  {author} {\bibinfo {author} {\bibfnamefont {Jonathan}\ \bibnamefont {Bauermann}}, \bibinfo {author} {\bibfnamefont {Giacomo}\ \bibnamefont {Bartolucci}}, \bibinfo {author} {\bibfnamefont {Christoph~A.}\ \bibnamefont {Weber}}, \ and\ \bibinfo {author} {\bibfnamefont {Frank}\ \bibnamefont {Jülicher}},\ }\bibfield  {title} {\enquote {\bibinfo {title} {{Theory of Reversed Ripening in Active Phase Separating Systems}},}\ }\href {\doibase 10.1103/f5x9-wp3g} {\bibfield  {journal} {\bibinfo  {journal} {Physical Review Letters}\ }\textbf {\bibinfo {volume} {135}},\ \bibinfo {pages} {148201} (\bibinfo {year} {2025}{\natexlab{a}})}\BibitemShut {NoStop}%
\bibitem [{\citenamefont {Bauermann}\ \emph {et~al.}(2025{\natexlab{b}})\citenamefont {Bauermann}, \citenamefont {Bartolucci}, \citenamefont {Boekhoven}, \citenamefont {Jülicher},\ and\ \citenamefont {Weber}}]{Bauermann.2025r8}%
  \BibitemOpen
  \bibfield  {author} {\bibinfo {author} {\bibfnamefont {Jonathan}\ \bibnamefont {Bauermann}}, \bibinfo {author} {\bibfnamefont {Giacomo}\ \bibnamefont {Bartolucci}}, \bibinfo {author} {\bibfnamefont {Job}\ \bibnamefont {Boekhoven}}, \bibinfo {author} {\bibfnamefont {Frank}\ \bibnamefont {Jülicher}}, \ and\ \bibinfo {author} {\bibfnamefont {Christoph~A.}\ \bibnamefont {Weber}},\ }\bibfield  {title} {\enquote {\bibinfo {title} {{Critical Transition between Intensive and Extensive Active Droplets}},}\ }\href {\doibase 10.1103/4nnd-tdky} {\bibfield  {journal} {\bibinfo  {journal} {Physical Review X}\ }\textbf {\bibinfo {volume} {15}},\ \bibinfo {pages} {041027} (\bibinfo {year} {2025}{\natexlab{b}})}\BibitemShut {NoStop}%
\bibitem [{\citenamefont {Fries}\ \emph {et~al.}(2025)\citenamefont {Fries}, \citenamefont {Diaz}, \citenamefont {Jardat}, \citenamefont {Pagonabarraga}, \citenamefont {Illien},\ and\ \citenamefont {Dahirel}}]{Fries.2025}%
  \BibitemOpen
  \bibfield  {author} {\bibinfo {author} {\bibfnamefont {Jacques}\ \bibnamefont {Fries}}, \bibinfo {author} {\bibfnamefont {Javier}\ \bibnamefont {Diaz}}, \bibinfo {author} {\bibfnamefont {Marie}\ \bibnamefont {Jardat}}, \bibinfo {author} {\bibfnamefont {Ignacio}\ \bibnamefont {Pagonabarraga}}, \bibinfo {author} {\bibfnamefont {Pierre}\ \bibnamefont {Illien}}, \ and\ \bibinfo {author} {\bibfnamefont {Vincent}\ \bibnamefont {Dahirel}},\ }\bibfield  {title} {\enquote {\bibinfo {title} {{Active droplets controlled by enzymatic reactions}},}\ }\href {\doibase 10.1098/rsif.2024.0803} {\bibfield  {journal} {\bibinfo  {journal} {Journal of the Royal Society Interface}\ }\textbf {\bibinfo {volume} {22}},\ \bibinfo {pages} {20240803} (\bibinfo {year} {2025})}\BibitemShut {NoStop}%
\bibitem [{\citenamefont {Alston}\ \emph {et~al.}(2022)\citenamefont {Alston}, \citenamefont {Parry}, \citenamefont {Voituriez},\ and\ \citenamefont {Bertrand}}]{Alston.2022}%
  \BibitemOpen
  \bibfield  {author} {\bibinfo {author} {\bibfnamefont {Henry}\ \bibnamefont {Alston}}, \bibinfo {author} {\bibfnamefont {Andrew~O.}\ \bibnamefont {Parry}}, \bibinfo {author} {\bibfnamefont {Raphaël}\ \bibnamefont {Voituriez}}, \ and\ \bibinfo {author} {\bibfnamefont {Thibault}\ \bibnamefont {Bertrand}},\ }\bibfield  {title} {\enquote {\bibinfo {title} {{Intermittent attractive interactions lead to microphase separation in nonmotile active matter}},}\ }\href {\doibase 10.1103/physreve.106.034603} {\bibfield  {journal} {\bibinfo  {journal} {Physical Review E}\ }\textbf {\bibinfo {volume} {106}},\ \bibinfo {pages} {034603} (\bibinfo {year} {2022})},\ \Eprint {http://arxiv.org/abs/2201.04091} {2201.04091} \BibitemShut {NoStop}%
\bibitem [{\citenamefont {Siggia}(1979)}]{Siggia.1979}%
  \BibitemOpen
  \bibfield  {author} {\bibinfo {author} {\bibfnamefont {Eric~D.}\ \bibnamefont {Siggia}},\ }\bibfield  {title} {\enquote {\bibinfo {title} {{Late stages of spinodal decomposition in binary mixtures}},}\ }\href {\doibase 10.1103/physreva.20.595} {\bibfield  {journal} {\bibinfo  {journal} {Physical Review A}\ }\textbf {\bibinfo {volume} {20}},\ \bibinfo {pages} {595--605} (\bibinfo {year} {1979})}\BibitemShut {NoStop}%
\bibitem [{\citenamefont {Fausti}\ \emph {et~al.}(2021)\citenamefont {Fausti}, \citenamefont {Tjhung}, \citenamefont {Cates},\ and\ \citenamefont {Nardini}}]{Fausti.2021}%
  \BibitemOpen
  \bibfield  {author} {\bibinfo {author} {\bibfnamefont {G.}~\bibnamefont {Fausti}}, \bibinfo {author} {\bibfnamefont {E.}~\bibnamefont {Tjhung}}, \bibinfo {author} {\bibfnamefont {M.~E.}\ \bibnamefont {Cates}}, \ and\ \bibinfo {author} {\bibfnamefont {C.}~\bibnamefont {Nardini}},\ }\bibfield  {title} {\enquote {\bibinfo {title} {{Capillary Interfacial Tension in Active Phase Separation}},}\ }\href {\doibase 10.1103/physrevlett.127.068001} {\bibfield  {journal} {\bibinfo  {journal} {Physical Review Letters}\ }\textbf {\bibinfo {volume} {127}},\ \bibinfo {pages} {068001} (\bibinfo {year} {2021})},\ \Eprint {http://arxiv.org/abs/2103.15563} {2103.15563} \BibitemShut {NoStop}%
\bibitem [{\citenamefont {Maire}\ \emph {et~al.}(2025)\citenamefont {Maire}, \citenamefont {Galliano}, \citenamefont {Plati},\ and\ \citenamefont {Berthier}}]{Maire.2025}%
  \BibitemOpen
  \bibfield  {author} {\bibinfo {author} {\bibfnamefont {Raphaël}\ \bibnamefont {Maire}}, \bibinfo {author} {\bibfnamefont {Leonardo}\ \bibnamefont {Galliano}}, \bibinfo {author} {\bibfnamefont {Andrea}\ \bibnamefont {Plati}}, \ and\ \bibinfo {author} {\bibfnamefont {Ludovic}\ \bibnamefont {Berthier}},\ }\bibfield  {title} {\enquote {\bibinfo {title} {{Hyperuniform Interfaces in Nonequilibrium Phase Coexistence}},}\ }\href {\doibase 10.1103/4b8v-4sbh} {\bibfield  {journal} {\bibinfo  {journal} {Physical Review Letters}\ }\textbf {\bibinfo {volume} {135}},\ \bibinfo {pages} {227102} (\bibinfo {year} {2025})},\ \Eprint {http://arxiv.org/abs/2507.03957} {2507.03957} \BibitemShut {NoStop}%
\bibitem [{\citenamefont {Cho}\ and\ \citenamefont {Jacobs}(2025)}]{Cho.2025}%
  \BibitemOpen
  \bibfield  {author} {\bibinfo {author} {\bibfnamefont {Yongick}\ \bibnamefont {Cho}}\ and\ \bibinfo {author} {\bibfnamefont {William~M.}\ \bibnamefont {Jacobs}},\ }\bibfield  {title} {\enquote {\bibinfo {title} {{Interfacial effects determine nonequilibrium phase behaviors in chemically driven fluids}},}\ }\href {\doibase 10.1073/pnas.2501145122} {\bibfield  {journal} {\bibinfo  {journal} {Proceedings of the National Academy of Sciences}\ }\textbf {\bibinfo {volume} {122}},\ \bibinfo {pages} {e2501145122} (\bibinfo {year} {2025})}\BibitemShut {NoStop}%
\bibitem [{\citenamefont {Bray}\ \emph {et~al.}(2001)\citenamefont {Bray}, \citenamefont {Cavagna},\ and\ \citenamefont {Travasso}}]{Bray.2001}%
  \BibitemOpen
  \bibfield  {author} {\bibinfo {author} {\bibfnamefont {Alan~J.}\ \bibnamefont {Bray}}, \bibinfo {author} {\bibfnamefont {Andrea}\ \bibnamefont {Cavagna}}, \ and\ \bibinfo {author} {\bibfnamefont {Rui D.~M.}\ \bibnamefont {Travasso}},\ }\bibfield  {title} {\enquote {\bibinfo {title} {{Interface fluctuations, Burgers equations, and coarsening under shear}},}\ }\href {\doibase 10.1103/physreve.65.016104} {\bibfield  {journal} {\bibinfo  {journal} {Physical Review E}\ }\textbf {\bibinfo {volume} {65}},\ \bibinfo {pages} {016104} (\bibinfo {year} {2001})},\ \Eprint {http://arxiv.org/abs/cond-mat/0106052} {cond-mat/0106052} \BibitemShut {NoStop}%
\bibitem [{\citenamefont {Solon}\ \emph {et~al.}(2018)\citenamefont {Solon}, \citenamefont {Stenhammar}, \citenamefont {Cates}, \citenamefont {Kafri},\ and\ \citenamefont {Tailleur}}]{Solon.2018}%
  \BibitemOpen
  \bibfield  {author} {\bibinfo {author} {\bibfnamefont {Alexandre~P.}\ \bibnamefont {Solon}}, \bibinfo {author} {\bibfnamefont {Joakim}\ \bibnamefont {Stenhammar}}, \bibinfo {author} {\bibfnamefont {Michael~E.}\ \bibnamefont {Cates}}, \bibinfo {author} {\bibfnamefont {Yariv}\ \bibnamefont {Kafri}}, \ and\ \bibinfo {author} {\bibfnamefont {Julien}\ \bibnamefont {Tailleur}},\ }\bibfield  {title} {\enquote {\bibinfo {title} {{Generalized thermodynamics of phase equilibria in scalar active matter}},}\ }\href {\doibase 10.1103/physreve.97.020602} {\bibfield  {journal} {\bibinfo  {journal} {Physical Review E}\ }\textbf {\bibinfo {volume} {97}},\ \bibinfo {pages} {020602} (\bibinfo {year} {2018})},\ \Eprint {http://arxiv.org/abs/1609.03483} {1609.03483} \BibitemShut {NoStop}%
\bibitem [{\citenamefont {Robinson}\ \emph {et~al.}(2025)\citenamefont {Robinson}, \citenamefont {Machon},\ and\ \citenamefont {Speck}}]{Robinson.2025}%
  \BibitemOpen
  \bibfield  {author} {\bibinfo {author} {\bibfnamefont {Joshua~F.}\ \bibnamefont {Robinson}}, \bibinfo {author} {\bibfnamefont {Thomas}\ \bibnamefont {Machon}}, \ and\ \bibinfo {author} {\bibfnamefont {Thomas}\ \bibnamefont {Speck}},\ }\bibfield  {title} {\enquote {\bibinfo {title} {{Universal limiting behavior of reaction-diffusion systems with conservation laws}},}\ }\href {\doibase 10.1103/1bdc-1bjb} {\bibfield  {journal} {\bibinfo  {journal} {Physical Review E}\ }\textbf {\bibinfo {volume} {111}},\ \bibinfo {pages} {065417} (\bibinfo {year} {2025})},\ \Eprint {http://arxiv.org/abs/2406.02409} {2406.02409} \BibitemShut {NoStop}%
\bibitem [{\citenamefont {Wittkowski}\ \emph {et~al.}(2014)\citenamefont {Wittkowski}, \citenamefont {Tiribocchi}, \citenamefont {Stenhammar}, \citenamefont {Allen}, \citenamefont {Marenduzzo},\ and\ \citenamefont {Cates}}]{Wittkowski.2014}%
  \BibitemOpen
  \bibfield  {author} {\bibinfo {author} {\bibfnamefont {Raphael}\ \bibnamefont {Wittkowski}}, \bibinfo {author} {\bibfnamefont {Adriano}\ \bibnamefont {Tiribocchi}}, \bibinfo {author} {\bibfnamefont {Joakim}\ \bibnamefont {Stenhammar}}, \bibinfo {author} {\bibfnamefont {Rosalind~J.}\ \bibnamefont {Allen}}, \bibinfo {author} {\bibfnamefont {Davide}\ \bibnamefont {Marenduzzo}}, \ and\ \bibinfo {author} {\bibfnamefont {Michael~E.}\ \bibnamefont {Cates}},\ }\bibfield  {title} {\enquote {\bibinfo {title} {{Scalar $\phi4$ field theory for active-particle phase separation}},}\ }\href {\doibase 10.1038/ncomms5351} {\bibfield  {journal} {\bibinfo  {journal} {Nature Communications}\ }\textbf {\bibinfo {volume} {5}},\ \bibinfo {pages} {4351} (\bibinfo {year} {2014})},\ \Eprint {http://arxiv.org/abs/1311.1256} {1311.1256} \BibitemShut {NoStop}%
\bibitem [{\citenamefont {AB}(2023)}]{AB.2023}%
  \BibitemOpen
  \bibfield  {author} {\bibinfo {author} {\bibfnamefont {COMSOL}\ \bibnamefont {AB}},\ }\bibfield  {title} {\enquote {\bibinfo {title} {{COMSOL Multiphysics® v. 6.1}},}\ }\href {www.comsol.com} {\bibfield  {journal} {\bibinfo  {journal} {Stockholm}\ } (\bibinfo {year} {2023})}\BibitemShut {NoStop}%
\bibitem [{\citenamefont {Inc.}(2024)}]{Inc..2024}%
  \BibitemOpen
  \bibfield  {author} {\bibinfo {author} {\bibfnamefont {Wolfram~Research}\ \bibnamefont {Inc.}},\ }\bibfield  {title} {\enquote {\bibinfo {title} {{Mathematica}},}\ }\href {https://www.wolfram.com/mathematica} {\bibfield  {journal} {\bibinfo  {journal} {Champaign, Illinois}\ } (\bibinfo {year} {2024})}\BibitemShut {NoStop}%
\bibitem [{\citenamefont {Raßhofer}\ and\ \citenamefont {Frey}(2025)}]{ZN2}%
  \BibitemOpen
  \bibfield  {author} {\bibinfo {author} {\bibfnamefont {Florian}\ \bibnamefont {Raßhofer}}\ and\ \bibinfo {author} {\bibfnamefont {Erwin}\ \bibnamefont {Frey}},\ }\bibfield  {title} {\enquote {\bibinfo {title} {{Local composition controls pattern formation in conserved active emulsions -- Additional material}},}\ }\href {\doibase 10.5281/zenodo.17790391} {\bibfield  {journal} {\bibinfo  {journal} {Zenodo}\ } (\bibinfo {year} {2025}),\ 10.5281/zenodo.17790391},\ \bibinfo {note} {digital repository containing Comsol simulation files and Mathematica notebooks used for linear stability analysis and the sharp-interface theory.}\BibitemShut {Stop}%
\end{thebibliography}
\end{document}